\documentclass[10pt,a4]{article}

\usepackage{geometry}
\usepackage{amsmath}
\usepackage{amssymb}
\usepackage{bm}                  % bold maths
\usepackage{physics}
\usepackage{graphicx}
\usepackage{mathtools} % loads amsmath and some very useful complements
\usepackage[colorlinks,linkcolor=black,citecolor=blue,urlcolor=blue]{hyperref}

\usepackage{verbatim}                   % used for commenting blocks of text
\usepackage{listings}
\usepackage[ruled,vlined]{algorithm2e}  % used for including algorithms
\usepackage{subcaption}                 % needed for using subfigures
\usepackage{tikz}                       % used for tree diagrams
\usepackage{tikz-qtree}
\usepackage{booktabs}       % professional-quality tables includes midrule etc.
\usepackage{fancyhdr}

\usepackage{cleveref}                   % must come after hyperref, see https://ctan.org/pkg/cleveref and https://www-uxsup.csx.cam.ac.uk/pub/tex-archive/macros/latex/contrib/cleveref/cleveref.pdf
\usepackage{appendix}                   % needed to get cross references to appendices labelled as appendices and not sections 

\usepackage{uniinput}
\usepackage[compat=0.8]{yquant}
\usetikzlibrary{fit,quotes,arrows,shapes.geometric}

\usepackage{ulem}

\yquantset{
   operators/every control box/.style={/yquant/operators/every rectangular box},
   operators/every control box/.append style={rounded corners=6pt}
}
\makeatletter
\yquant@langhelper@declare@command@uncontrolled%
   {controlbox}%
   \yquant@register@get@allowmultitrue%
   {%
      \expandafter\yquant@prepare@injection%
         \expandafter{\yquant@lang@attr@value}%
         {/yquant/operators/every control box}%
   }
\makeatother

\usepackage{nomencl}
\makenomenclature

\definecolor{codegreen}{rgb}{0,0.6,0}
\definecolor{codegray}{rgb}{0.5,0.5,0.5}
\definecolor{codepurple}{rgb}{0.58,0,0.82}
\definecolor{backcolour}{rgb}{0.95,0.95,0.92}

\lstdefinestyle{mystyle}{
    backgroundcolor=\color{backcolour},   
    commentstyle=\color{codegreen},
    keywordstyle=\color{magenta},
    numberstyle=\tiny\color{codegray},
    stringstyle=\color{codepurple},
    basicstyle=\ttfamily\footnotesize,
    breakatwhitespace=false,         
    breaklines=true,                 
    captionpos=b,                    
    keepspaces=true,                 
    numbers=left,                    
    numbersep=5pt,                  
    showspaces=false,                
    showstringspaces=false,
    showtabs=false,                  
    tabsize=2
}
\newif\ifnames
\namestrue   % set \namestrue to show names etc.

\fancypagestyle{firstpage}
{
   \fancyhf{}
   \chead{}
}

\graphicspath{{figures/}}

\usepackage{authblk}

\begin{document}

% Front page
% ----------
%\title{Sparse matrix block encoding for unstructured CFD meshes}
\title{Sparse block encoding with low subnormalisation for quantum linear solvers on unstructured meshes}

\author[1]{Leigh Lapworth\thanks{\href{mailto:leigh.lapworth@rolls-royce.com}{leigh.lapworth@rolls-royce.com}}\ \,}
\author[2]{Jarrett Smalley\ \,}
\author[3]{Christoph Sünderhauf\thanks{\href{mailto:christoph.sunderhauf@riverlane.com}{christoph.sunderhauf@riverlane.com}}\ \,}

\affil[1,2]{Rolls-Royce plc, P.O. Box 31, Derby, DE24 8BJ, UK}
\affil[3]{Riverlane, St. Andrews House, 59 St. Andrews Street, Cambridge CB2 3BZ, UK}

\renewcommand\Authands{ and }

\maketitle

% abstract
% --------
\begin{abstract}
Unstructured meshes are widely used in modelling and simulation in Computational Fluid Dynamics (CFD) due to their
geometric flexibility. However, they require a block encoding circuit describing the system, and efficiency is paramount. Here, we present a new block encoding methodology for 
unstructured mesh solvers based on decomposing the system matrix into a sum
of 1-sparse matrices.
This block encoding is engineered to reduce subnormalisation even in the absence of structure by hiding the contribution of large matrix elements behind others. We investigate three methods of 1-sparse decomposition.
Our circuit constructions  can optionally take advantage of limited matrix bandwidth and repeated values. For example we consider digitisation (reduced precision), where the number of multiplexed rotation gates is equal to the number of non-unique
matrix entries.

We evaluate our methodologies using 2D and 3D CFD matrices for unstructured
aerofoil meshes, structured cavity, and duct meshes, with and without
matrix preconditioning.
Our results show that we are able to achieve a block encoding with a 
near constant subnormalisation factor as the number of matrix preconditioning infill levels increases.
This significantly reduces the subnormalised condition number and, in one case, reduces the degree of the QSP polynomial approximation to $1/x$ by 100,000.
Reduced precision digitisation is shown to become more effective as the number of preconditioning 
infill levels is increased and, in one case, reduces the number of multiplexed rotations from 2,785,576 to
546 whilst maintaining an infidelity of $2.6\times10^{-5}$ relative to the classical
solution.
The effect of digitisation on QSVT solutions is demonstrated using 2D and 3D unstructured mesh aerofoil cases. Due to its flexibility and automatic applicability, we believe our 1-sparse decomposition block encoding can offer benefits to other quantum applications, too.
\end{abstract}

\thispagestyle{firstpage}

% Contents
% --------
\setcounter{tocdepth}{1}
%\tableofcontents
%\newpage

% sections
% --------

\section{Introduction}
\label{sec-intro}
Computational methods for engineering model the domain of interest
using a discrete set of points spanning the domain.
In a large class of methods, the points are connected to form a mesh.
Mesh types divide into structured, lattice type, meshes and unstructured
meshes. Structured meshes offer ease of programming and can achieve good
geometric flexibility via curvilinear coordinates and multi-block meshing.
Unstructured meshes offer the greatest flexibility and are widely used in
Finite Element Analysis (FEA) and general purpose Computational Fluid Dynamics (CFD) codes. It is unstructured meshes that are of interest in this work.

Many non-linear system solvers, such as CFD,
adopt a predictor-corrector approach, where the predictor step involves the
solution of a large system of linear equations. This is usually the most
compute intensive part of the solver.
The system is expressed as the matrix equation 
\begin{equation}
    Ax=b
    \label{eqn-Ax=b}
\end{equation}
where $A \in \mathbb{R}^{N \times N}$ and $x,b \in \mathbb{R}^{N}$.
Typically, $A$ has maximal row and column sparsities $s_r$ and $s_c$ that do not depend on $N$. 
For CFD, and other engineering applications, $A$ is rarely unitary and must,
therefore, be encoded as a block within a larger unitary matrix, $U_A$:
\begin{equation}
  U_A = 
  \begin{pmatrix}
    A/\alpha & * \\
    *   & *
  \end{pmatrix}
  \label{eqn-UA}
\end{equation}
where $\alpha\ge 1$ is the subnormalisation factor \cite{lin2022lecture} and the starred blocks depend on the technique used to
construct the unitary. The condition number, $\kappa_s$, of the encoded 
matrix is related to the original matrix by:
\begin{equation}
    \kappa_s = \frac{\alpha}{\lambda_{\min}}
    \label{eqn-kappa-s}
\end{equation}
where $\lambda_{\min}$ is the minimum singular value of $A$.

Whilst Quantum Linear Equation Solvers (QLES) offer complexity advantages
over classical solvers, these rely on assumptions that are often not valid in
engineering applications. 
The time complexity $O(\log(N)s^2\kappa^2/\epsilon)$ advantage of the
HHL algorithm \cite{harrow2009quantum} is lost for cases where $\kappa$ scales 
at least linearly with $N$, as is often the case.
The query complexity advantages, $O(\kappa\log (\kappa/\epsilon))$ of
Quantum Singular Value Transformation \cite{gilyen2019quantum, martyn2021grand,sunderhauf_matrix_2025} 
and $O(\kappa\log (1/\epsilon))$ of the adiabatic inspired methods 
(\cite{costa2022optimal,dalzell2024shortcut,jennings2025randomized})
exclude the complexity of the oracle which, except
in simple cases such as Toeplitz matrices, also scales with $N$. 
The new square root sparsity block encoding of \cite{bravo2026optimal, patel2026near} is a promising avenue for reducing oracle costs.

In classical linear algebra, it is common to precondition \Cref{eqn-Ax=b} by solving the system:
\begin{equation}
    PAx =Pb
    \label{eqn-PAx=Pb}
\end{equation}
where $\kappa(PA) < \kappa(A)$ and ideally $\kappa(PA) \ll \kappa(A)$.

From the above considerations, it is highly unlikely that an unpreconditioned
QLES will achieve a practical complexity advantage over a preconditioned
classical solver.
Preconditioned QLES methods have been reported by 
\cite{clader2013preconditioned, golden2022quantum,wan2018asymptotic, shao2018quantum, zhou2017efficient, hosaka2023preconditioning, tong2021fast}.
However, as demonstrated empirically in \cite{lapworth2025preconditioned} and proven theoretically by \cite{pechan2026block}, 
implementing QLES preconditioning via multiplication within a quantum circuit 
\cite{gilyen2019quantum, dalzell2023quantum}
does not, nor cannot, reduce $\kappa_s$.
If preconditioning is essential for QLES, it must be implemented as 
a classical pre-processing step.

The second requirement for a complexity advantage from oracle based QLES methods
is to find oracles that have polylogarithmic complexity in $N$.
Digital sampling techniques  \cite{babbush2018encoding,lemieux2024quantum,sanders2019black} offer a means
to reduce the cost of expensive multiplexed rotation gates but the scaling
with matrix rank is unclear. 
The related technique of using Walsh series approximations \cite{alonso2025quantum} has shown good accuracy in encoding the matrix entries using 64 Walsh coefficients.

Recent ``beyond $\kappa$" methods \cite{li2025new,dalzell2026faster} provide upper
bounds for an effective condition number, to be used in a QLES, that is 
lower than $\kappa_s$ by truncating the lowest eigenvalues.
Empirical analyses of truncating the smallest eigenvalues in the context of HHL
\cite{lapworth2022implicit} and of using $\kappa_{eff} < \kappa_s$ for generating QSP phase factors \cite{lapworth2026probability,goldfriend2026approximate} have shown that truncating the lowest eigenvalues can be effective for CFD but must be considered within the context of the outer non-linear iterations.
In this work, the empirically derived $\kappa_{\textit{eff}} \sim 0.8 \kappa_s$ is used.

Thus, desirable criteria for a block encoding are
\begin{itemize}
    \item Low gate and qubit count, and
    \item Low subnormalisation $α$
\end{itemize}
with an approximate trade-off expressed by the figure of merit ``gate count $\times$ subnormalisation'' \cite{sunderhauf2024block}. A further important, often overlooked desirable factor is
\begin{itemize}
    \item Low preprocessing cost.
\end{itemize}
In fact, classical preprocessing costs can be in direct competition to the other desirable factors. For example, methods relying on singular value decompositions can have costs equivalent to, or greater than, the classical solver.
Some preprocessing tasks require expert judgement, e.g. choosing from a range of encoding schemes, see next subsection, or setting the degree of the QSP polynomial. 
In industrial usage, these are usually calibration tasks that, once completed, can be applied to
an entire class of use cases. Classical methods are similarly calibrated.

\if 0
{\color{red}
In fact, classical preprocessing cost can be in direct competition to the other desirable factors. For example, in order to achieve the optimal theoretically possible subnormalisation $α=||A||_\text{spec}$, the spectral norm of the matrix must be precomputed classically, which can be similarly expensive to outright solving the problem classical.} 
{\color{blue} The preprocessing cost includes any expert time spent: Many schemes have been proposed that improve a block encoding by exploiting some structure present in the respective matrix families considered by the respective authors, see the next subsection for examples. Yet, experts manually inspecting each matrix is not a scalable proposition for preprocessing. {\bf I am not sure about this statement. The formulation of an encoding strategy is a one-off task. We are also arguing that our 1-sparse can be applied to any matrix.}}
\fi

In this paper, we show a new ``1-sparse decomposition'' block encoding. It is based on decomposing the matrix of interest into a sum of 1-sparse matrices, and an efficient circuit implementation thereof.
An advantage of our block encoding to some other schemes is that it can be performed completely automatically, reducing preprocessing cost as explained above.
We analyse its performance on a range of example matrices based on unstructured CFD meshes.
We use CFD matrices where the
main diagonal has been scaled to unity. This reduces their condition numbers and tends to equalise the off-diagonal entries. The latter is important for our analysis of digitisation, which lowers the precision of the matrix entries to reduce the number of unique entries.

% previous work
% -------------
\subsection{Previous work}
\label{subsec-intro-prev}

A number of techniques for block-encoding have sought to exploit the structure 
of the matrix being encoded
\cite{clader2013preconditioned,nguyen2022block,camps2024explicit, sunderhauf2024block,lapworth2025preconditioned}, such as Toeplitz, circulant,
binary tree and banded-diagonal matrices.
A trend in this work has been to find encodings where the subnormalisation factor, $\alpha$,
is low and, ideally, independent of the matrix dimension, $N$. For example, by addressing
the matrix using column and diagonal query oracles.

\begin{table*}[htb]
  \centering
  \begin{tabular}{c | c c c c c c c c}
    \toprule
    Block encoding & Anc. qubits& Multiplexed Rot gates & Subnorm \\
    \midrule
    Sparse access \cite{gilyen2018quantum} & 
     $n+3$ & $2Z$ & $\sqrt{s_rs_c}||A||_{\max}$\\
    Frobenius norm \cite{chakraborty_2019,clader2022quantum} & $n+1$ & $N+Z$ & $||A||_F$\\
    Q-norm \cite{clader2022quantum} &  $n+2$ & $4Z+P$ & $\mu_p(A)$\\
    Digitised Q-norm \cite{clader2022quantum} &  $n+b+2$ & $2(Z+U)+P$ & $\mu_p(A)$\\
    FABLE \cite{camps2022fable} & $n+1$ & $Z$ & $N$\\
    Coherent Perms \cite{setty2026block} & $z+1$ & $2Z$ & $\sum_{k=0}^{U-1} |\psi_k|$\\
    Dictionary \cite{yang2025dictionary} & $\mathcal{O}(n^2U)$ & $2U$ &
    $\sum_{k=0}^{U-1} |A_k|$\\
    \midrule
    \textit{This work} \\
    \Cref{subsec-index-ij} &  $n+s+1$& $2S+Z$&
    $\sum_{k=0}^{S-1} ||A_k||_{\max}$\\
    \Cref{subsec-diag-offset,subsec-diag-counter} &  $l+s+1$& $2S+Z$&
    $\sum_{k=0}^{S-1} ||A_k||_{\max}$\\
    \Cref{subsec_list_rot} &  $l+s+b+1$& $2S+U$ &
    $\sum_{k=0}^{S-1} ||A_k||_{\max}$\\
    \bottomrule
  \end{tabular}
  \caption{\centering Comparison of methods for block encoding a matrix of size $N\times N$ with $Z$ non-zero entries of which there are $U\le Z$ unique values and maximal row and column sparsities $s_r$ and $s_c$.
  $S$ is the number of matrices in a 1-sparse decomposition.
  $n = \lceil log_2N \rceil$,  $s=\lceil log_2S \rceil$, 
  $z=\lceil log_2 Z \rceil$,and
  $l=\lceil log_2L \rceil$, where $L$ is the length of an auxiliary list, e.g. diagonal offsets.
  $b$ is the number of bits used to digitise rotation angles. For Q-norm encoding
  $\mu_p(A)$ is defined in \Cref{eqn-pnorm} and $P$ is the number of additional rotations arising from permutation operations in the row and column state loaders, see \Cref{subsubsec-qnorm}.
  }
  \label{tab-be-compare}
\end{table*}

Previous work on general purpose matrix block encoding is compared with our 1-sparse encoding in \Cref{tab-be-compare}.
Later in the manuscript, we consider both how previous methods scale with matrix size and with the number of infill levels
used in the preconditioner. 
Here, circuit depth and the number of ancilla qubits are based on the logical circuit. In this section, we use the
notation $Z$ for the number of non-zeros in the matrix and $U$ for the number of
unique values in the matrix.

To provide concrete values for comparison, \Cref{tab-be-unstr} gives the subnormalisation factors
for the 2D and 3D unstructured mesh cases used in the digitised QSVT evaluations
(\Cref{subsubsec-2d-qsvt,subsubsec-3d-qsvt}).
Our 1-sparse and the q-norm encoding give significantly lower subnormalisation factors than
the other methods, with q-norm being the lowest. Below, the 1-sparse and the q-norm 
encodings are compared in greater detail than the other methods.

\begin{table*}[htb]
  \centering
  \begin{tabular}{c | c c c c c c c c}
    \toprule
    Case & Sparse & Frobenius & Q-norm & FABLE & Coherent & Dictionary & 1-sparse \\
         & access & norm & $(p=0.5)$& & Perms & &  \\
    \midrule
    AFT-1022(4) & 26.0 & 33.4 & 2.48 & 1,024 & 666.4   & 19.8  & 3.76\\
    AFH-2040(0) & 20.0 & 47.9 & 2.33 & 2,048 & 1,693.1 & 164.4 & 3.90\\
    \bottomrule
  \end{tabular}
  \caption{\centering Comparison of subnormalisation factors for the digitised unstructured mesh
  cases AFT-1022 (\Cref{subsubsec-2d-qsvt}) and AFH-2040 (\Cref{subsubsec-3d-qsvt}).
  Both use the SPAI preconditioner and the number in brackets is the number of infill levels.}
  \label{tab-be-unstr}
\end{table*}

\subsubsection{Sparse access}
The sparse access model \cite{gilyen2018quantum} requires three oracles for the row and column 
indices and for the matrix values. This leads to $\alpha=\sqrt{s_rs_c}||A||_{\max}$ which is independent of the matrix dimension.
For 2D triangular meshes, the pressure correction matrix has row and column sparsities of 4.
However, these increase significantly when preconditioning is used as shown in 
\Cref{tab-be-unstr} where 4 levels of infill have been used in the 2D case.
The sparse access model can be implemented with $b$ bit precision, reducing the number of
rotation gates to $U$ and increasing the number of ancilla qubits to $n+b+3$.
The number of rotation gates is lower than our equivalent $2S+U$, but given that $S \ll U$, this does not offset the large difference in subnormalisation factors.

\subsubsection{Frobenius and q-norm}
\label{subsubsec-qnorm}
Encoding based on a multiplexed state preparation unitary
\cite{chakraborty_2019,clader2022quantum} has $\alpha=||A||_F$, the Frobenius norm.
For a CFD matrix with unit diagonal, this leads to $\alpha\ge \sqrt{N}$.

An extension of this encoding is the \textit{q-norm} model \cite{clader2022quantum} using 2 multiplexed state preparation unitaries, which gives
$\alpha=\mu_p(A)$ where
\begin{equation}
   \mu_p(A) = \sqrt{S_{2p}(A)S_{2(1-p)}(A^T)}
   \label{eqn-pnorm}
\end{equation}
and $S_q=\max_i \sum_j |a_{ij}|^{q}$ with $0\le q\le 2$. For CFD matrices, where the diagonal entry is the negative row-wise sum of the off-diagonal entries, this can result in a low subnormalisation factor. This encoding is less affected by preconditioning infill levels as $PA$ becomes a better approximation to the unit matrix.

The q-norm encoding is the closest contender to the 1-sparse decomposition; as such detailed comparisons of the two approaches are in order. Here, we focus on comparing rotation gate counts and subnormalisation factors for various practical example use-cases,
including an implementation on one of our cases.
\Cref{app-multiplex-ssp} gives a theoretical comparison of the two encoding methods' circuits.

\begin{figure}[ht]
  \centering
  \begin{subfigure}[b]{0.49\textwidth}
    \centering
    \includegraphics[width=.9\textwidth]{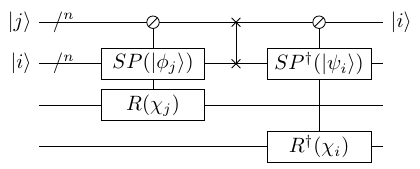}
    \vspace*{0.4in}
    \captionsetup{justification=centering}
    \caption{Circuit diagram for $p=1/2$.}
  \label{fig:clader-qnorm}
  \end{subfigure}
  \begin{subfigure}[b]{0.49\textwidth}
      \centering
      \includegraphics[width=0.99\textwidth]{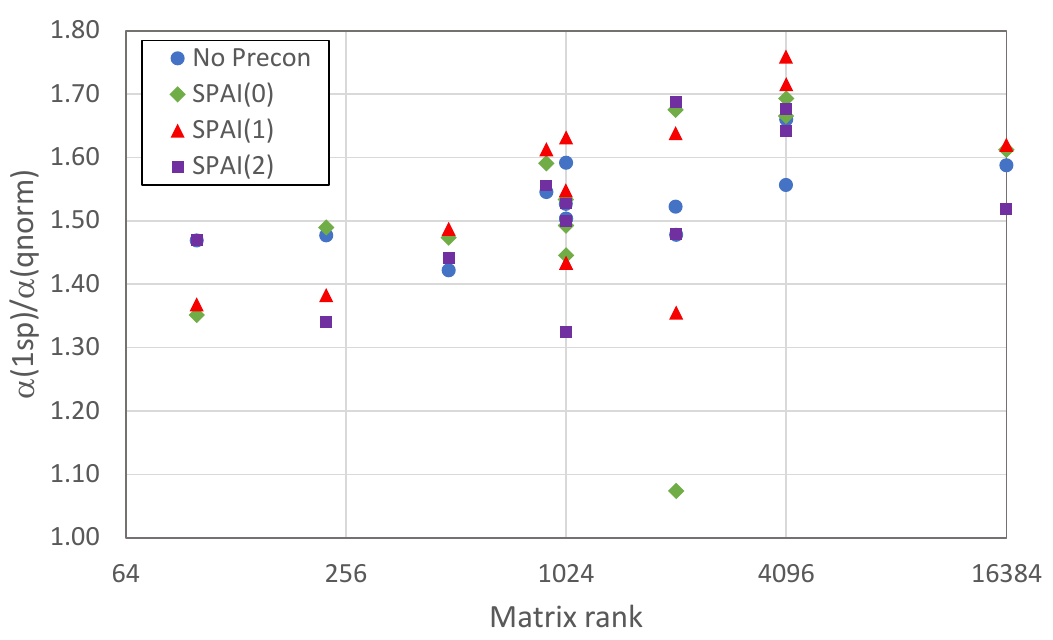}
      \caption{\small Ratio of subnormalisation factors: 1-sparse/q-norm.}
      \label{fig:qnorm ratio}
  \end{subfigure}
  \caption{\centering Q-norm block encoding \cite{clader2022quantum}.
     $SP(\ket{\phi_j})$ and $SP(\ket{\psi_i})$ are state loaders for the columns
    and rows of the matrix respectively.
    The rotations by angles $\chi_i$ and $\chi_j$ apply row and column 
    normalisation factors.}
  \label{fig_qnorm}
\end{figure}

In q-norm encoding, the matrix is loaded twice, each time with additional rotations by angles $\chi$ for each row and column, see \Cref{fig:clader-qnorm}. However, the state loader 
\cite{mottonen2004transformation}
for each row and column is a circuit of controlled rotations each of which incurs an added $\chi$ rotation.
In principle this leads to $2Z$ rotation gates for state loading and $2Z$ for the
$\chi$ rotations of which there is a maximum of $2N$ unique angles. In practice, the state loader contains a significant number of additional rotations by $\pi$.

For sparse states, the loader can be considered as two parts: load the amplitudes and permute them to the correct position. The $\pi$ rotations (i.e. \textsc{NOT} gates) are the permutation operators.
Using the state loader of \cite{mottonen2004transformation} the $\pi$ rotations arise naturally through the binary tree decomposition \cite{araujo2021divide}. 
These are equivalent to the permutations in the sparse state loading method \cite{rupprecht2026sparse} discussed in \Cref{subapp-ssp}.
The permutation operations are part of the multiplexed controls acting on the $\chi$ rotations and their number is denoted as $P$ in \Cref{tab-be-compare}. 

To give a concrete example, we analyse the AFQ-100 test case (\Cref{fig:naca_0012_crse3_quad}), where there are $Z=480$ non-zeros in the matrix. Using the optimal q-norm value $p=0.5$, the total
number of rotations for state loading is 2,751 of which 1,791 are rotations by $\pi$.
Adding the 960 non-$\pi$ state loading rotations to the 2,751 $\chi$ rotations gives a
total number of non-$\pi$ rotations of 3,711 $\simeq 8Z$, double the expected value of $4Z$ and, hence, $P \simeq 4Z$ in this case.
However, since there are only $2N$ unique $\chi$ angles digitising them in the manner of
\Cref{subsec_list_rot}, reduces the total number of rotation
gates to $2(Z+U)+P$. Still much larger than the $2S+U$ of the 1-sparse method.

A further consideration is the use of techniques to reduce the number of unique values in the
matrix, which we investigate in \Cref{subsec_digit_smooth,subsec_qsvt}. In 1-sparse encoding there is a 1:1 correspondence between the reduction in unique values and rotation gates.
For state preparation, this is not the case and the reduction in rotation gates may only be 
$1/10^{th}$ of the reduction in unique values \cite{lapworth2025preconditioned}.

\Cref{fig:qnorm ratio} shows the ratio of the 1-sparse to q-norm subnormalisation factors for all the test cases used here.
Apart from one outlier, the ratio ranges from 1.32 to 1.76, showing that q-norm consistently gives 
the lowest values.
However, the reduction in QSVT phase factors that q-norm gains from its lower $\alpha$ is more than
offset by the double loading of the matrix and the high number of multiplexed rotations by $\pi$.

\subsubsection{Diffusion and Prep-Select}
Block encoding based on pre and post application of a diffusion operator  \cite{lin2022lecture} formed the basis of the FABLE method \cite{camps2022fable}
where the multiplexed rotation gates are permuted and applied, without controls, using Gray code
ordering. Both methods have $s=N$ which makes them inappropriate for large matrices.

An extension of diffusion operators is to use \textsc{Prep-Select} encoding as used in some of
the previously mentioned structured mesh block encoding techniques, using the \textsc{Prep} operator to load the matrix entries as ordered state amplitudes.
The coherent permutations method \cite{setty2026block} loads entries as diagonals and then offsets elements to their correct position and deletes unneeded entries.
Isolated entries are directly inserted in their correct positions.
The subnormalisation factor is the sum of the unique amplitudes along each
diagonal and any directly inserted entries.

An alternative is the dictionary based approach \cite{yang2025dictionary} designed for matrices with a small number, $U$, of unique entries, $A_k$, that are repeated many times. For each unique entry, the dictionary lists their
positions in the matrix. The \textsc{Prep} operator loads each unique value,
leading to $\alpha=\sum_{k=0}^{U-1}|A_k|$. By definition our 1-sparse encoding is always lower than this as our $\alpha$ is calculated from a subset of these values, see \Cref{tab-be-unstr}.

With regard to the digitisation used in \Cref{tab-be-unstr}, the
dictionary approach benefits directly from the reduction in the number of
unique values. For the coherent permutations method, the benefit is far less due to the large number of short diagonals in the unstructured matrix.
The modified method of \cite{setty2026block} which encodes a diagonal's maximal element, and the  corresponding difference for the remaining element may benefit diagonals with similar valued entries and reduce subnormalisation.
This is bounded below by the sum of the maximum entries along each diagonal,
which is 47.3 and 109.6 for the two cases in \Cref{tab-be-unstr}.
Without digitisation or using the modification, both the dictionary and coherent permutation methods have subnormalisation factors of the same order as FABLE.

The main contribution of the \textsc{Prep-Select} methods \cite{setty2026block,yang2025dictionary}
is their low circuit depth implementations.
Focussing on the permutation operations to move state loaded entries to their correct
position in the matrix, \cite{setty2026block} provides a coherent permutation methodology that
replaces a sequence of permutations with a single set of multiplexed \textsc{NOT} gates.
Focussing on the column oracle, \cite{yang2025dictionary} used the 
sparse-access input model \cite{zhang2024circuit} to create low circuit depth oracles.
Effectively, this swaps the depth and width of the circuit.

Whilst not the focus of our work, the nature of these low-circuit depth methods is such that they can be implemented
within our 1-sparse encoding. This is particularly so for the digitised version in \Cref{fig:sparse-count-rot-gen} where the majority of multiplexed controls involve only NOT gates as used in \cite{yang2025dictionary}. The number of multiplexed rotation gates, $U$, is far lower than the $Z$ multiplexed NOT gates. The multiplexed permutation operators are also amenable to the method of \cite{setty2026block}.

\subsubsection{1-sparse decomposition}
Our method requires a decomposition into a linear combination of 1-sparse matrices.
An efficient decomposition of a Hamiltonian with column sparsity of $s_c$ into a sum of 1-sparse Hamiltonians was
developed by \cite{berry2007efficient} using the colouring scheme of \cite{ahokas2004improved}.
The decomposition required at most $6s_c^2$ 1-sparse matrices and has been implemented within an
HHL solver for electromagnetic scattering \cite[Supplementary Material]{clader2013preconditioned}.
An improved approach for bipartite matrices \cite{kothari2014efficient,berry2014exponential} reduced
the maximum number of 1-sparse matrices to $s_c^2$.
The encoding of Toeplitz matrices \cite{sunderhauf2024block} and banded diagonal matrices
\cite{lapworth2025preconditioned} can be interpreted as a decomposition into $\mathcal{O}(s_c)$ 1-sparse 
matrices.

% our work
% --------
\subsection{Our contribution}
\label{subsec-intro-us}
We present a method for block encoding arbitrary matrices using 
decomposition into $S$ 1-sparse matrices where the subnormalisation factor is $\alpha=\sum_{k=0}^{S-1} ||A_k||_{\max}$. The theoretical developments are supported by detailed practical evaluations, where we show that preconditioning and digitisation lead to time-efficient circuits.

\subsubsection{Complexity}
In terms of the number of rotation gates our method uses $S$ rotations each for the \textsc{Prep} and \textsc{Unprep} operators and $Z$, the number of non-zero entries,
for data loading. If the $Z$ non-zero entries can be approximated by $U \ll Z$ unique entries, the total number of rotation gates is $2S+U$ where, typically, $S \ll U$.
As shown in \Cref{subsubsec-qnorm,sec:comparison}, the 1-sparse method gives the lowest subnormalisation factors apart from q-norm 
encoding \cite{clader2022quantum} but has shorter circuits and, hence, the lowest time-complexity for QLES applications.

\subsubsection{1-sparse construction}
We provide three methods for constructing the 1-sparse matrices, which must be performed in classical preprocessing.
The most effective demonstrates that $α$ is largely independent of the number
of the 1-sparse matrices for a range of CFD test matrices (see also \Cref{sec-spai-front}).
We also demonstrate that the number of 1-sparse matrices has $\mathcal{O}(s_r)$ scaling for a range of matrices for structured and unstructured CFD meshes. 

\subsubsection{Circuit construction}
From an initial formulation using an oracle based on addressing the matrix entries by their row and column indices \cite{lin2022lecture}, we provide 
three refinements to improve the method.
The first is to address the matrix entries by their offset from the main
diagonal which can be seen as a generalisation of the authors' 
banded diagonal block encoding for structured meshes \cite{lapworth2025preconditioned}.
We couple this with classical bandwidth reduction preprocessing to reduce the
number of encoding qubits.
The second refinement is to list the offsets rather than address them directly.
Whilst the list could be considered as a simplified dictionary, the entries
can be addressed by their bit patterns avoiding the overhead of a dictionary.
The third refinement is to, additionally, use a list of non-unique rotation angles supporting the use of digitised rotations \cite{gilyen2018quantum, clader2022quantum}.

\subsubsection{Preconditioning}
\label{subsubsec-our-precon}
We show that for a range of unstructured and structured mesh testcases,
the 1-sparse encoding subnormalisation factor is largely independent of the number of infill levels used in the preconditioner. 
This is not the case for banded diagonal encoding where the benefit of more infill levels is eroded by an accompanying increase in subnormalisation.
This is an important result if preconditioned quantum linear solvers are to beat
preconditioned classical ones.

\subsubsection{Reduced precision digitisation}
\label{subsubsec-our-digit}
Using a discriminator $L_2(x -\hat{x}) \lesssim 1\times 10^{-2}$, where $x$ is
the classical solution of the original linear system and $\hat{x}$ is the solution with the modified matrix, we compare our previously published binning technique \cite{lapworth2025preconditioned} with a simple digitisation to reduce the bitwise precision of the matrix entries.
We show that digitisation becomes more effective, i.e. fewer unique rotation angles,
with more preconditioning infill levels.
Coupled with near constant subnormalisation, this is a major finding for the
time complexity of preconditioned quantum linear equation solvers.

\subsubsection{Practical examples}
Our main body of work is to demonstrate our 1-sparse block-encoding on a range of
practical use cases.
This includes new unstructured mesh test cases for 2D and 3D aerofoils. 
As discussed in the introduction, central to these is the use of preconditioning
and, in particular, the SPAI preconditioner \cite{benson1984parallel,chow1998approximate}.
The practical analysis leads to the findings given in 
\Cref{subsubsec-our-precon,subsubsec-our-digit}. 
These provide concrete examples of the scope for digitised rotation angles in
practical applications.
We show that reductions in the number of
non-unique angles by up to a factor of 5,000 are achievable by digitisation.

\subsubsection{QSVT}
Finally we demonstrate, in emulation, the first QSVT calculations for 2D and 3D aerofoils with unstructured meshes. 
In each case, we compute QSVT solutions using the original matrix and, separately,
a digitised matrix that meets the above $L_2$ criterion.
Equivalent error norms can be achieved with 9 and 12 bit digitisation in the 2D and 3D cases respectively.
We believe these are the first practical evaluations
of the use of digitised angles within a quantum linear equation solver.

\subsubsection{Layout of manuscript}
The following sections are organised as follows.

\Cref{sec-theory} develops the 1-sparse decomposition block-encoding and 1-sparse decomposition methodologies. 
 \Cref{app-multiplex-ssp} presents a more theoretical discussion of the same encoding.
 \Cref{sec-app-deriv} gives detailed derivations of the generalised versions of the encodings.

\Cref{sec-testcases} introduces the test case matrices including a new set of 2D and 3D unstructured mesh cases for the flow around an aerofoil.
    \Cref{subsec-rcm} demonstrates the use of classical bandwidth reduction for
    the unstructured mesh matrices.
    \Cref{subsec-precon} demonstrates the use of the SPAI preconditioner on the 
    2D unstructured meshes.

\Cref{sec_results} presents the results for 2D and 3D unstructured
    aerofoil cases without preconditioning and with SPAI preconditioning with 0,1 and 2 levels of infill. 
    The unstructured encoding is also compared with the banded-diagonal encoding for the structured meshes.
    \Cref{subsec_digit_smooth,subsec_nqubits} analyse the use of digitised rotation angles.
    \Cref{subsec_qsvt} presents the QSVT results for 2D and 3D aerofoil cases.
    \Cref{sec-spai-front} provides a theoretical justification for the behaviour of the SPAI preconditioner observed in our results.

\Cref{sec-conclusions} discusses the results and draws conclusions.

% theory
% ------
\section{Theory and circuit construction}
\label{sec-theory}

In this section, we introduce the unstructured mesh block encoding through a set of $4 \times 4$
test matrices that are used to illustrate the successive refinements we have made.

A more theoretical exposition of the block encodings based on Toffoli counts is given in \Cref{app-multiplex-ssp}. We also provide explicit circuits and detailed derivations of the general circuits in \Cref{sec-app-deriv}.

The principle of the unstructured mesh encoding is to express the $N\times N$ matrix $A$ as 
a sum of $S$ 1-sparse matrices:

\begin{equation}
    A = \sum_{k=0}^{S-1} A_k
    \label{eqn_1sp_sum}
\end{equation}

This leads to a subnormalisation factor, $\alpha$, where:
\begin{equation}
    \alpha = \sum_{k=0}^{S-1} ||A_k||_{\max}
    \label{eqn_1sp_alpha}
\end{equation}

This is well suited to CFD meshes where the sparsity is controlled by the mesh 
topology and tends to be relatively consistent throughout the mesh, leading to 
consistent row and column sparsities throughout the matrix.

The set of 1-sparse matrices are loaded as a Linear Combination of Unitaries
(LCU) where the \textsc{Prepare} operator, $P$ loads:
\begin{equation}
    P\ket{0} = \frac{1}{\sqrt{\alpha}}\sum_{k=0}^{S-1} \sqrt{||A_k||_{\max}} \ket{k}
    \label{eqn_1sp_prep}
\end{equation}

There is some arbitrariness in how the set of 1-sparse matrices are constructed
leading to different values for $\alpha$ and $S$. The sparsity pattern also
leads to different choices in constructing the \textsc{Select} operator
for the LCU. These are discussed in the remainder of this section.

% row-column indexing
% -------------------
\subsection{Encoding based on row-column indexing}
\label{subsec-index-ij}

\begin{figure}[htb]
  \centering
  \includegraphics[width=.9\textwidth]{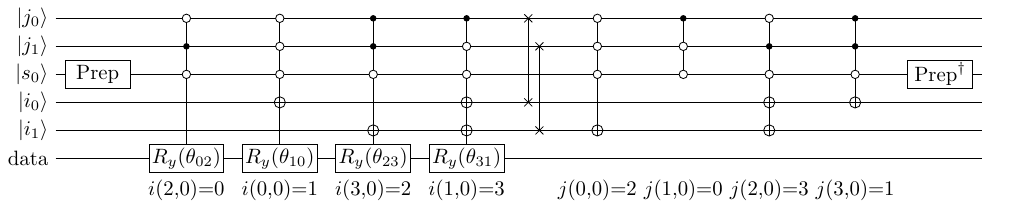}
  \captionsetup{justification=centering}
  \caption{Circuit for loading 4x4 1-sparse matrix in \Cref{eqn-4x4-asym} using unstructured encoding. The function $i(j,k)$ returns the
  row index for the entry in $j^{th}$ column of the $k^{th}$ 1-sparse
  matrix. $j(i,k)$ is the inverse function. In this example, $S=1$ and the $s$ register is retained for completeness.}
  \label{fig:sparse-ij-4x4}
\end{figure}

We illustrate the main features of the unstructured mesh encoding 
using the $4 \times 4$ 1-sparse matrix with an asymmetric sparsity pattern:

\begin{equation}
  \begin{pmatrix}
    0 & 0 & a & 0\\
    b & 0 & 0 & 0\\
    0 & 0 & 0 & c \\
    0 & d & 0 & 0
  \end{pmatrix}
  \label{eqn-4x4-asym}
\end{equation}
where $a,b,c,d$ are arbitrary entries with absolute values $\le 1$. 
This matrix can be block-encoded using the circuit shown in \Cref{fig:sparse-ij-4x4}.

There are four registers in \Cref{fig:sparse-ij-4x4}:
\begin{itemize}
    \item \textbf{$j$-register (system register)}. For the compute step, the $j$ register contains the bit patterns
    for each column of the 1-sparse matrices.
    For the uncompute step, this is flipped and it contains the bit patterns 
    for each row.
    \item \textbf{$s$-register (flag register)}. This contains the bit pattern for $k$, the 
    index for each 1-sparse matrix. In this example, there is only one 
    1-sparse matrix and this register could be omitted, but has been
    retained to show the overall structure of the encoding.
    \item \textbf{$i$-register (ancilla register)}. For the compute step, the $i$ register 
    implements the oracle $i(j,k)$ which gives the row index for the $j^{th}$ 
    column of the $k^{th}$ 1-sparse matrix. 
    For the uncompute step, it implements the oracle $j(i,k)$ which gives the 
    column index for the $i^{th}$ row of the $k^{th}$ 1-sparse matrix.
    \item \textbf{Data qubit (flag qubit)}. The data qubit applies the rotation
    gates based on the values $\theta_{ij} = 2\cos^{-1}(a^{(k)}_{ij})$ for the
    $(i,j)^{th}$ entry in the $k^{th}$ 1-sparse matrix.
\end{itemize}

Note that the uncompute step for $j(1,0)=0$ is shown for completeness and
can be omitted as it has no effect.

The generalised version of this circuit is described in \Cref{subsec-app-rowcol} where
it can be seen that the $i$ and $j$ registers are of size $\lceil \log_2 N\rceil$,
the $s$ register is of size $\lceil\log_2 S \rceil$.

% diag offset
% -----------
\subsection{Encoding based on offset from the diagonal}
\label{subsec-diag-offset}
\begin{figure}[htb]
  \centering
  \includegraphics[width=.99\textwidth]{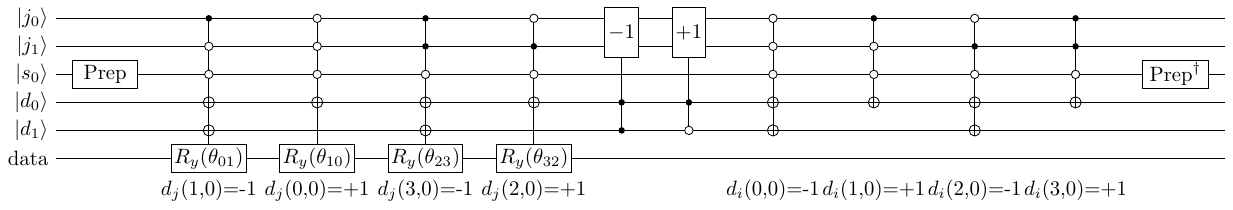}
  \captionsetup{justification=centering}
  \caption{Circuit for loading 4x4 1-sparse matrix in \Cref{eqn-4x4-asym2} using unstructured encoding with diagonal offsets. 
  The function $d_j(j,k)$ returns the
  diagonal offset for the entry in $j^{th}$ column of the $k^{th}$ 1-sparse
  matrix.
  $d_i(i,k)$ returns the diagonal offset for the entry in $i^{th}$ row of the $k^{th}$ 1-sparse matrix.}
  \label{fig:sparse-diag-4x4}
\end{figure}

As will be discussed in relation to \Cref{fig_rcm_afq_489} there are
two distinct types of unstructured matrix.
On the left in \Cref{fig_rcm_afq_489} is the \textit{raw} unstructured
matrix where encoding based on $(i,j)$ index is most effective.
On the right in \Cref{fig_rcm_afq_489} the matrix has been permuted to have a much lower
bandwidth where encoding based on offset from the diagonal is most
effective. 
The corresponding circuit is illustrated in \Cref{fig:sparse-diag-4x4} for the matrix:
\begin{equation}
  \begin{pmatrix}
    0 & a & 0 & 0\\
    b & 0 & 0 & 0\\
    0 & 0 & 0 & c \\
    0 & 0 & d & 0
  \end{pmatrix}
  \label{eqn-4x4-asym2}
\end{equation}

The $i$-ancilla register and oracle in the previous encoding are replaced
by a diagonal, $d$, register and oracles $d_j$ and $d_i$ that give
the diagonal offsets for the $j^{th}$ column and $i^{th}$ row
respectively.
The diagonal offsets are signed, with negative values corresponding to
super-diagonals and positive values to sub-diagonals.
Negative values are encoded using two's complement.

Although the circuits in \Cref{fig:sparse-ij-4x4} and 
\Cref{fig:sparse-diag-4x4} contain the same numbers of qubits, it can
be seen that for larger matrices with only $-1$ and $+1$ diagonals
the size of the diagonal register remains constant.
The generalised version of this circuit is described in \Cref{subsec-app-diagcol}

% diag list
% -----------
\subsection{Encoding based on listing offsets}
\label{subsec-diag-counter}

\begin{figure}[htb]
  \centering
  \includegraphics[width=.99\textwidth]{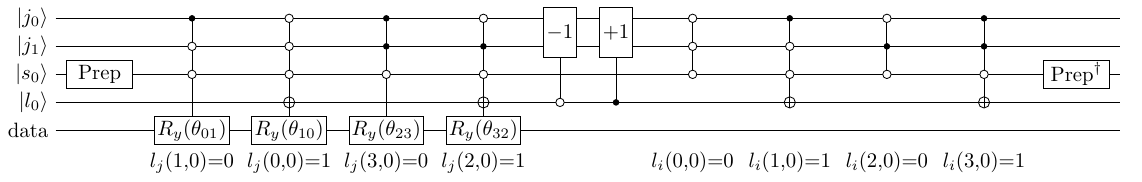}
  \captionsetup{justification=centering}
  \caption{Circuit for loading 4x4 1-sparse matrix in \Cref{eqn-4x4-asym2} using a list of diagonal offsets. 
  $l_j(j,k)$ returns the
  position in the list of offsets for the entry in $j^{th}$ column of the $k^{th}$ 1-sparse
  matrix.
  $l_i(i,k)$ returns the position in the list of offsets for the entry in $i^{th}$ row of the $k^{th}$ 1-sparse matrix.}
  \label{fig:sparse-count-4x4}
\end{figure}

A feature of the encoding in \Cref{fig:sparse-diag-4x4} is that the
diagonal register requires 2 qubits, even though there are only two
distinct diagonals.
The size of the diagonal register was determined by the range of
the offsets not the number of offsets.
This leads to a refinement of the encoding based on addressing the
offsets by their position in a list of offsets.
For the matrix in \Cref{eqn-4x4-asym2}, the list $l$ contains two
offsets: $\{-1, 1\}$.
The circuit for implementing this list is shown in \Cref{fig:sparse-count-4x4}.

The generalised version of this circuit is described in \Cref{subsec-app-listcol}.
Encoding via a list of offsets can be seen as the generalisation of 
the diagonal encoding method for structured 
meshes matrices \cite{lapworth2025preconditioned}.
In practice, the offset method works best for structured meshes
where the size of the $s$ and $l$ registers remain constant as the size
of the mesh is increased.
For unstructured mesh matrices of the type on the right in \Cref{fig_rcm_afq_489},
the list of offsets often contains most/all of the values between the lower and
upper bandwidth values and the two diagonal encoding methods are equivalent.

%
% listing rotation angles
% -----------------------
\subsection{Encoding based on listing rotation angles}
\label{subsec_list_rot}

\begin{figure}[htb]
  \centering
  \includegraphics[width=.99\textwidth]{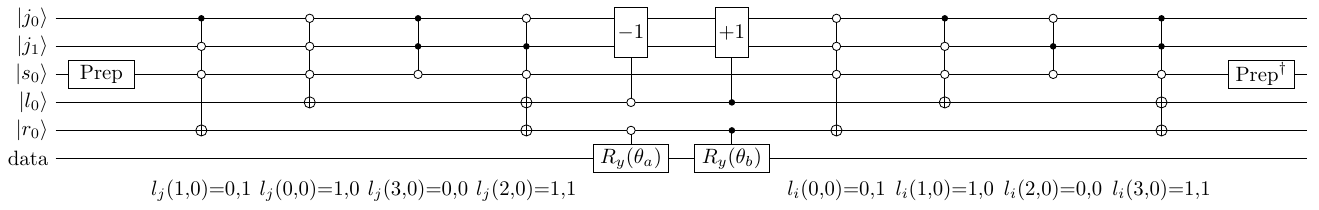}
  \captionsetup{justification=centering}
  \caption{Circuit for loading 4x4 1-sparse matrix in \Cref{eqn-4x4-list-ry} using a list of diagonal offsets and a list of rotation angles. The $l_i$ and $l_j$ operators are as above.
  The bit patterns in the $r$ register gives the  position in the list of
  rotation angles.}
  \label{fig:sparse-count-rot-4x4}
\end{figure}

The model of listing offsets can be extended to the angles of the multiplexed rotations.
This is of benefit when the matrix has a number of repeated entries, as in \Cref{eqn-4x4-list-ry} and shown in \Cref{fig:sparse-count-rot-4x4}.

\begin{equation}
  \begin{pmatrix}
    0 & b & 0 & 0\\
    a & 0 & 0 & 0\\
    0 & 0 & 0 & a \\
    0 & 0 & b & 0
  \end{pmatrix}
  \label{eqn-4x4-list-ry}
\end{equation}

Here, an additional register, $r$ of size $\lceil \log_2 R\rceil$ where $R$ is the number of
distinct rotation angles,
is used
to list the unique rotation angles, with each angle implemented only once.

\begin{figure}[htb]
  \centering
  \includegraphics[width=.7\textwidth]{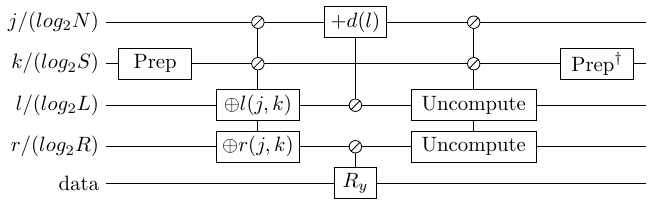}
  \captionsetup{justification=centering}
  \caption{Generalisation of \Cref{fig:sparse-count-rot-4x4}. $j$ is the column register, $k$ is 1-sparse register, $l$ is the diagonal offset register and $r$ is the rotation angle register. The
  matrix has $N$ rows, $S$ 1-sparse matrices with $L$ diagonal offsets and $R$ unique rotation angles. Note that uncomputation can use measurement based procedures for speed up, see \Cref{app-multiplex-ssp}.}
  \label{fig:sparse-count-rot-gen}
\end{figure}

\Cref{fig:sparse-count-rot-gen} shows a general circuit for block encoding a matrix with
$N$ rows, $S$ 1-sparse matrices and lists of $L$ diagonal offsets and $R$ unique rotation angles.
Implementing rotations via a list can be seen to be of the same family as the
alias sampling \cite{babbush2018encoding}, 
rejection sampling \cite{lemieux2024quantum,sanders2019black} and phase gradient \cite{o2025quantum} approaches.
These methods reduce the T-gate complexity by digitising the rotation angles using a prescribed number of bits.
Listing the rotations gives more flexibility, as they give savings for repeated values. This way, our block encoding encompasses savings from repeated values as seen in the dictionary block encoding \cite{yang2025dictionary} or in \cite{setty2026block}. 
Angle \textit{binning} as used in \cite{lapworth2025preconditioned} and reducing the digital precision can also
be implemented to further reduce the number of unique values.

As \Cref{subsec_digit_smooth} will show, the number of qubits needed for 
binning the angles or digitising them is broadly similar.
Putting the digitised angles into a list, and using the circuit in
\Cref{fig:sparse-count-rot-gen}, may still be of 
benefit as this avoids the need for a \textit{comparator} in the sampling approaches.

%
% 1-sparse decomposition
% ----------------------
\subsection{Decomposing into 1-sparse matrices}
\label{subsec_1sp_pop}

Converting a $s_r$ row-sparse matrix into a sum of 1-sparse matrices is a non-unique task.
Here we consider three alternative approaches. Each approach uses tags to keep track of which 
entries in the matrix have been allocated and, for each 1-sparse matrix, which columns have
been populated. Additionally, the max-search approach needs a tag for which rows have been
populated.

\begin{itemize}
    \item \textbf{Row-search.} The set of 1-sparse matrices is assembled by looping over
    the rows of the original matrix. If the first untagged entry on each row can be entered into
    the current 1-sparse matrix, this is done and the search moves to the next row. If not the next
    entry is checked until the end of the row is reached.
    If all entries on the row have been allocated, it is skipped.
    When the last row of the original matrix has been checked, populating the next 1-sparse matrix
    begins and repeats until all the entries of the original matrix have been allocated.
    This approach uses compressed sparse row (CSR) but the approach also can be applied to 
    compressed sparse column (CSC).
    \item \textbf{Max-search.} The entries in the CSR/CSC storage are converted into 
    coordinate (COO) storage and the entries are listed in descending order of absolute value. 
    The COO entries are searched in order and each untagged entry is added to the 1-sparse matrix,
    if allowed. If the 1-sparse matrix has been filled or the last COO entry has been
    reached, the next 1-sparse matrix is begun.
    \item \textbf{Rowmax-search.} This is a hybrid of the row and max search approaches, where each row is separately reordered to have its entries in
    descending order of absolute value. This is more efficient as the reordering complexity is 
    $\mathcal{O}(Nd\log (d))$ instead of $\mathcal{O}(Nd \log(Nd))$ for max-search.
\end{itemize}

Row-search is more efficient as it avoids the sorting needed for max-search. The latter
seeks to reduce the subnormalisation factor by collecting the largest entries into the same
1-sparse matrices.
By construction, max-search produces matrices in order of descending max-norm.
This gives the option to trim the number of matrices by omitting those at the end of
the list, or to use randomised sampling of those matrices.

Unexpectedly, in preliminary testing, rowmax searching mostly gave subnormalisation factors similar to or higher than row searching despite being more expensive. Therefore, this work focuses on and compares row-search and max-search. We expect that future 1-sparse decomposition heuristics might yield slightly better results.

%
% Testcases
% ---------
\section{Test cases}
\label{sec-testcases}

Two sets of testcases are considered:  unstructured meshes for the flow around the NACA-0012 
aerofoil and previously used structured meshes for the flow in lid-driven cavities and the flow 
through a square cross-section ducts. 
Both 2 and 3 dimensional cases are considered.
All cases are for incompressible flow with a constant viscosity.
The aerofoil cases are for \textit{pseudo inviscid} flow, i.e., slip-walls and low laminar viscosity.
The evaluations use matrices exported from the non-linear solver after initial transients in the early iterations have subsided.

% Matrix structures
% -----------------
\subsection{Matrix types}
\label{subsec-mat-types}

The system of equations to be solved results from linearising the
Navier-Stokes equations to find incremental updates to the 
non-linear flow field:
\begin{equation}
   \begin{split}
   \mathbf{u}^{n+1} &=  \mathbf{u}^{n} + \delta\mathbf{u}\\
   p^{n+1} &= p^n+\delta p 
  \end{split}
  \label{eqn-navier}
\end{equation}
where  $\mathbf{u}$ is the vector of fluid velocities $(u,v,w)$, 
$p$ is the fluid pressure, $n$ is the non-linear iteration counter
and $\delta\mathbf{u}$ and $\delta p$ are the
solutions of the linearised equations.

The most direct approach is to couple the equations in a single implicit
system \cite{darwish2009coupled} which, in 3D, has the form:
\begin{equation}
    \begin{pmatrix}
    A_u & 0   & 0   & -D_x & 0\\
    0   & A_v & 0   & -D_y & 0\\
    0   & 0   & A_w & -D_z & 0\\
    M_x & M_y & M_z &  A_p & 0 \\
    0   & 0   & 0   &  0 &  \gamma I
  \end{pmatrix}
  \begin{pmatrix}
    \delta u \\
    \delta v \\
    \delta w \\
    \delta p \\
    *
  \end{pmatrix}
  =
  \begin{pmatrix}
    R_u \\
    R_v \\
    R_w \\
    R_p \\
    0  
  \end{pmatrix}
  \label{eqn-implicit-3d}
\end{equation}
where
$A_u, A_v,A_w$, are sub-matrices containing the discrete convection and
diffusion operators for the $u, v, w$ momentum equations;
$D_x, D_y,D_z$, $M_x, M_y,M_z$ and $R_u, R_v,R_w$ are, respectively,
the discrete pressure gradient, discrete velocity gradient and the momentum
equation residuals in the $x,y,z$ directions; and,
$A_p$ is the Poisson-like smoothing operator from the continuity equation
\cite{darwish2009coupled}.
The lower right block in \Cref{eqn-implicit-3d} is padding to ensure
that the dimension of the matrix is a power of 2:
the $*$ denotes a dummy variable and 
$\gamma$ is a constant scaling term based on the maximum diagonal entry in
$\{A_u, A_v, A_w, A_p\}$.

Two-dimensional coupled systems have the same structure as \Cref{eqn-implicit-3d}
with the rows and columns associated to the $z,w$ directions deleted. 

An alternative to coupling the equations is to solve each equation in
\Cref{eqn-navier} separately by forming a Poisson type
correction equation for the pressure \cite{patankar1972calculation}:
\begin{equation}
    A_p\delta p = R_p
    \label{eqn-pcorr}
\end{equation}
If the dimensions of $A_p$ are not a power of 2, padding is applied
as above.

For QLES, coupled systems are preferred as they generally have lower
condition numbers and achieve non-linear convergence in fewer iterations.
They are often memory bound on classical systems and, hence, well placed
to take advantage of the much larger quantum address space.

For the structured mesh cases, both coupled and pressure correction
matrices are analysed.
For the unstructured mesh cases, only pressure correction matrices
are available.

% Unstructured aerofoil meshes
% ----------------------------
\subsection{Unstructured meshes}

\begin{figure}[htb]
  \centering
  \captionsetup{justification=centering}
    \begin{subfigure}[b]{0.49\textwidth}
      \centering
      \includegraphics[width=.99\textwidth]{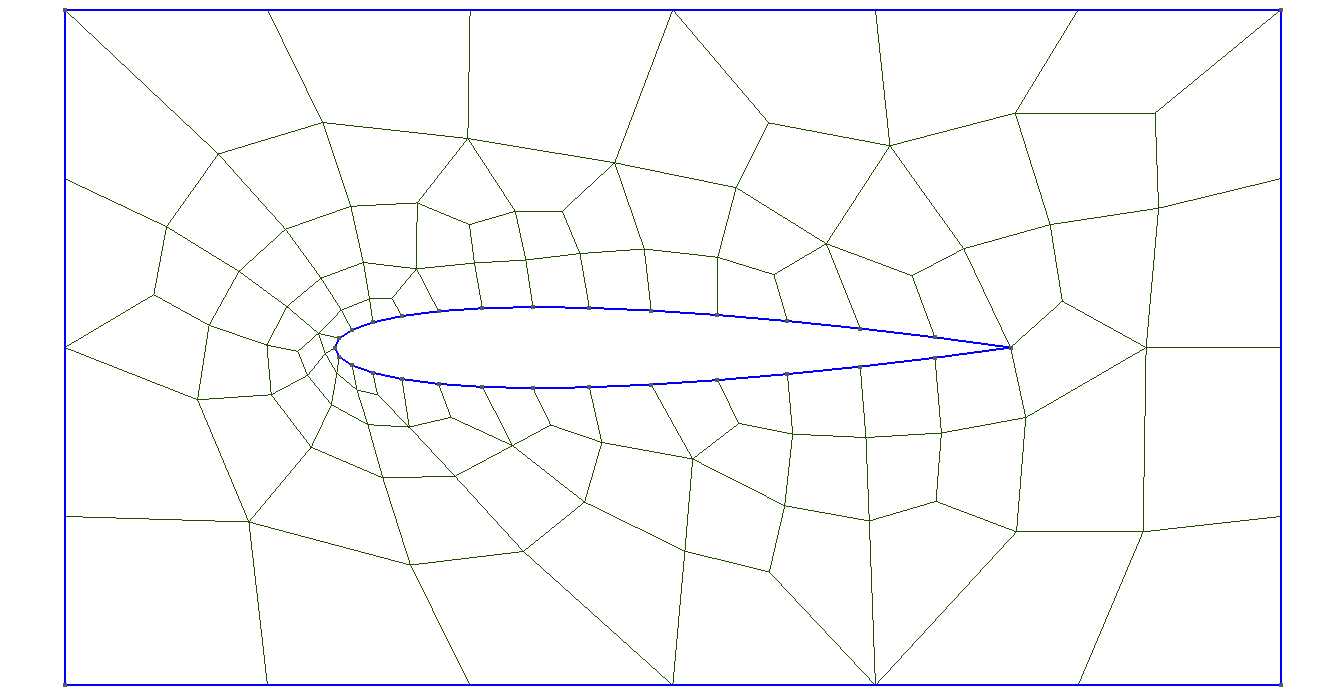}
      \caption{AFQ-100: mesh with 100 quadrilaterals.}
      \label{fig:naca_0012_crse3_quad}
  \end{subfigure}
    \begin{subfigure}[b]{0.49\textwidth}
      \centering
      \includegraphics[width=.99\textwidth]{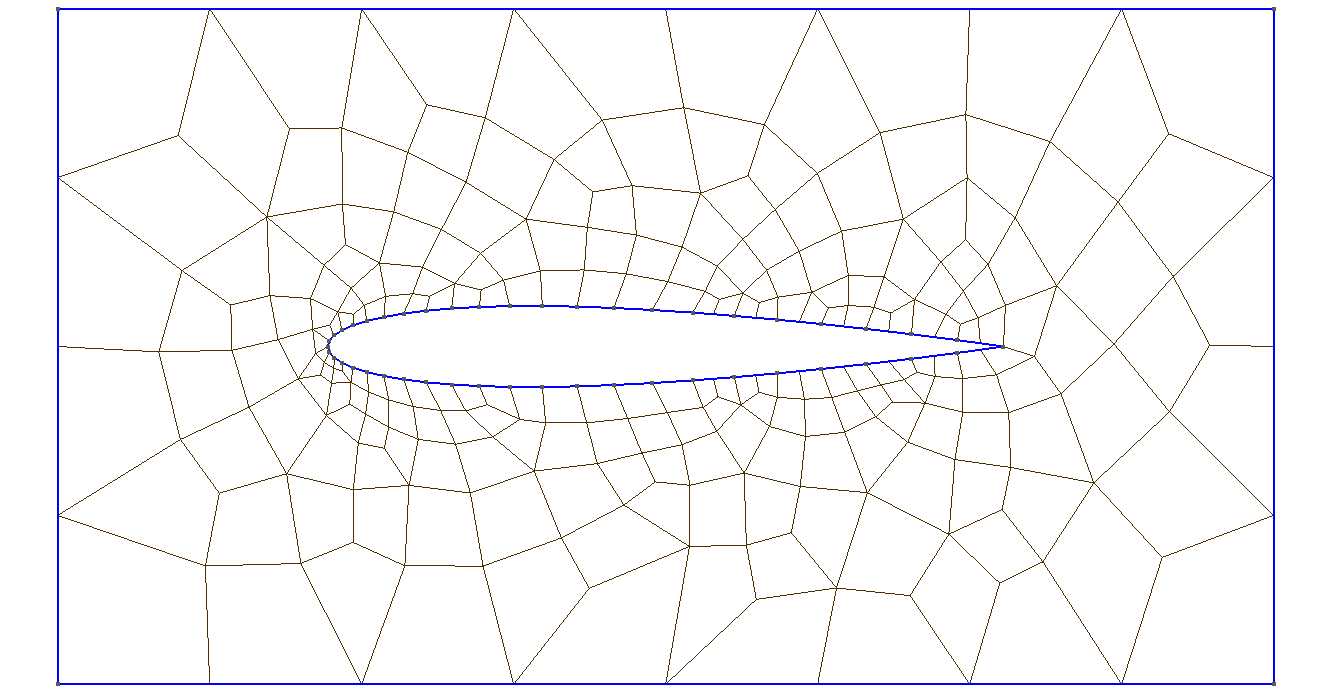}
      \caption{AFQ-226: mesh with 226 quadrilaterals.}
      \label{fig:naca_0012_crse2_quad}
  \end{subfigure}

    \begin{subfigure}[b]{0.49\textwidth}
      \centering
      \includegraphics[width=.99\textwidth]{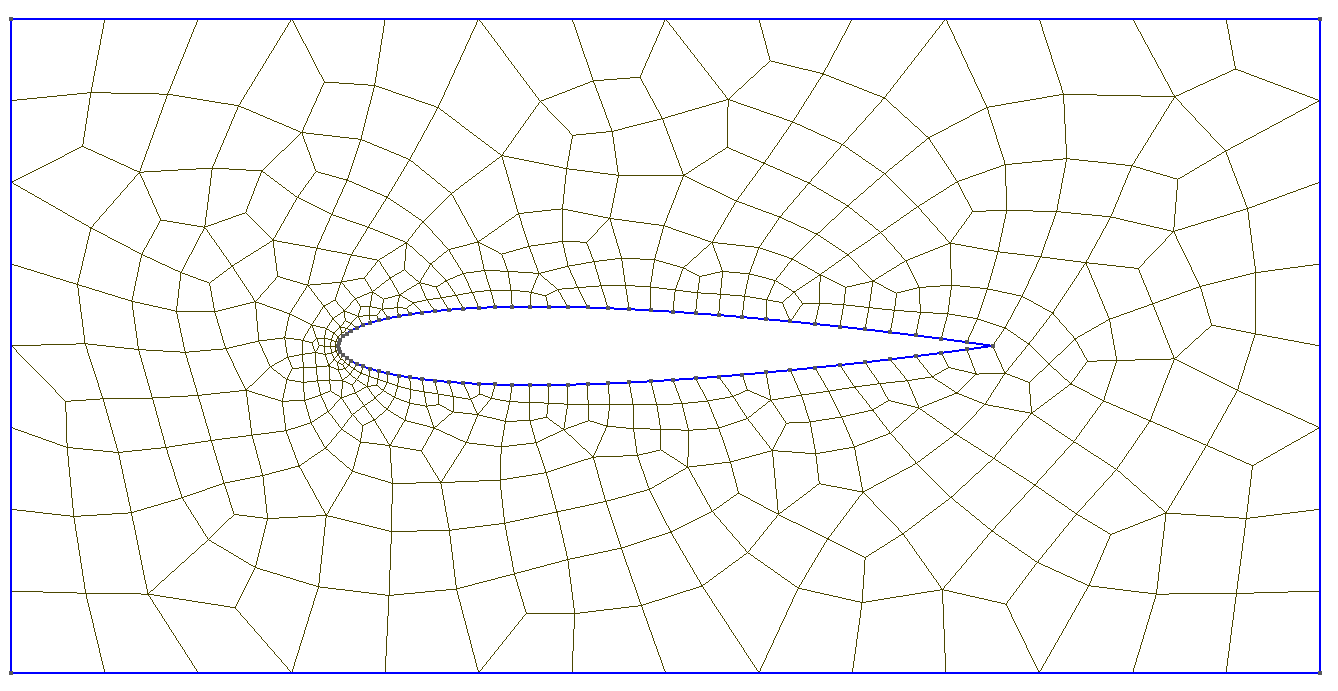}
      \caption{AFQ-489: mesh with 489 quadrilaterals.}
      \label{fig:naca_0012_crse_quad}
  \end{subfigure}
 \begin{subfigure}[b]{0.49\textwidth}
      \centering
      \includegraphics[width=0.99\textwidth]{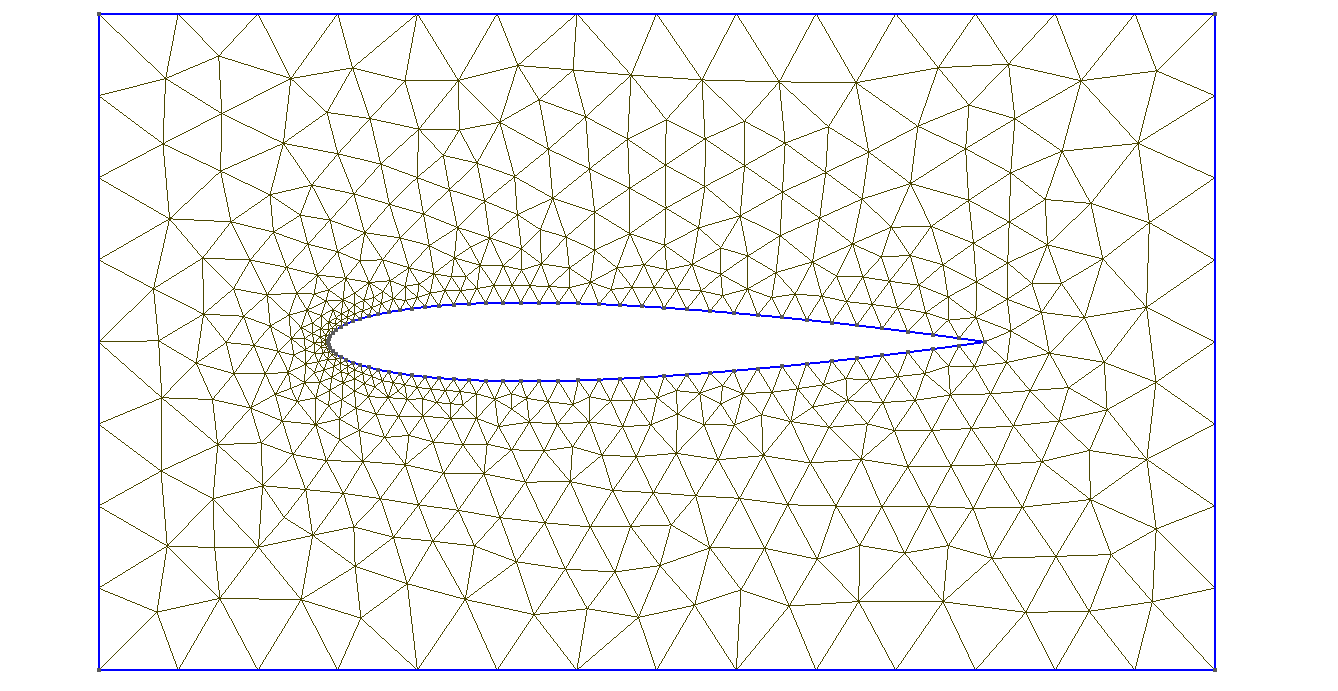}
      \caption{AFT-1022: mesh with 1,022 triangles.}
      \label{fig:naca_0012_crse_tri}
  \end{subfigure}
  \caption{Unstructured meshes for the 2 dimensional NACA 0012 aerofoil. 
  Domains are adjusted to generate meshes that are close to powers of 2.}
  \label{fig:naca_0012_2D}
\end{figure}

The 2D unstructured meshes and their identifiers are shown in 
\Cref{fig:naca_0012_2D}.
The meshes have been tailored to fit as closely as possible within the power of 2 
boundaries: 128, 256, 512 and 1024. They provide a combination of unstructured triangular and quadrilateral meshes.

The 3D unstructured meshes are extruded 2D meshes and are designated:
AFH-904, AFH-2040 and AFP-4088. Where AFH and AFP denote hexahedral and prism
meshes respectively, and the digits are the number of cells in the mesh.
Again, the meshes have been tailored to fit within the power of 2 
boundaries: 1024, 2048 and 4096.

% Structured meshes
% -----------------
\subsection{Structured meshes}
The structured mesh test cases are listed in \Cref{tab-str-testcases}
and are taken from previous work
\cite{lapworth2022hybrid, lapworth2022implicit,lapworth2026probability}.
Both pressure correction and implicit coupled matrices are examined.
The latter are of interest due to their block banded-diagonal structure
which enables some of the off-diagonal blocks to be combined to reduce the number 
of 1-sparse matrices needed.

\begin{table*}[t]
  \centering
  \begin{tabular}{c c c c c c c c c}
    \toprule
    Identifier & Test case & Dim & Matrix type  & CFD mesh size  \\
    \midrule
     CAP-1024 \cite{lapworth2022hybrid} & Lid driven cavity & 2D & Pressure Correction    & 32x32  \\
     CAC-1024 \cite{lapworth2022implicit} & Lid driven cavity & 2D & Coupled implicit       & 16x16  \\
     CAC-1024-RCM \cite{lapworth2022implicit}  & Lid driven cavity & 2D & Coupled Implicit + RCM & 16x16 \\
     DUP-4096 \cite{lapworth2026probability}    & Square duct      & 3D & Pressure Correction    & 16x16x16 \\ 
     DUC-2048 \cite{lapworth2026probability}    & Square duct      & 3D & Coupled implicit       & 8x8x8 \\
     DUC-16384 \cite{lapworth2026probability}   & Square duct      & 3D & Coupled implicit       & 16x16x16 \\ 
    \bottomrule
  \end{tabular}
  \caption{\centering Structured mesh test cases}
  \label{tab-str-testcases}
\end{table*}

% Bandwidth reduction
% -------------------
\subsection{Bandwidth reduction}
\label{subsec-rcm}

Bandwidth reduction is a common technique in classical computing. 
As highlighted in \Cref{subsec-diag-offset}, this is also
a useful technique to reduce the number of qubits for block encoding. 
The most common bandwidth reduction technique is Reverse Cuthill-McKee (RCM) \cite{cuthill1969reducing,george1981computer}. In this, the linear system matrix, $A$, provides the
adjacency matrix for the dual mesh, i.e. the graph formed by connecting the centres of the
mesh cells. The degree of the mesh is the number neighbours for each cell. 
The Cuthill-McKee algorithm starts by finding the cell with the lowest degree. 
This is the first entry, i.e. root, in the reordering list. The remaining cells are added to the list via a front propagating from the root.
This has the character of a space-filling curve. Populating the list is an $\mathcal{O}(N)$ operation.
Once the list is complete, it is reversed since this reduces fill-in when using classical
techniques such as lower-upper decomposition.
For non-symmetric matrices, RCM is applied to $(A+A^T)/2$.
The output of the RCM algorithm is a permutation matrix, $P$, that transforms the linear system $Ax=b$ to:
\begin{equation}
    \begin{split}
        PAP^Ty &= Pb \\
        x &= P^{T}y
    \end{split}
\end{equation}

\begin{figure}[htb]
  \centering
  \includegraphics[width=.67\textwidth]{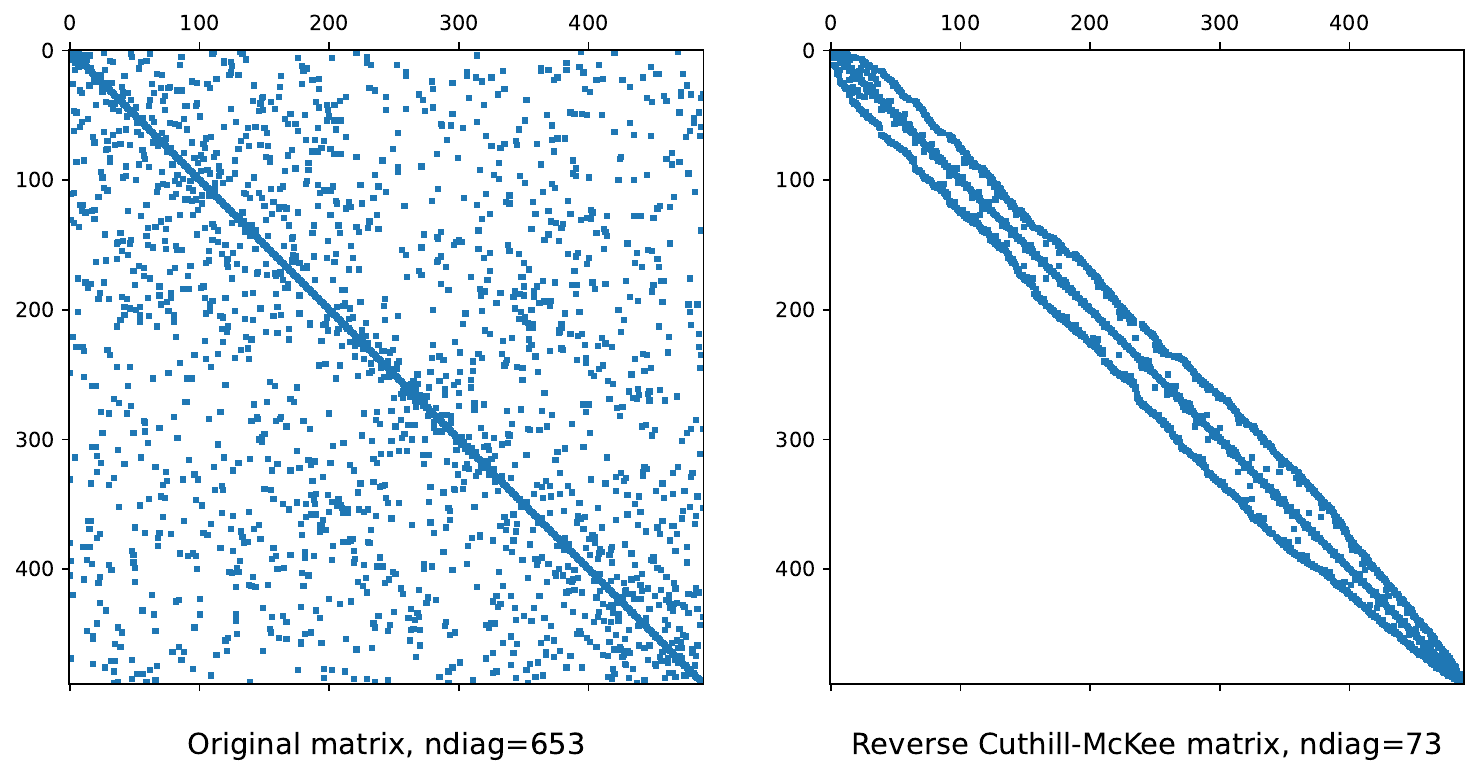}
  \captionsetup{justification=centering}
  \caption{Application of Reverse Cuthill-McKee \cite{cuthill1969reducing,george1981computer} to
  the AFQ-489 test matrix.}
  \label{fig_rcm_afq_489}
\end{figure}

The application of RCM to the AFQ-489 test matrix is illustrated in 
\Cref{fig_rcm_afq_489}.
Here, the number of diagonals with at least one entry has been reduced from 653 to 73.
This means the unstructured encoding based on diagonal offset requires two fewer
qubits than row-column encoding. Although this is a small reduction, more significant savings
can be made for larger matrices.

% Preconditioning
% ---------------
\subsection{Preconditioning}
\label{subsec-precon}

As with classical solvers, the matrix condition number, $\kappa$, has a primary influence
on the resources needed by a Quantum Linear Equation Solver (QLES).
In many practical applications, $\kappa$ scales at least linearly with the size
of the matrix.
In QLES, the situation is compounded by the fact that effective condition 
number, $\kappa_s$, is scaled by a subnormalisation factor greater than unity
(\Cref{eqn-kappa-s}).
It is, therefore, highly unlikely that an unpreconditioned QLES will be faster than a 
preconditioned classical solver.

Unfortunately, as demonstrated empirically in \cite{lapworth2025preconditioned} and confirmed theoretically by
\cite{pechan2026block}, implementing QLES preconditioning via multiplication within a
quantum circuit does not, nor cannot, reduce the effective condition number.
Hence, preconditioning must be applied as a classical pre-processing task.
A common choice is the Sparse Approximate Inverse
(SPAI) preconditioner \cite{benson1984parallel,chow1998approximate} that can be efficiently
parallelised and which only needs to be applied once at the start of the QLES.

\begin{figure}[htb]
  \centering
  \includegraphics[width=.49\textwidth]{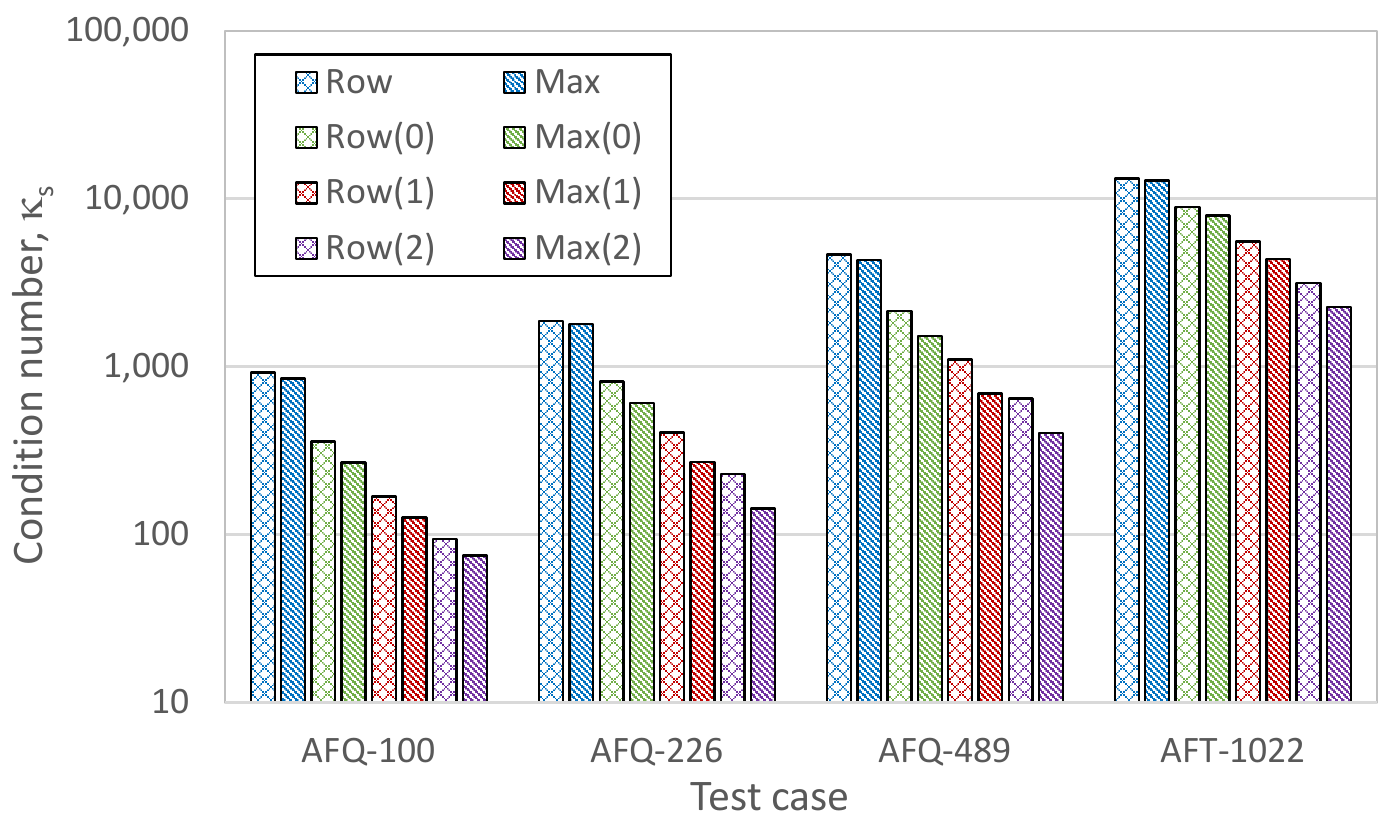}
  \captionsetup{justification=centering}
  \caption{Subnormalised condition numbers for SPAI preconditioning with varying levels
  of infill. All cases have Reverse Cuthill-McKee reordering. The number in brackets shows
  the number SPAI infill levels. The cases without a bracketed number are without preconditioning.}
  \label{fig:kappa-spai-rcm}
\end{figure}

\Cref{fig:kappa-spai-rcm} shows the variation in $\kappa_s$ for the 2D
unstructured cases with SPAI preconditioning and up to 2 levels of infill.
For the cases without SPAI preconditioning, Jacobi diagonal scaling is applied which
can reduce the condition number by a factor of 2-3 due to the larger variations
in cell size in the unstructured meshes.

Since RCM and preconditioning are classical preprocessing steps, subsequent references to the
linear system matrix, $A$, should be understood as the reordered and/or preconditioned matrix.
The labelling of the results should make it clear when these have been applied.

\section{Results}
\label{sec_results}

In all the figure legends in this section, the number in brackets gives the number of SPAI infill levels. Legends without a bracketed number show results without preconditioning.
Unless otherwise stated, the results have used max-search.

\begin{figure}[ht]
  \centering
  \begin{subfigure}[b]{0.49\textwidth}
      \centering
      \includegraphics[width=0.99\textwidth]{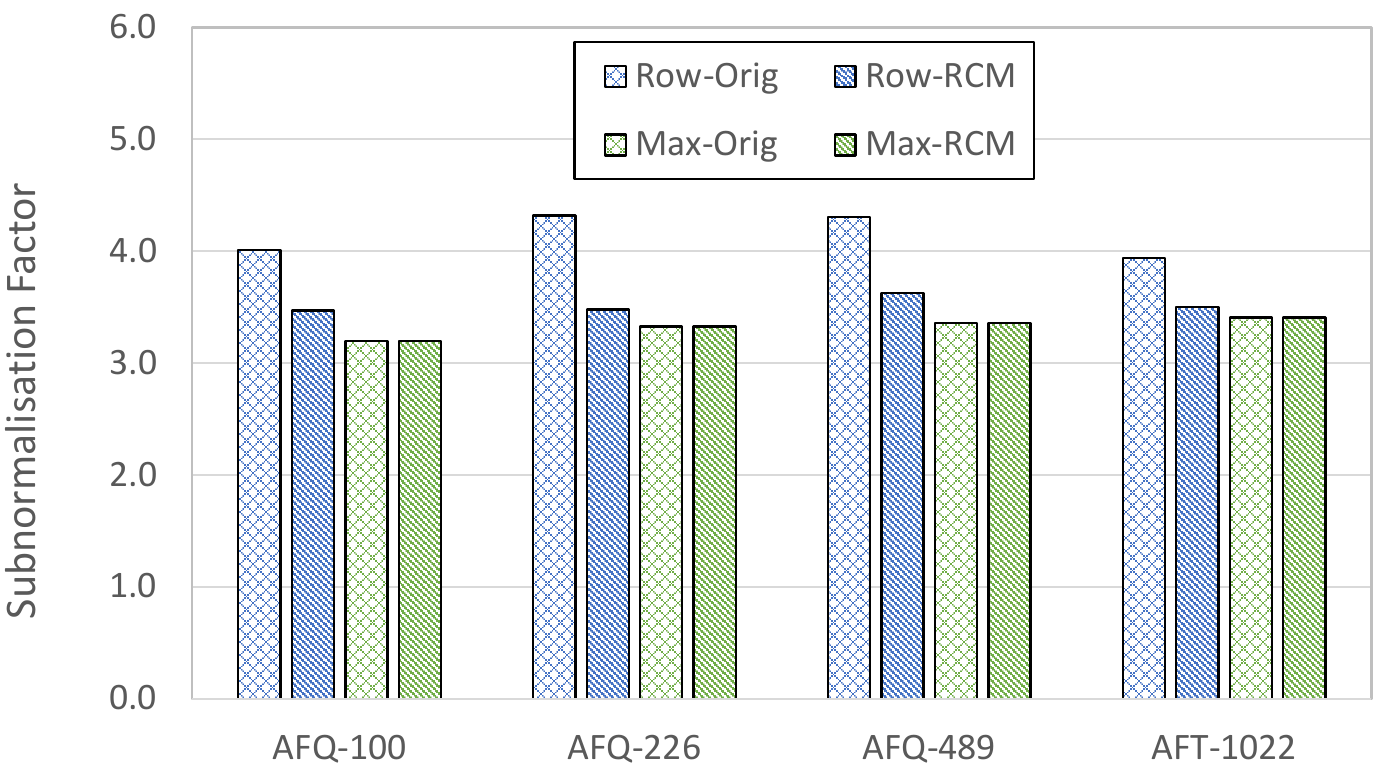}
      \caption{\small Subnormalisation factor.}
      \label{fig_subn_orig_rcm}
  \end{subfigure}
  \begin{subfigure}[b]{0.49\textwidth}
      \centering
      \includegraphics[width=0.99\textwidth]{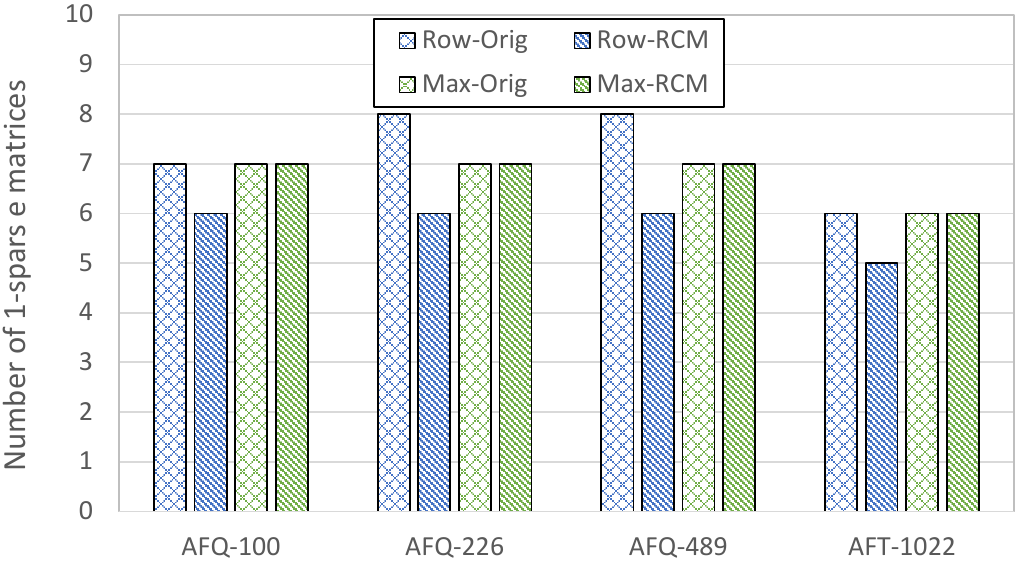}
      \caption{\small Number of 1-sparse matrices.}
      \label{fig_n1sp_orig_rcm}
  \end{subfigure}
  \caption{\centering 2D unstructured meshes.
  Influence of search scheme and bandwidth reduction on the 1-sparse decomposition.}
  \label{fig_orig_rcm}
\end{figure}

% 2D unstructured meshes
% ----------------------
\subsection{2D unstructured meshes}

Firstly, we consider the effect of bandwidth reduction on the 1-sparse decomposition.
As shown in \Cref{fig_subn_orig_rcm}, the subnormalisation factors for row-search are
higher than for max-search, particularly when bandwidth reduction has not been applied.
When RCM is applied, the two search schemes are much closer. As expected, the 
bandwidth reduction enables the row-search to march more effectively along diagonal
segments of the matrix. 
The reduction in subnormalisation factor is largely due to the row-search needing
fewer 1-sparse matrices after the bandwidth reduction (\Cref{fig_n1sp_orig_rcm}).
In fact, the row-search is more compact than the max-search when RCM is used.
The max-search results are independent of the matrix ordering - this is not guaranteed as,
although the max-search works with the same ordered list of entries, the $(i,j)$ locations of those
entries are different.

Although there is an effect on the number of 1-sparse matrices for each test case,
the number of non-zero entries is unaffected and, hence, so is the number of multiplexed
rotation gates needed to block-encode the CFD matrix.

\begin{figure}[ht]
  \centering
  \begin{subfigure}[b]{0.49\textwidth}
      \centering
      \includegraphics[width=\textwidth]{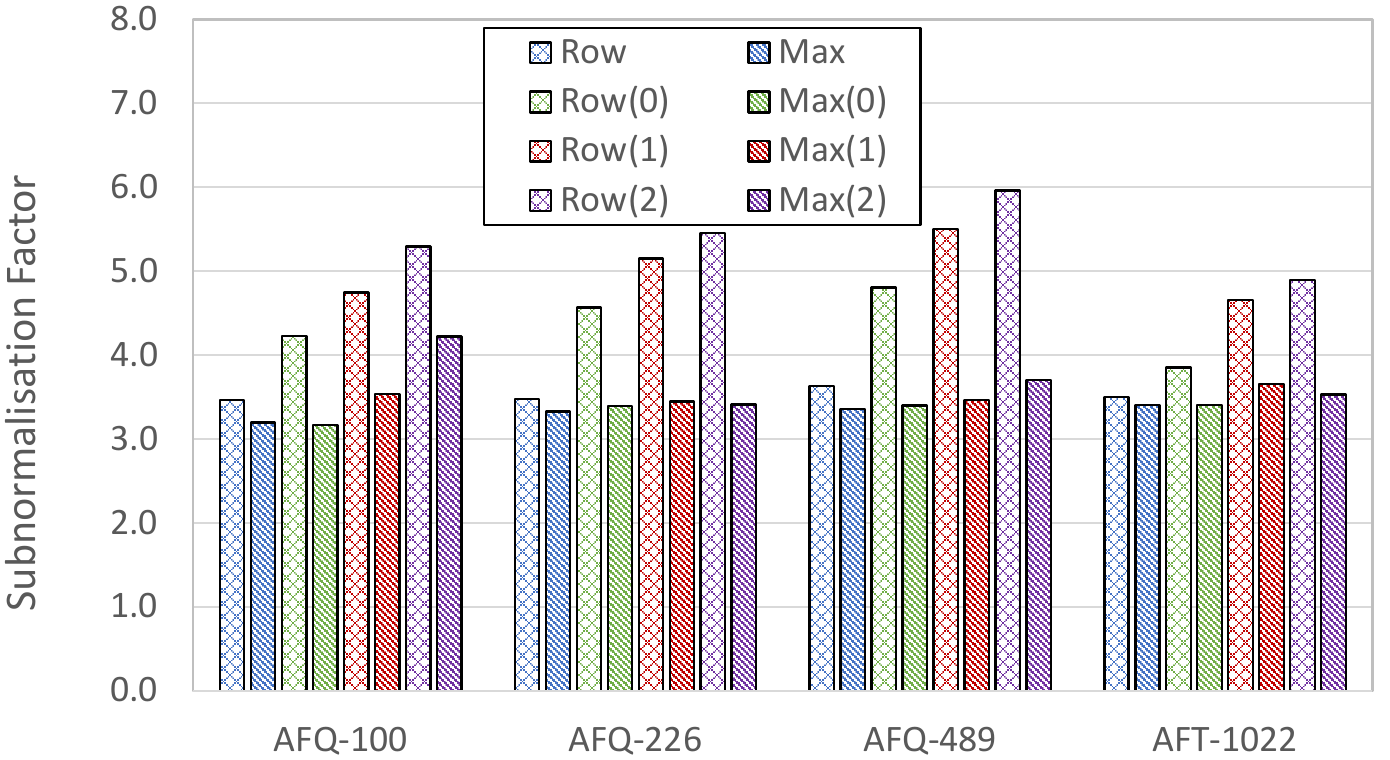}
      \caption{\small Subnormalisation factor.}
      \label{fig_subn_af2D}
  \end{subfigure}
  \begin{subfigure}[b]{0.49\textwidth}
      \centering
      \includegraphics[width=\textwidth]{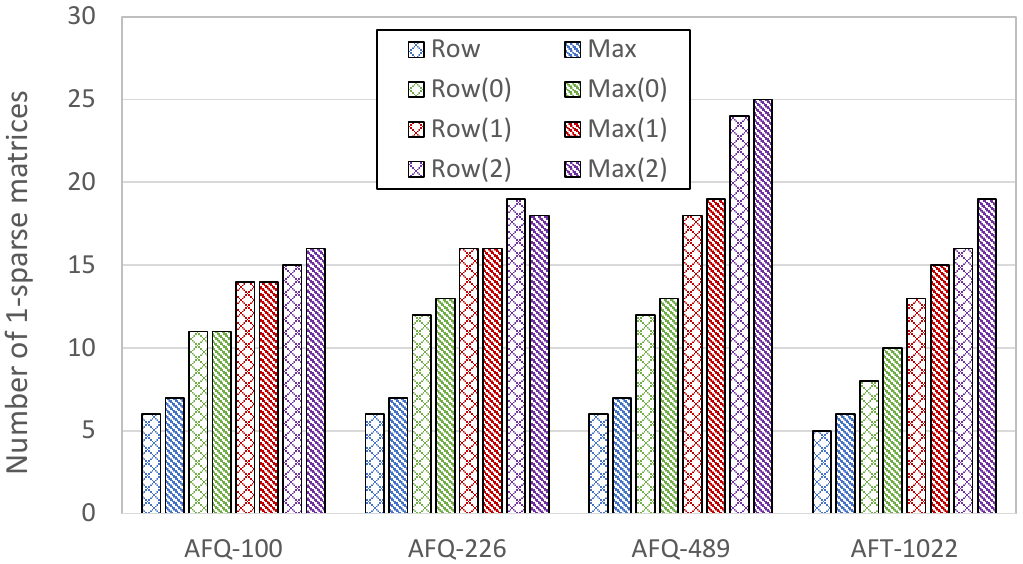}
      \caption{\small Number of 1-sparse matrices.}
      \label{fig_n1sp_af2D}
  \end{subfigure}
  \caption{\centering 2D unstructured meshes.
  Influence of search scheme and the number of SPAI infill levels 
  on the 1-sparse decomposition.  All cases have Reverse Cuthill-McKee reordering.}
  \label{fig_af2D}
\end{figure}

\Cref{fig_af2D} compares the row and max search approaches when SPAI preconditioning is applied.
Here, a more consistent trend appears.
Row-search shows a roughly linear increase in subnormalisation factor as the number
of infill levels is increased. 
Whereas, max-search shows less variation, other than for the 
smallest case, and the subnormalisation factor remains broadly constant as the infill level
increases. 
This is despite the number of 1-sparse
matrices growing with the number of infill levels for both search options 
(\Cref{fig_n1sp_af2D}). We give an explanation for this observation in \Cref{sec-spai-front}.

For the triangular mesh case (AFT-1022), each cell has only 3 neighbours which reduces
the row sparsity relative to the quadrilateral meshes and leads to a corresponding 
reduction in the number of 1-sparse matrices.
For row-search, this leads to reduction in the subnormalisation factor, whereas it
is largely unchanged for max-searching.

%
% 2D Structured meshes
% ---------------------
\subsection{2D structured meshes}
\label{subsec_1sp_structured}

\begin{figure}[ht]
  \centering
  \begin{subfigure}[b]{0.49\textwidth}
      \centering
      \includegraphics[width=0.99\textwidth]{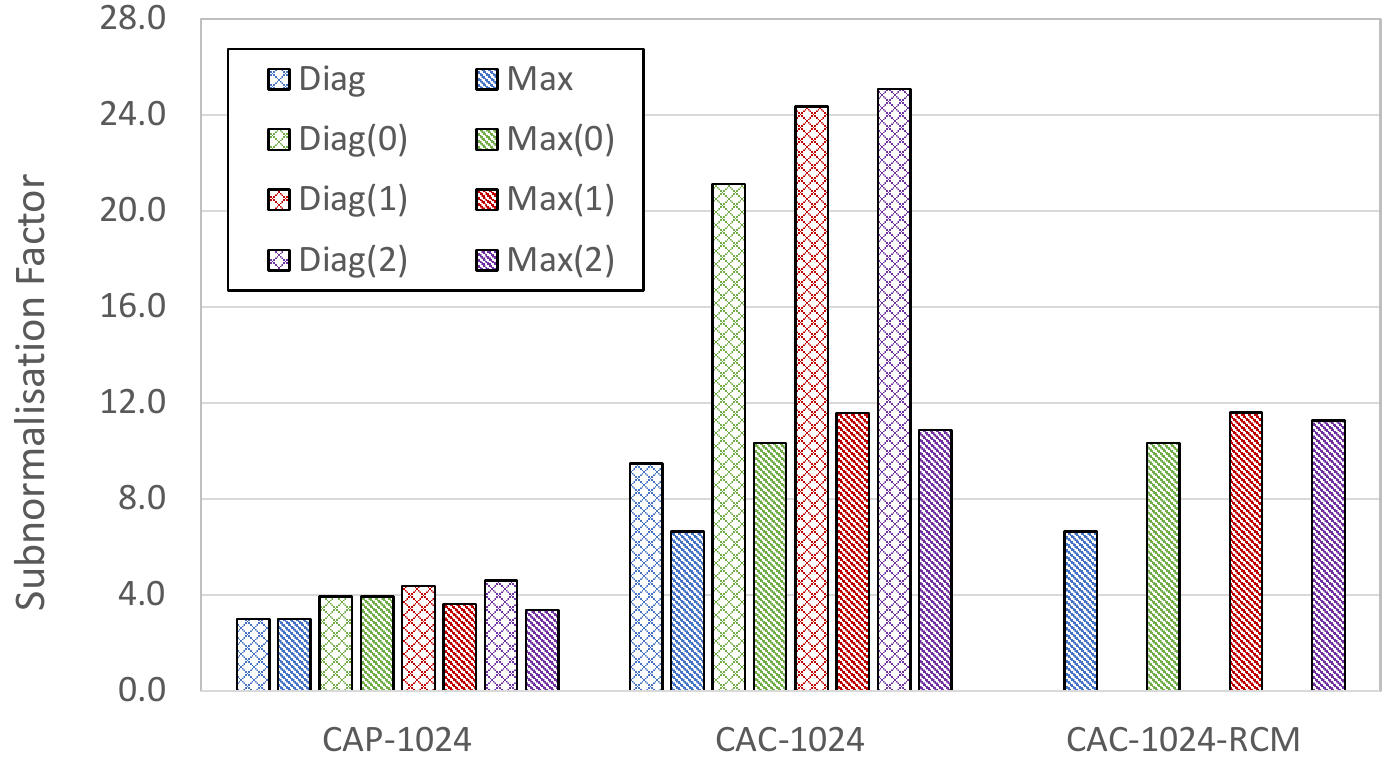}
      \caption{\small Subnormalisation factor.}
      \label{fig_subn_cav2d}
  \end{subfigure}
  \begin{subfigure}[b]{0.49\textwidth}
      \centering
      \includegraphics[width=0.99\textwidth]{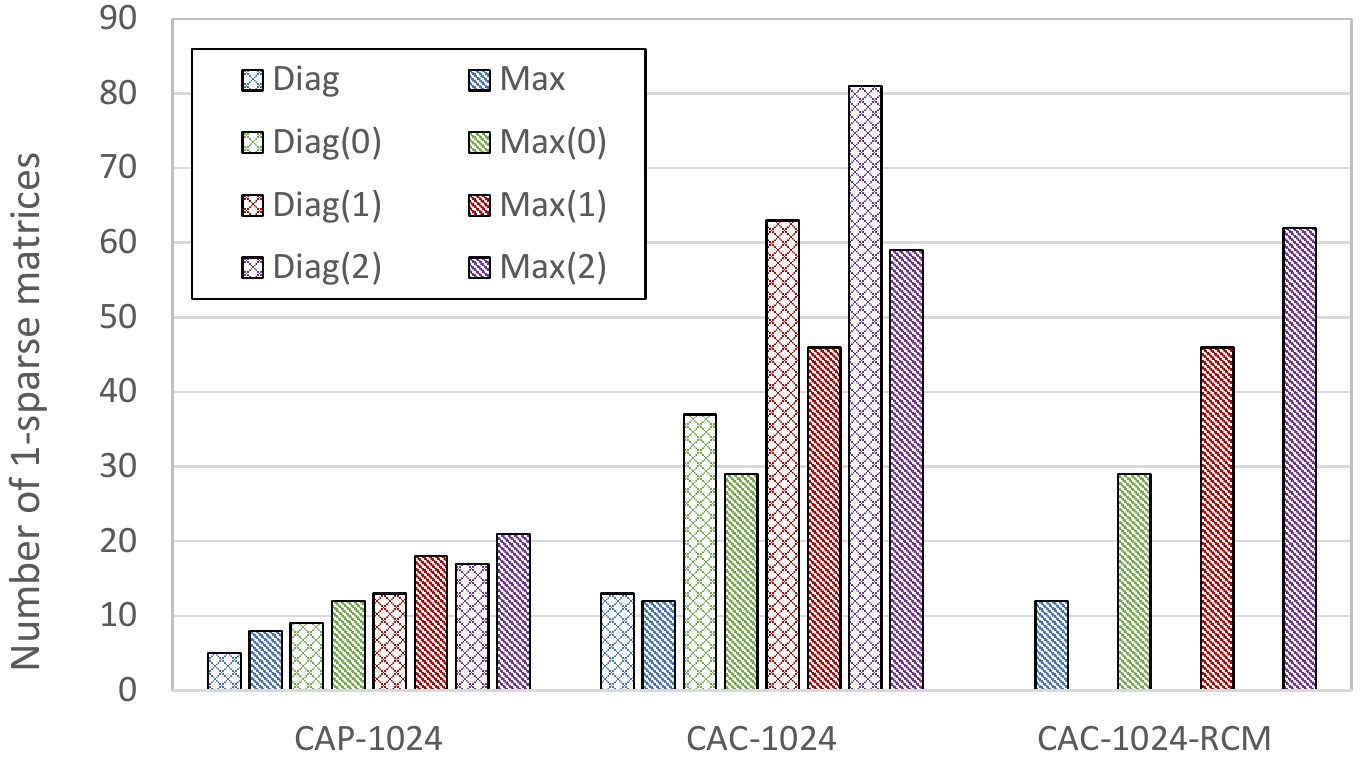}
      \caption{\small Number of diagonals/1-sparse matrices.}
      \label{fig_n1sp_cav2d}
  \end{subfigure}
  \caption{\centering 2D structured mesh cases. 
  Comparison of banded diagonal and 1-sparse encoding with the number of SPAI infill levels.}
  \label{fig_cav2d}
\end{figure}

In \Cref{fig_cav2d}, we consider whether the unstructured block-encoding offers a benefit over
the banded diagonal encoding for structured meshes.
Only the 1-sparse results with max-search are presented as the behaviour with row-search
is similar to that seen in \Cref{fig_af2D}.
For the size of the \textsc{Prepare} register we compare the number of diagonals with the number
of 1-sparse matrices.

The first block, CAP-1024, in \Cref{fig_cav2d} is for a pressure correction matrix.
Here, the unstructured encoding gives a slight reduction in the
subnormalisation factor at the expense of slightly more 1-sparse matrices than there are diagonals. Overall, the differences are marginal and the extra qubits needed for
the unstructured encoding are not justified.

For the coupled implicit case, CAC-1024, there is a more marked difference.
The unstructured encoding gives around a factor of 2 reduction in subnormalisation factor
when SPAI preconditioning is applied, for all levels of infill.
The number of 1-sparse matrices in the unstructured encoding is also less than the number
of diagonals in the banded-diagonal encoding. 
This is a result of being able to combine diagonals from the different
sub-matrices, as shown in matrix structure in \Cref{eqn-implicit-3d}.

The final block on the right in \Cref{fig_cav2d} shows the effect on applying RCM bandwidth reduction to the CAC-1024 coupled matrix. This destroys the banded-diagonal structure of the matrix, creating
a large number of partial diagonals and making banded-diagonal encoding inappropriate.
As in \Cref{fig_orig_rcm}, the unstructured encoding with max-search shows only a marginal 
difference between the original and reordered matrices.

% 3D unstructured meshes
% ----------------------
\subsection{3D unstructured meshes}
\label{subsec-3d-unstr-meshes}

\begin{figure}[ht]
  \centering
  \begin{subfigure}[b]{0.49\textwidth}
      \centering
      \includegraphics[width=0.99\textwidth]{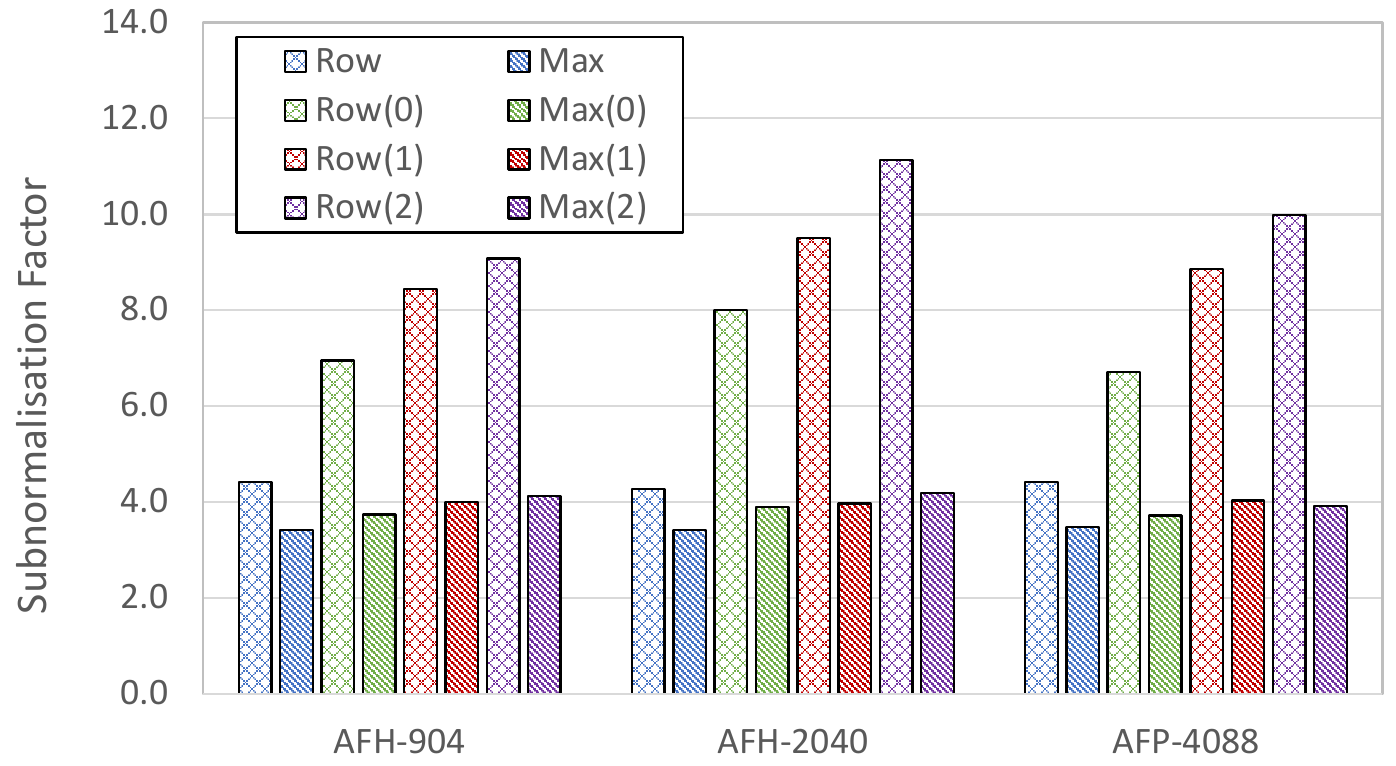}
      \caption{\small Subnormalisation factor.}
      \label{fig_subn_af3d}
  \end{subfigure} 
  \begin{subfigure}[b]{0.49\textwidth}
      \centering
      \includegraphics[width=0.99\textwidth]{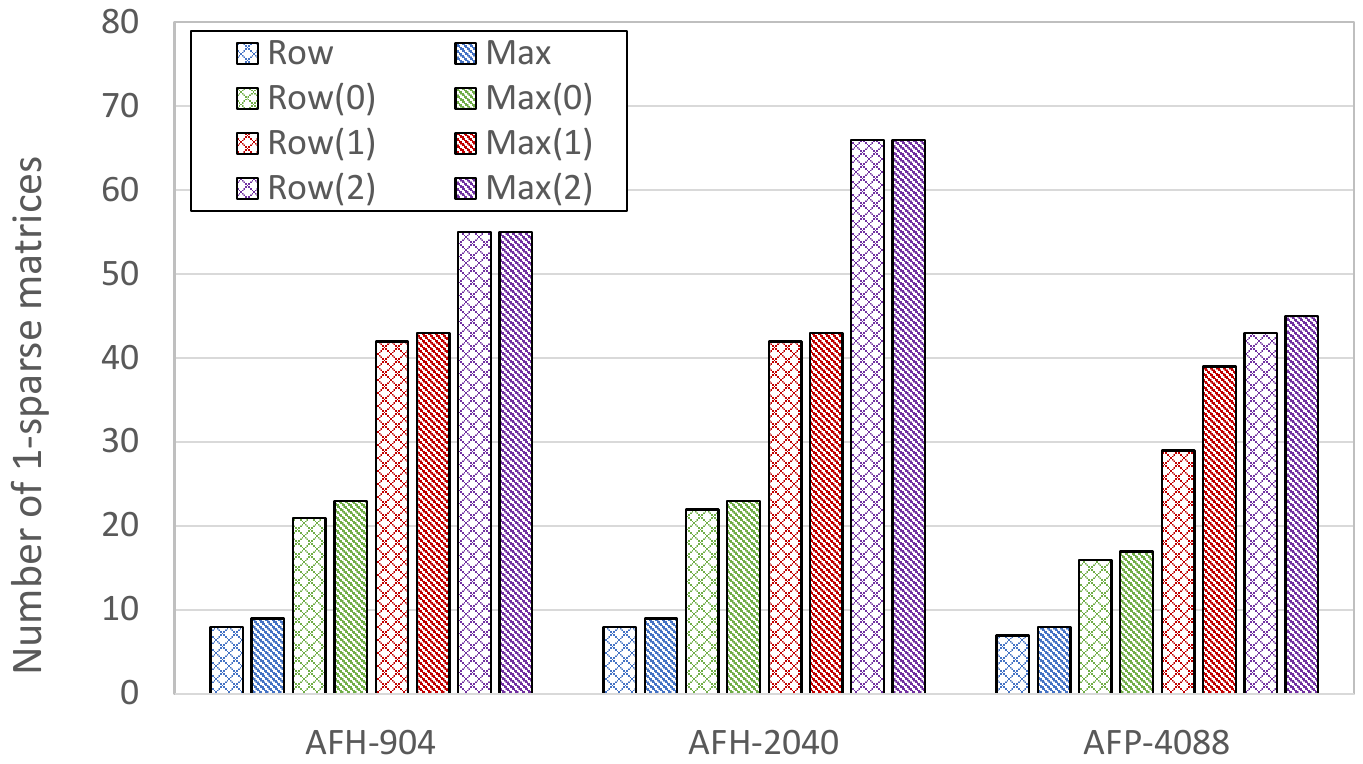}
      \caption{\small Number of 1-sparse matrices.}
      \label{fig_n1sp_af3d}
  \end{subfigure}
  \caption{\centering 3D unstructured mesh cases.
  Influence of search scheme and the number of SPAI infill levels 
  on the 1-sparse decomposition.  All cases have Reverse Cuthill-McKee reordering.}
  \label{fig_af3d}
\end{figure}

\Cref{fig_af3d} compares row and max search when SPAI preconditioning is applied to the
3D unstructured testcases.

As in the 2D unstructured cases, the three 3D cases are very consistent.
For the subnormalisation factor, max-search gives a more pronounced benefit, achieving a reduction of a factor of 
almost 3 relative to row-search for the SPAI(2) cases.
They also remain roughly constant as the number of infill levels increases.

\Cref{fig_n1sp_af3d} shows a marked increase in the number of 1-sparse matrices relative to
the 2D case in \Cref{fig_n1sp_af2D}.
Without preconditioning, the increase is commensurate with the increase in the difference
stencil from 4 neighbours (quadrilateral) to 6 neighbours (hexahedron).
However, the ratios increase with the number of SPAI infill levels.
This is due to the SPAI \textit{front} advancing in three dimensions instead of two.

% 3D structured meshes
% ----------------------
\subsection{3D structured meshes}
\label{subsec-3d-str-meshes}

\begin{figure}[ht]
  \centering
  \begin{subfigure}[b]{0.49\textwidth}
      \centering
      \includegraphics[width=0.99\textwidth]{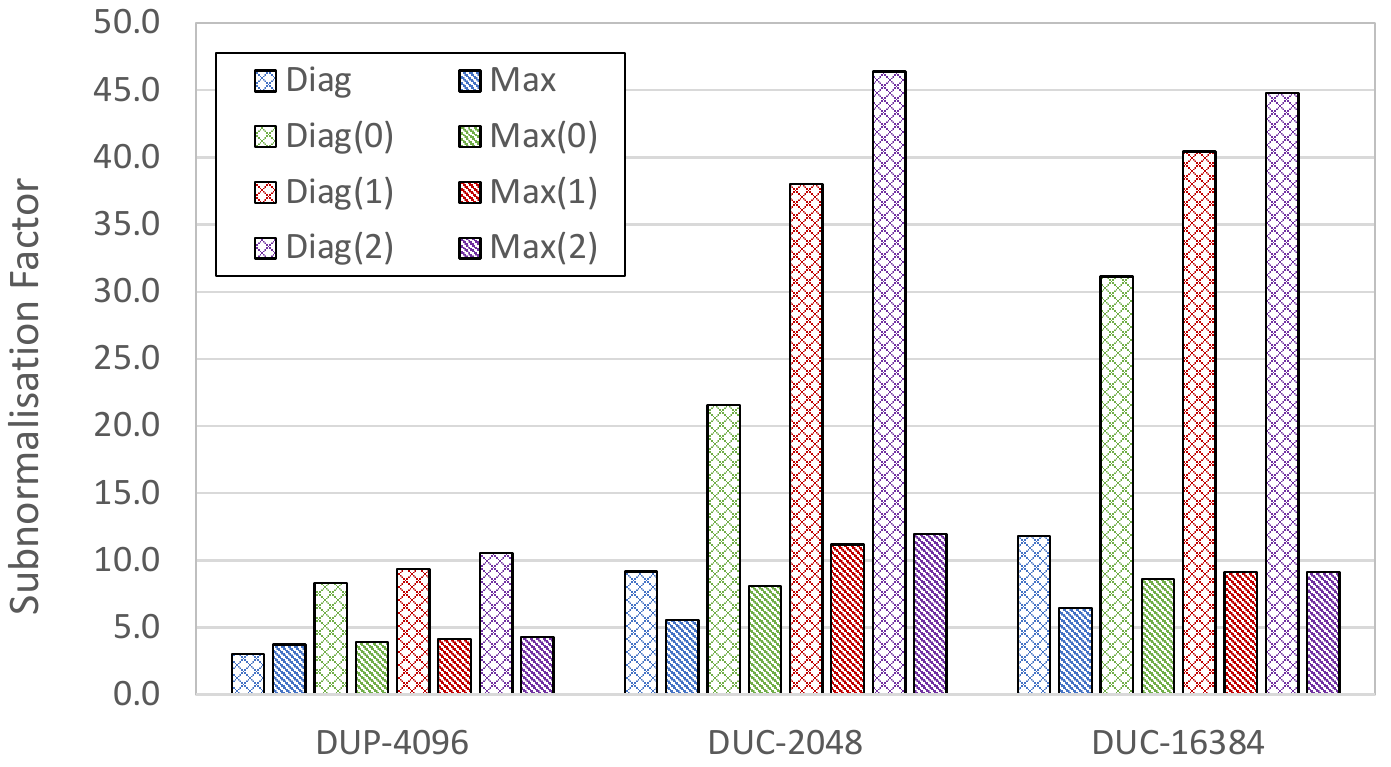}
      \caption{\small Subnormalisation factor.}
      \label{fig_subn_duct3d}
  \end{subfigure}
  \begin{subfigure}[b]{0.49\textwidth}
      \centering
      \includegraphics[width=0.99\textwidth]{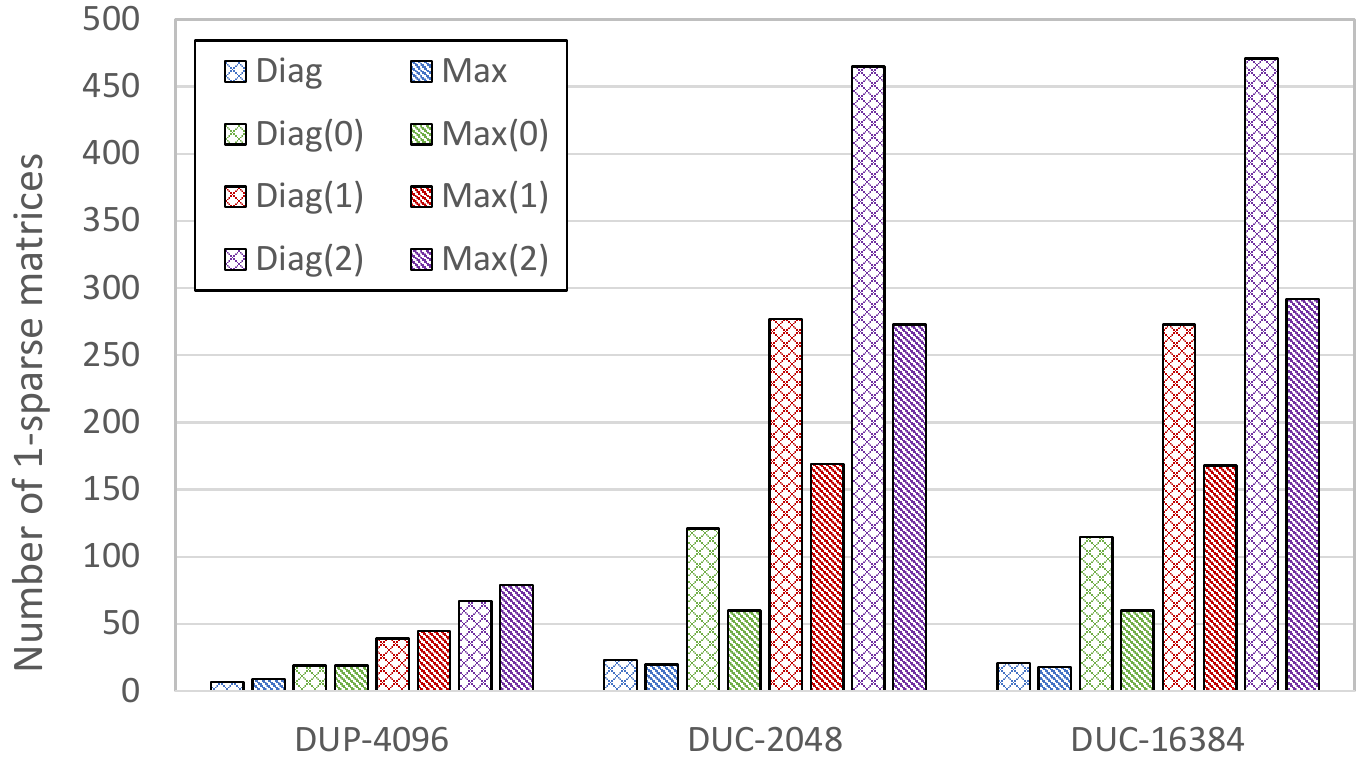}
      \caption{\small Number of 1-sparse matrices.}
      \label{fig_n1sp_duct3d}
  \end{subfigure}
  \caption{\centering  3D structured mesh cases.
  Comparison of banded diagonal and 1-sparse encoding with the number of
SPAI infill levels.}
  \label{fig_duct3d}
\end{figure}

\Cref{fig_duct3d} compares the unstructured mesh max-search encoding with the banded
diagonal encoding for the 3D structured mesh test case.
The benefits of the unstructured encoding over the diagonal encoding are
much higher than in the 2D structured mesh results (\Cref{fig_cav2d}).
As in the 3D unstructured results, this is due to the SPAI \textit{front} 
advancing in 3 dimensions instead of 2.

In the coupled implicit cases (DUC-2048 and DUC-16384), the rapid increase in
subnormalisation factor with SPAI infill levels, effectively eliminates any benefit
of preconditioning with banded diagonal block encoding. 
The unstructured block encoding restores that benefit with a reduction of
almost a factor of 5 for the DUC-16384 case with SPAI(2) preconditioning.

An indicator of the 3D \textit{front} effect is the rapid rise in the number of diagonals and, to a lesser extent the number of 1-sparse matrices with the infill level.
For the DUC-16384 case with SPAI(2) preconditioning, the ratio of the number of diagonals to the subnormalisation factor for banded diagonal encoding is about 10.
For the 1-sparse encoding, it is about 30.

% scaling
% -------
\subsection{Scaling}
\label{subsec_scaling}

Considering all the results in this section, the 1-sparse encoding has demonstrated a 
consistent trend of maintaining low subnormalisation factors as the number of
preconditioning infill levels is increased.

\begin{figure}[htb]
  \centering
  \includegraphics[width=.49\textwidth]{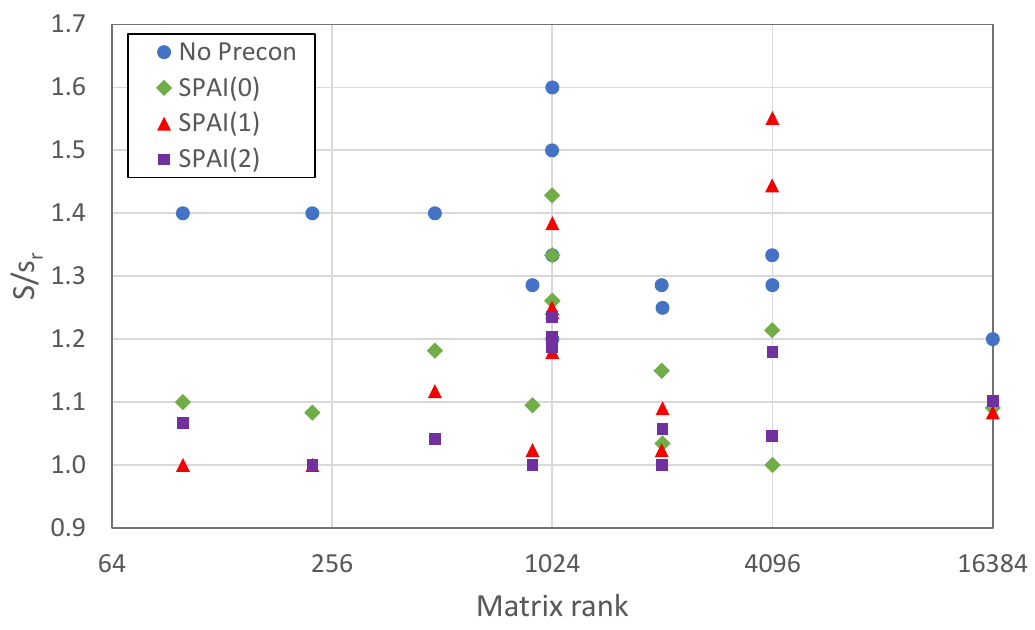}
  \captionsetup{justification=centering}
  \caption{Ratio of the number of 1-sparse matrices ($S$) to the maximum row
  sparsity ($s_r$) for all the test cases. Max-search results only.}
  \label{fig:complexity}
\end{figure}

\Cref{fig:complexity} shows the ratio $S/s_r$, where $s_r$ is the maximum row sparsity,
for all the test cases using
max-search. The row-search results are not included as these have the same or
fewer 1-sparse matrices.
All the ratios are $\lesssim 1.6$. 
Except for 2 cases, the non-preconditioned matrices have the largest $S/s_r$ values.
The lowest ratios generally correspond to the unstructured mesh test cases.
With 2 levels of infill, all but one of the unstructured cases has $S/s_r<1.1$.
These results confirm that $S=\mathcal{O}(s_r)$ with a low prefactor.

% rotation angles
% ---------------
\subsection{Digital precision vs. binning the rotation angles}
\label{subsec_digit_smooth}

Digitising the multiplexed rotations means truncating their binary expansion to a fixed bit length 
$d$:
% $1+\log_2 D$:
\begin{equation}
    a_{ij} \mapsto \lfloor a_{ij}\times D \rfloor /D, \qquad i,j \in A
    \label{eqn-digit}
\end{equation}
where $D= 2^d$.
Since the matrix entries are signed and $|a_{ij}| \le 1$, the maximum number of non-zero unique angles is $2D$.

Here, we compare digitisation with the smoothing based on the non-overlapping bins technique
\cite{lapworth2025preconditioned}. The smoothing+binning technique involves a free parameter that sets the
maximum width of the bins. These are both relevant to the encoding in
\Cref{subsec_list_rot} and the number of multiplexed rotations depends on the
number of unique angles.

In this work, we use \Cref{eqn-digit} to reduce the precision of the rotation angles, this gives
a clearer comparison with the binning technique which retains 64-bit precision angles.

\begin{table}[ht]
  \centering
  \begin{tabular}{c c c c c c c c c c}
    \toprule
    \multicolumn{2}{c}{Test case} & \multicolumn{4}{c}{Smoothing+binning} & \multicolumn{3}{c}{Digitisation} \\
    \cmidrule(r){1-2}
    \cmidrule(r){3-6}
    \cmidrule(r){7-9}
    Test case &$n_{nz}$  & $s_f$ & $n_{l}$ & $\epsilon$ & $n_{q}$ & $D$ & $n_b$ &  $\epsilon$ & $n_{q}$ \\
    \midrule
    AFP-4088(0) &48,279 & $\pm0.25\%$ & 1,698 & $9.2\times10^{-3}$ & 11 & 2,048 &704 & $4.6\times10^{-3}$ & 10 \\
    AFP-4088(1) &84,436 & $\pm0.9\%$  & 824   & $9.1\times10^{-3}$ & 10 & 1,024 &324 & $1.2\times10^{-2}$ & 9 \\
    AFP-4088(2) &123,866& $\pm0.25\%$ & 874   & $7.2\times10^{-3}$ & 10 & 1,024 &263 & $1.3\times10^{-2}$ & 9 \\
    \\
    DUP-4096(0) &68,783 & $\pm0.25\%$ & 1,094 & $5.7\times10^{-3}$ & 11 & 2,048 &674 & $3.4\times10^{-3}$ & 10 \\
    DUP-4096(1) &131,671& $\pm1.5\%$  & 335   & $1.4\times10^{-2}$ & 9  & 1,024 &313 & $1.7\times10^{-2}$ & 9 \\
    DUP-4096(2) &210,591& $\pm1.5\%$  & 424   & $1.1\times10^{-2}$ & 9  & 2,048 &452 & $8.0\times10^{-3}$ & 9 \\
    \\
    DUC-2048(0) &60,779 & $\pm0.05\%$ & 8,452 & $2.3\times10^{-2}$ & 14 & 1,024 &837 & $1.8\times10^{-2}$ & 10 \\
    DUC-2048(1) &165,086& $\pm0.3\%$  & 4,828 & $3.5\times10^{-3}$ & 13 & 1,024 &738 & $1.1\times10^{-2}$ & 10 \\
    DUC-2048(2) &243,402& $\pm1\%$    & 1,752 & $1.1\times10^{-2}$ & 11 & 512 &360  & $1.4\times10^{-2}$ & 9 \\
    \bottomrule
  \end{tabular}
  \caption{\centering Comparison of smoothing and digitisation of the original matrix entries. Test cases are labelled by name and number of SPAI infill levels.
  $n_{nz}$ is the number of non-zeros in the matrix, $s_f$ is the smoothing factor, $n_{l}$ is the
  number of entries in the list of unique values, $n_b$ is the number of unique bit patterns and $\epsilon$ is the $L_2$ norm of the difference with the
  solution to the original matrix. $n_{q}=\lceil\log_2 n_b\rceil$ (or $\lceil\log_2 n_l\rceil$) is number of qubits for the digital precision (or list) registers, i.e. for the $r$ register in \Cref{fig:sparse-count-rot-gen}. 
  For digitisation via, say, phase gradients, $\log_2D$ qubits would be
  needed.}
  \label{tab-smooth-digit}
\end{table}

\Cref{tab-smooth-digit} compares the two approaches for selected 3D test cases.
To perform a like-with-like comparison, the solutions to the original system
$Ax=b$ and the digitised/binned system $\tilde{A} \tilde{x}=b$ are normalised
and the $L_2$ norm of their difference, $\epsilon$, is evaluated.
For each case, the smoothing factor and the digitisation
constant are adjusted to give $\epsilon \sim 10^{-2}$.
The adjustment can be quite sensitive and the upper bound for selecting the number
of digitisation bits was set to  $\epsilon \le 2\times10^{-2}$.

For the pressure correction cases (AFP-4088 and DUP-4096) the number of qubits needed
for both techniques is the same or similar.
For the coupled implicit case (DUC-2048), digitisation consistently requires fewer
qubits than binning.

An interesting feature of digitisation, is shown in the $n_b$ column in
\Cref{tab-smooth-digit}. This gives the number of unique bit patterns and hence
rotation angles.
In the DUP-4096 case, the number of unique angles and bit patterns is broadly
similar.
In the DUC-2048 case, the number of unique bit patterns is far fewer, e.g.,
with SPAI(2) there are 360 unique bit patterns relative to 1,752 unique angles.
This is over a factor of 4 reduction in the multiplexed rotation angles using the
encoding in \Cref{subsec_list_rot}. 
The reduction in unique values is a common feature with \Cref{tab-duc-16384-nqubits,tab-afp-15568-nqubits}
which we discuss in \Cref{sec-conclusions}.

Coupled with the reduction in subnormalisation factor, digitised rotation angles, 
whether implemented via a list or digital register, 
mean that unstructured encoding with max-search gives a large
reduction in the number of multiplexed rotations needed for linear solvers 
based on block encoding.

% qubit counts
% ------------
\subsection{Comparison of qubit counts}
\label{subsec_nqubits}

Here, we consider the number of qubits for the high level block encodings
described in \Cref{sec-theory} using the largest structured and unstructured
mesh test cases. We begin with the structured mesh cases to compare the 
banded diagonal and 1-sparse encodings with and without digital rotation.
The comparison of the two 1-sparse approaches are broadly similar for
the unstructured meshes.

% Structured meshes
% -----------------
\subsubsection{Structured meshes}
\label{subsubsec-qcount-str}

\Cref{tab-duc-16384-nqubits} gives a breakdown of the number of qubits needed to
encoded the implicit coupled 16x16x16 structured mesh test case. 
The banded diagonal structured mesh encoding is compared with unstructured mesh
encoding using a list of diagonal offsets, with and without digitised rotation angles.

\begin{table}[ht]
  \centering
  \begin{tabular}{c c c c c c c c c}
    \toprule
    Test case  & $n_c$ & $n_{sp}$ & $n_d$ & $n_{rot}$ & $n_{total}$ & $\Sigma_{rot}$ & $\kappa_s$\\
    \midrule
    \multicolumn{8}{c}{\textit{Structured banded-diagonal encoding}}\\
    DUC-16384(0) & 14 & 0 & 7 & 1 & 22 & 526,364   & 21,053\\
    DUC-16384(1) & 14 & 0 & 9 & 1 & 24 & 1,685,744 & 16,015\\
    DUC-16384(2) & 14 & 0 & 9 & 1 & 24 & 2,785,576 & 13,536\\
    \multicolumn{8}{c}{\textit{1-sparse unstructured offset-list encoding}}\\
    DUC-16384(0) & 14 & 6 & 7 & 1 & 28 & 526,364   & 5,812\\
    DUC-16384(1) & 14 & 8 & 9 & 1 & 32 & 1,685,744 & 3,849\\
    DUC-16384(2) & 14 & 9 & 9 & 1 & 33 & 2,785,576 & 2,766\\
    \multicolumn{8}{c}{\textit{1-sparse unstructured offset-list encoding + digital rotations}}\\
    DUC-16384(0) & 14 & 6 & 7 & 11 & 38 & 1,164 & 5,812\\
    DUC-16384(1) & 14 & 8 & 9 & 11 & 42 & 791 & 3,849\\
    DUC-16384(2) & 14 & 9 & 9 & 11 & 43 & 546 & 2,766\\
    \bottomrule
  \end{tabular}
  \caption{\centering Qubit counts for structured mesh DUC-16384 test case. 
  $n_{c}, n_{sp}, n_d, n_{rot}$ are the numbers of qubits in the
  column, 1-sparse LCU, diagonal and rotation registers respectively, 
  $n_{total}$ is the total number of qubits.
  $\Sigma_{rot}$ is the total number of rotation gates and $\kappa_s$ is the
  subnormalised condition number.
  The unstructured encodings use max-search and a list of diagonal offsets.}
  \label{tab-duc-16384-nqubits}
\end{table}

Firstly, the number of diagonals is the same for both structured and unstructured
mesh block encodings as they are both using the same list of diagonals.
The unstructured mesh block encodings carry the overhead of the 1-sparse register
which increases with the number of infill levels as shown in \Cref{fig_n1sp_duct3d}.
For SPAI(2) preconditioning this increases the total number of qubits from
24 to 33.
Adding the list register for digitised rotation angles adds a further 10 qubits
increasing the total number to 43 for the SPAI(2) case.

Applying digitisation gives a drastic reduction in the number of unique
rotation angles. In the SPAI(2) case, there is a factor of over 5,000 reduction in angles, whilst maintaining a normalised $L_2$ accuracy of $7.2 \times 10^{-3}$, equivalent to an infidelity of $2.6 \times 10^{-5}$.
Results using binning are not included as, from \Cref{tab-smooth-digit}, digitisation
is the same or better.
The benefits of the almost constant subnormalisation factor with infill level is also
evident in the more marked reductions in $\kappa_s$ relative to the banded diagonal
encoding.

% Untructured meshes
% -----------------
\subsubsection{Unstructured meshes}
\label{subsubsec-qcount-unstr}

\Cref{tab-afp-15568-nqubits} gives the qubit breakdown for a large 3D unstructured mesh case, AFP-15568, with 15,568 prism elements.
Although this matrix is approximately the same size as the DUC-16384 case, it is for
a pressure correction matrix which accounts for the differences in the $\Sigma_{rot}$ and 
$\kappa_s$ entries in \Cref{tab-duc-16384-nqubits,tab-afp-15568-nqubits}.

\begin{table}[ht]
  \centering
  \begin{tabular}{c c c c c c c c c}
    \toprule
    Test case  & $n_c$ & $n_{sp}$ & $n_d$ & $n_{rot}$ & $n_{total}$ & $\Sigma_{rot}$ & $\kappa_s$\\
    \midrule
    \multicolumn{8}{c}{\textit{1-sparse unstructured offset-list encoding}}\\
    AFP-15568(0) & 14 & 5 & 11 & 1 & 31 & 205,899 & 239,778\\
    AFP-15568(1) & 14 & 6 & 11 & 1 & 32 & 386,149 & 143,865\\
    AFP-15568(2) & 14 & 6 & 12 & 1 & 33 & 614,852 & 87,846\\
    \multicolumn{8}{c}{\textit{1-sparse unstructured offset-list encoding + digital rotations}}\\
    AFP-15568(0) & 14 & 5 & 11 & 13 & 43 & 2,587 & 239,778\\
    AFP-15568(1) & 14 & 6 & 11 & 13 & 44 & 2,882 & 143,865\\
    AFP-15568(2) & 14 & 6 & 12 & 12 & 44 & 1,873 & 87,846\\
    \bottomrule
  \end{tabular}
  \caption{\centering Qubit counts for unstructured mesh AFP-15568 test case. 
  Column headers as in \Cref{tab-duc-16384-nqubits}.}
  \label{tab-afp-15568-nqubits}
\end{table}

Relative to the structured mesh encoding, there are more diagonals due to the 
unstructured nature of the mesh and a larger number of digitised rotation angles.
The latter is due to the face normals for the prism cells 
pointing in a range of directions whereas, the structured mesh the face
normals are aligned.

% QSVT
% ------------
\subsection{QSVT results}
\label{subsec_qsvt}

Here we show two linear system solutions for a 2D and 3D aerofoil mesh 
with and without digitisation using an emulated QSVT solver. 

% 2D Testcase
% -----------
\subsubsection{2D unstructured aerofoil mesh}
\label{subsubsec-2d-qsvt}

The 2D case uses SPAI preconditioning with 4 levels of infill for the AFT-1022
triangular mesh.
The unstructured block encoding requires 26 1-sparse matrices and gives
a subnormalisation factor, $\alpha=3.76$.
To minimise the number of qubits for the emulation, the encoding shown in
\Cref{subsec-diag-offset} is used, this requires a total of 26 qubits for the
QSVT solution.
The pressure correction matrix has $\lambda_{min} \simeq 0.003$, leading to 
$\kappa_s \simeq 1,250$. From previous studies \cite{lapworth2024qles}, it has been shown that QSVT is tolerant
to approximate values of $\kappa_s$ and the QSVT polynomial approximation is 
based on $\kappa_s=1,000$ and $\epsilon=10^{-4}$. The corresponding polynomial was constructed via the algorithm in \cite{sunderhauf_matrix_2025} and the associated phase factors using the 
open-source
\textsc{NLFT/QSP}
\footnote{\href{https://github.com/LorenzoLaneve/nlft-qsp}{https://github.com/LorenzoLaneve/nlft-qsp}}
software package of \cite{laneve2025generalized}.

Two QSVT calculations were performed to emulate the effect of 
8 and 9 digitisation qubits.
The number of unique bit patterns were 76 and 137 relative to a total
number of entries of 15,662.
The normalised errors, as defined in \Cref{subsec_digit_smooth}, 
were $\epsilon_8=9.3\times10^{-2}$, $\epsilon_9=1.2\times10^{-2}$ for digitisation and 
$\epsilon=8.6\times10^{-3}$ for the non-digitised QSVT.
The $\epsilon_8$ and $\epsilon_9$ values show the cliff-edge effect when too few
digitisation qubits are used.

% figure and text for comparing differences with and without digitisation.
\begin{figure}[ht]
  \centering
  \begin{subfigure}[T]{0.45\textwidth}
      \centering
      \includegraphics[width=0.99\textwidth,trim={0 3.9cm 0 4.5cm},clip]{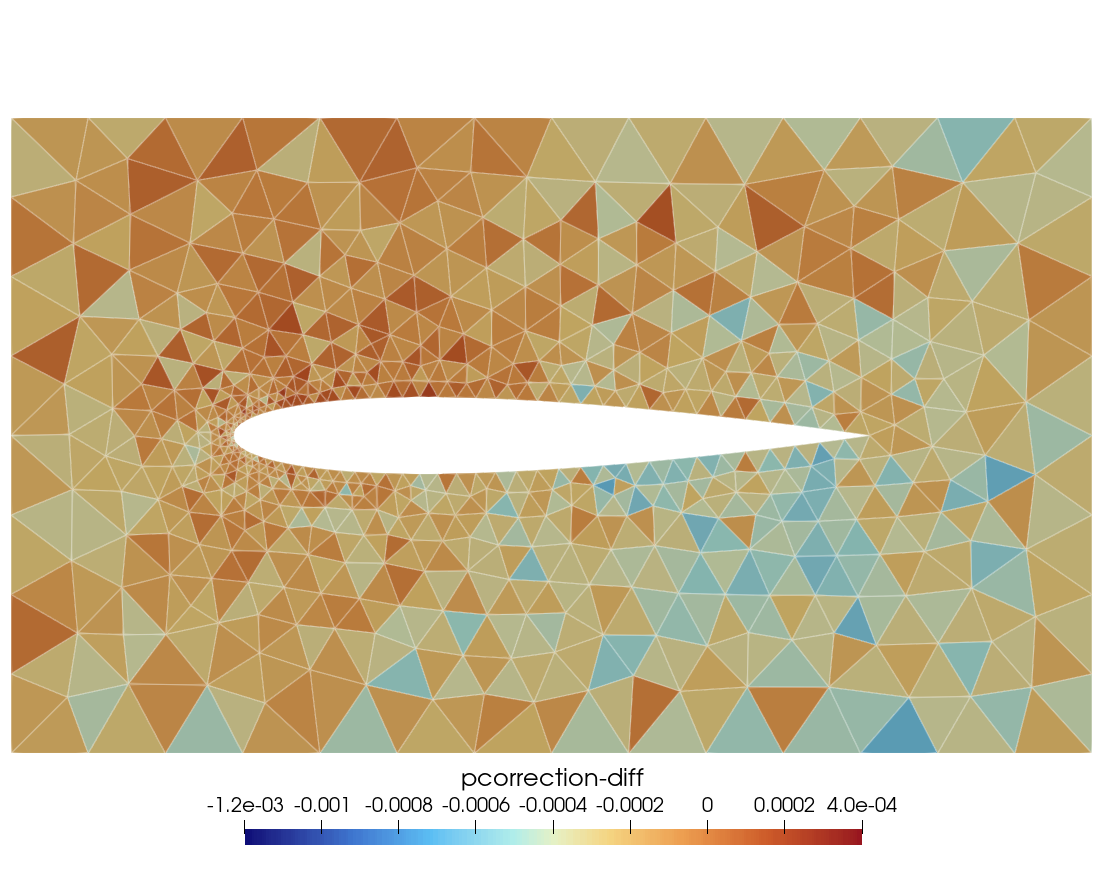}
      \caption{\small No digitisation}
      \label{fig_aft-1022-b0-diff2}
  \end{subfigure} 
  \begin{subfigure}[T]{0.075\textwidth}
        \includegraphics[width=0.99\textwidth]{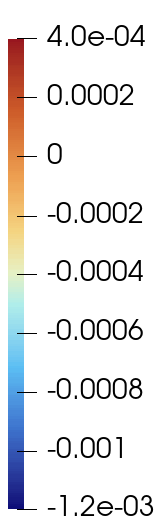}
  \end{subfigure}
  \begin{subfigure}[T]{0.45\textwidth}
      \centering
      \includegraphics[width=0.99\textwidth,trim={0 3.9cm 0 4.5cm},clip]{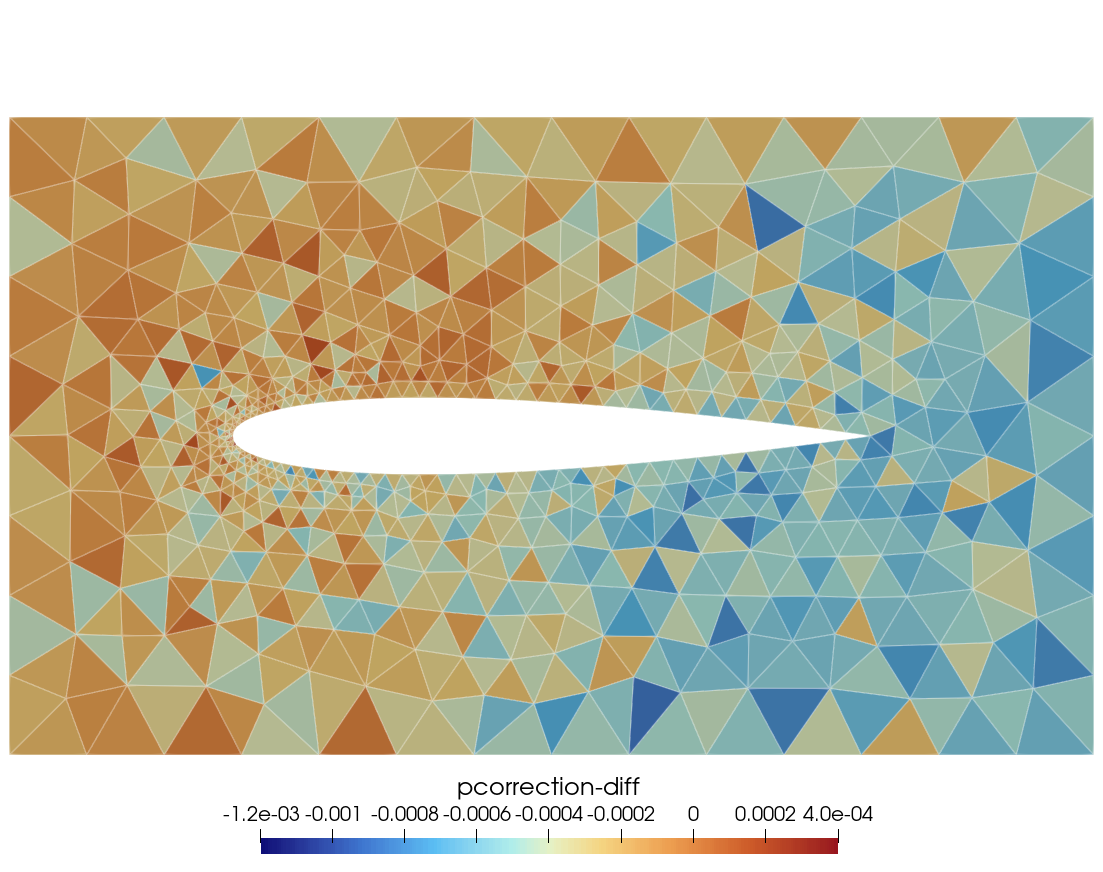}
      \caption{\small 9-bit digitisation.}
      \label{fig_aft-1022-b9-diff2}
  \end{subfigure}
  \caption{\centering AFT-1022 2D aerofoil testcase. Contours of the difference between classical and QSVT solutions with and without digitisation.}
  \label{fig_aft-1022-diff}
\end{figure}

\Cref{fig_aft-1022-b0-diff2,fig_aft-1022-b9-diff2} show the local differences of the QSVT
solutions with the classical solution for the non-digitised and 9-bit digitised encodings
respectively.
The lack of smoothness in the figures is due to the fact we are solving a correction equation
of the form $A\delta x=\delta b$ and $\delta b$ is not smooth.
The digitised solution has higher differences near the exit boundary, this is not unexpected as the
pressure correction boundary condition is applied at point on the inlet boundary.
The 9-bit digitised solution shows good overall agreement with the non-digitised solution and reduces the number of rotation gates by a factor of over 100.

% 3D Testcase
% -----------
\subsubsection{3D unstructured aerofoil mesh}
\label{subsubsec-3d-qsvt}

The 3D case uses SPAI preconditioning without any infill for the AFH-2040
hexahedral mesh.
The unstructured block encoding requires 23 1-sparse matrices and gives
a subnormalisation factor, $\alpha=3.90$.
As above, the encoding shown in
\Cref{subsec-diag-offset} is used, requiring a total of 28 qubits for the
QSVT solution.
The pressure correction matrix has $\lambda_{min} \simeq 0.0003$, leading to 
$\kappa_s \simeq 13,000$. 
The QSVT calculation was performed with $\kappa_s=10,000$ and $\epsilon=10^{-5}$, requiring a
polynomial of degree 101,267.

\begin{figure}[ht]
  \centering
  \begin{subfigure}[b]{0.49\textwidth}
      \centering
      \includegraphics[width=0.99\textwidth,trim={0 0 0 2.5cm},clip]{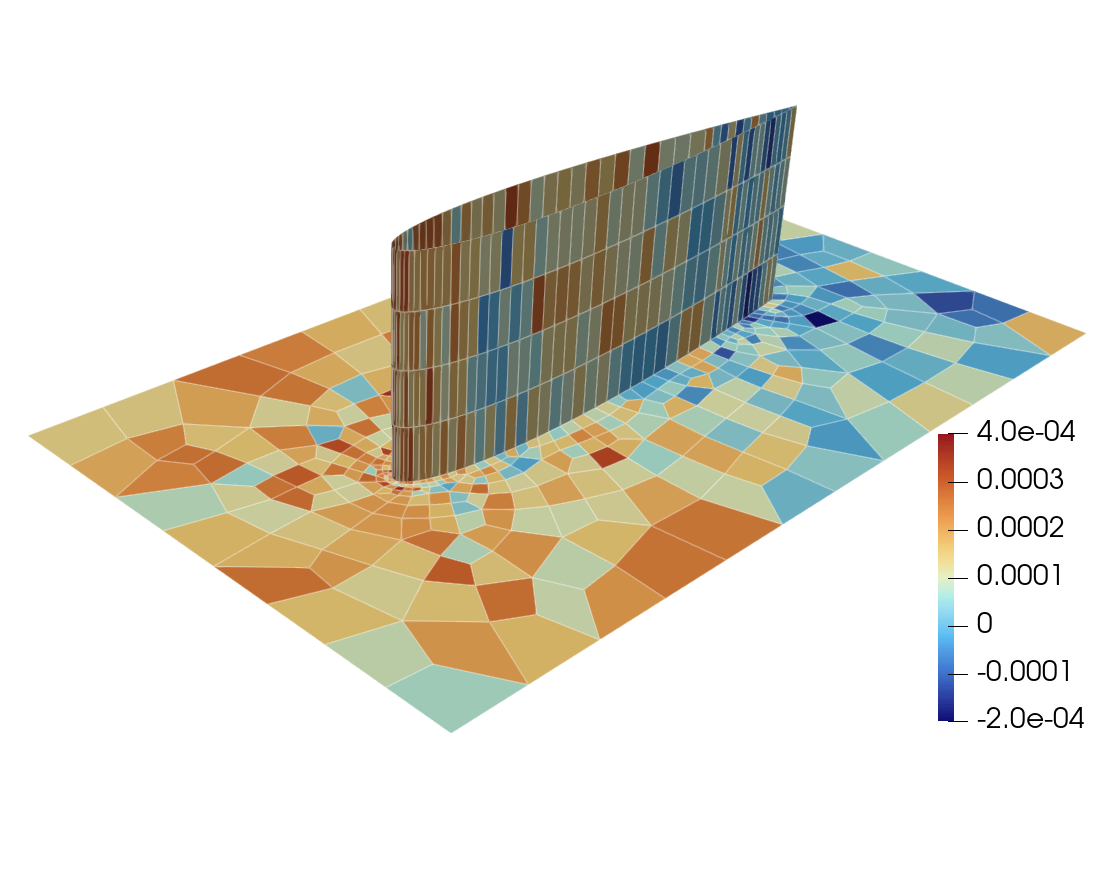}
      \caption{\small No digitisation.}
      \label{fig_afh-2040-b0-diff}
  \end{subfigure}
  \begin{subfigure}[b]{0.49\textwidth}
      \centering
      \includegraphics[width=0.99\textwidth,trim={0 0 0 2.5cm},clip]{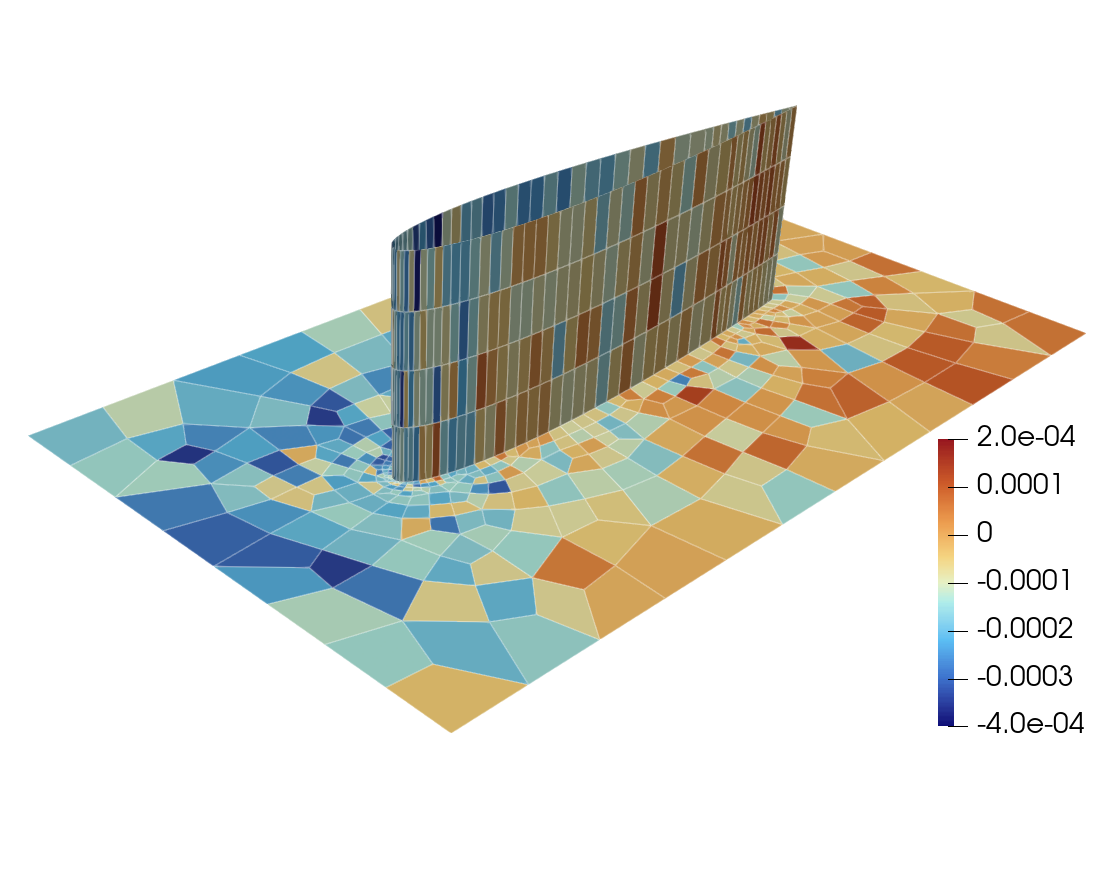}
      \caption{\small 12-bit digitisation.}
      \label{fig_afh-2040-b12-diff}
  \end{subfigure} 
  \caption{\centering AFH-2040 3D aerofoil testcase. 
  Contours of the difference between classical and QSVT solutions with and without digitisation.}
  \label{fig_afh-2040-b0}
\end{figure}

\Cref{fig_afh-2040-b0-diff} shows the local differences between the QSVT
solution and the classical solution without digitisation. The $L_2$ difference between the normalised QSVT and classical
state vectors is $7.0\times10^{-3}$.
\Cref{fig_afh-2040-b12-diff} shows the difference between the QSVT and classical solutions using 12-bit digitisation.
Note that the contour scales have the same range but are offset by $2.0\times10^{-4}$. Although the contour patterns are different, the
two solutions have the same levels of error. This is confirmed by the $L_2$ error of
$7.0\times10^{-3}$  relative to the classical solution.
The 12-bit digitisation creates 1,137 unique bit patterns relative to a total of 30,123 entries.
As in the 2D case, there is a cliff-edge effect with 11-bit digitisation giving a $L_2$ difference of $5.8\times10^{-2}$.

\section{Discussion and Conclusions}
\label{sec-conclusions}

We have demonstrated a new unstructured mesh block-encoding technique
which coupled with max-search for constructing the
1-sparse matrices has shown significant benefits.
Primarily, giving near constant subnormalisation factors as the number of 
preconditioning infill levels increases. 

This gives us what we consider to be three key ingredients
necessary for QLES to show a quantum advantage:
\begin{itemize}
    \item Efficient preconditioning as a classical preprocessing step.
    \item Block-encodings that do not negate the benefits of preconditioning.
    \item The use of digitisation to drastically reduce the number of multiplexed
    rotation gates, particularly as the number of preconditioning infill levels is increased.
\end{itemize}
Together these lead to orders of magnitude improvement in the time-efficiency of quantum linear equation solvers such as QSVT.

A recurring feature across the test cases is that the unstructured mesh 
encoding using max-search gives near constant 
subnormalisation factors as the number of preconditioning infill levels 
increases.
In \Cref{subsec-3d-unstr-meshes,subsec-3d-str-meshes} the advancing front
nature of SPAI was posited as the reason for this behaviour, which we now expand.
For a given cell, each new level of SPAI infill adds an extra layer of cells (front) to its difference stencil.
By construction, when the matrix $PA$ is computed, the matrix entries are zero for those
cells fully enclosed by the front. The main diagonal entry is also 1 by construction.
From \Cref{sec-spai-front}, the minimisation used by SPAI ensures that as the front expands, the matrix entries on the
front reduce in magnitude, giving the observed behaviour.

%Although the range in the size of the test case matrices analysed is relatively
%small, there is some consistency in the number of digitisation qubits needed to
%meet the accuracy requirement of $\epsilon \lesssim 10^{-2}$.

We also see a trend that as the number of infill levels increases, the number
of digitised rotation angles reduces, suggesting that there is an equalising effect
of increasing infill, see \Cref{sec-spai-front}.
This can be seen more clearly in the smoothing results for cases AFP-4088(0) and AFP-4088(2) in 
\Cref{tab-smooth-digit}. These have the same smoothing factor, $s_f$, and give
a like for like comparison of the effect of infill. 
Adding 2 more levels of infill increases the number of matrix entries by a factor of 2.5 but reduces the number of unique values by a factor of almost 2.
Hence, the average number of entries per bin is increased by a factor of 5.
Since the size of the bins is directly related to the smoothness of the ordered list of matrix entries, the same effect is seen with digitisation which does not use bins.

The 2-3 orders of magnitude reductions in the number of rotation gates with
digitisation reduces both the number of multiplexed rotations and number of 
control qubits as the latter depends on the number of unique angles rather 
than the number of non-zeros, $n_{nz}$, in the matrix.
As shown in \Cref{fig:sparse-count-rot-gen}, the total number of multiplexed NOT
operators remains as $2n_{nz}$, but these are much cheaper to implement than multiplexed rotations.
The broad equivalence to digitising matrix entries or rotation angles gives the
freedom to apply rotation gates directly, as done here, or use a sampling technique
\cite{babbush2018encoding,lemieux2024quantum,sanders2019black}.

Beyond focussing on rotation gates, we have also discussed our method in terms of Toffoli/T gates in \Cref{app-multiplex-ssp}. We have seen that it can give some advantage over the q-norm encoding. In practice, this could depend on the specific matrix and it may be prudent to automatically try different block encoding schemes and select the best.

Whilst the techniques used here do not rely on parallelisation using circuit chunking
and reassembly via SWAP gates \cite{low2024trading,goldstein2026compas}, 
they do trade qubits for circuit depth.
One of the largest benefits of a low subnormalisation factor is the
reduction in the degree of the QSP polynomial.
Taking the SPAI(2) cases for structured and unstructured encoding from
\Cref{tab-duc-16384-nqubits}, the latter requires 100,000 fewer phase angles.

Finally, all techniques to improve quantum solvers must be viewed in the context of
whether the same techniques can be applied to classical solvers.
Digitisation using, say, 9 bits is not a classical benefit as 
CPUs are optimised for IEEE compliant 32 and 64 bit variables.
SPAI offers the same benefit to both, although many classical solvers avoid
large matrix vector multiplications in favour of \textit{matrix-free} implementations.
The latter are easier to parallelise and are less prone to being memory-bound.
There is no classical equivalent to encoding, say, 2,785,576 matrix entries using
just 546 multiplexed rotations.

A further consideration is that the length of the block-encoding circuit depends only
on the number of non-zeros in the matrix and not their locations.
This is counter to classical computing, where unstructured solvers carry a
significant overhead due to the need for indirect addressing of cell neighbours via a connectivity table.
This may be one area where we can point to an advantage
from the inherent matrix mechanics underlying quantum computation.

\section{Acknowledgements}

\ifnames
Authors L.L. and J.S thank Paul Mannix of PsiQuantum and Angus Kan formerly of PsiQuantum, now of ApexQuantum,
for informative discussions on rejection sampling under grant reference SQPOC01 of the UK SPARQ Quantum Computing Programme.
We also thank Chukwudubem Umeano of Rolls-Royce for his useful comments on our manuscript. C.S.~thanks Aleksei Ivanov of Riverlane for early discussions on the block-encoding method.

The permission of Rolls-Royce plc and Riverlane to publish this work is gratefully acknowledged.
\fi

% References
% ----------
\addcontentsline{toc}{section}{References}
\bibliography{references}

@book{ahokas2004improved,
  title={Improved algorithms for approximate quantum Fourier transforms and sparse Hamiltonian simulations},
  author={Ahokas, Graeme Robert},
  year={2004},
  publisher={University of Calgary}
}

@article{alonso2025quantum,
author={Alonso-Linaje, Guillermo and Azad, Utkarsh and Soni, Jay and Smalley, Jarrett and Lapworth, Leigh and Arrazola, Juan Miguel},
	title={Quantum compilation framework for data loading},
	journal={Quantum Science and Technology},
	url={http://iopscience.iop.org/article/10.1088/2058-9565/ae5fc7},
	year={2026}
}

@article{araujo2021divide,
  title={A divide-and-conquer algorithm for quantum state preparation},
  author={Araujo, Israel F and Park, Daniel K and Petruccione, Francesco and da Silva, Adenilton J},
  journal={Scientific Reports},
  volume={11},
  number={1},
  pages={1--12},
  year={2021},
  publisher={Nature Publishing Group}
}

@article{babbush2018encoding,
  title={Encoding electronic spectra in quantum circuits with linear T complexity},
  author={Babbush, Ryan and Gidney, Craig and Berry, Dominic W and Wiebe, Nathan and McClean, Jarrod and Paler, Alexandru and Fowler, Austin and Neven, Hartmut},
  journal={Physical Review X},
  volume={8},
  number={4},
  pages={041015},
  year={2018},
  publisher={APS}
}

@article{benson1984parallel,
  title={Parallel algorithms for the solution of certain large sparse linear systems},
  author={Benson, M and Krettmann, J{\"u}rgen and Wright, M},
  journal={International journal of computer mathematics},
  volume={16},
  number={4},
  pages={245--260},
  year={1984},
  publisher={Taylor \& Francis}
}

@article{berry2007efficient,
  title={Efficient quantum algorithms for simulating sparse Hamiltonians},
  author={Berry, Dominic W and Ahokas, Graeme and Cleve, Richard and Sanders, Barry C},
  journal={Communications in Mathematical Physics},
  volume={270},
  number={2},
  pages={359--371},
  year={2007},
  publisher={Springer}
}

@inproceedings{berry2014exponential,
  title={Exponential improvement in precision for simulating sparse Hamiltonians},
  author={Berry, Dominic W and Childs, Andrew M and Cleve, Richard and Kothari, Robin and Somma, Rolando D},
  booktitle={Proceedings of the forty-sixth annual ACM symposium on Theory of computing},
  pages={283--292},
  year={2014}
}

@article{bravo2026optimal,
  title={An Optimal Quantum Linear Systems Algorithm},
  author={Bravo-Prieto, Carlos and Harrow, Aram W and Kothari, Robin},
  journal={arXiv preprint arXiv:2609.35660},
  year={2026}
}

@inproceedings{camps2022fable,
  title={Fable: Fast approximate quantum circuits for block-encodings},
  author={Camps, Daan and Van Beeumen, Roel},
  booktitle={2022 IEEE International Conference on Quantum Computing and Engineering (QCE)},
  pages={104--113},
  year={2022},
  organization={IEEE}
}

@article{camps2024explicit,
  title={Explicit quantum circuits for block encodings of certain sparse matrices},
  author={Camps, Daan and Lin, Lin and Van Beeumen, Roel and Yang, Chao},
  journal={SIAM Journal on Matrix Analysis and Applications},
  volume={45},
  number={1},
  pages={801--827},
  year={2024},
  publisher={SIAM}
}

@inproceedings{chakraborty_2019,
  author =	{Chakraborty, Shantanav and Gily\'{e}n, Andr\'{a}s and Jeffery, Stacey},
  title =	{{The Power of Block-Encoded Matrix Powers: Improved Regression Techniques via Faster Hamiltonian Simulation}},
  booktitle =	{46th International Colloquium on Automata, Languages, and Programming (ICALP 2019)},
  pages =	{33:1--33:14},
  series =	{Leibniz International Proceedings in Informatics (LIPIcs)},
  ISBN =	{978-3-95977-109-2},
  ISSN =	{1868-8969},
  year =	{2019},
  volume =	{132},
  editor =	{Baier, Christel and Chatzigiannakis, Ioannis and Flocchini, Paola and Leonardi, Stefano},
  publisher =	{Schloss Dagstuhl -- Leibniz-Zentrum f{\"u}r Informatik},
  address =	{Dagstuhl, Germany},
  URL =		{https://drops.dagstuhl.de/entities/document/10.4230/LIPIcs.ICALP.2019.33},
  URN =		{urn:nbn:de:0030-drops-106092},
  doi =		{10.4230/LIPIcs.ICALP.2019.33}
}

@article{chow1998approximate,
  title={Approximate inverse preconditioners via sparse-sparse iterations},
  author={Chow, Edmond and Saad, Yousef},
  journal={SIAM Journal on Scientific Computing},
  volume={19},
  number={3},
  pages={995--1023},
  year={1998},
  publisher={SIAM}
}

@article{clader2013preconditioned,
  title={Preconditioned quantum linear system algorithm},
  author={Clader, B David and Jacobs, Bryan C and Sprouse, Chad R},
  journal={Physical Review Letters},
  volume={110},
  number={25},
  pages={250504},
  year={2013},
  publisher={APS}
}

@article{clader2022quantum,
  title={Quantum resources required to block-encode a matrix of classical data},
  author={Clader, B David and Dalzell, Alexander M and Stamatopoulos, Nikitas and Salton, Grant and Berta, Mario and Zeng, William J},
  journal={IEEE Transactions on Quantum Engineering},
  volume={3},
  pages={1--23},
  year={2022},
  publisher={IEEE}
}

@article{costa2022optimal,
  title={Optimal scaling quantum linear-systems solver via discrete adiabatic theorem},
  author={Costa, Pedro CS and An, Dong and Sanders, Yuval R and Su, Yuan and Babbush, Ryan and Berry, Dominic W},
  journal={PRX quantum},
  volume={3},
  number={4},
  pages={040303},
  year={2022},
  publisher={APS}
}

@inproceedings{cuthill1969reducing,
  title={Reducing the bandwidth of sparse symmetric matrices},
  author={Cuthill, Elizabeth and McKee, James},
  booktitle={Proceedings of the 1969 24th national conference},
  pages={157--172},
  year={1969}
}

@article{dalzell2023quantum,
  title={Quantum algorithms: A survey of applications and end-to-end complexities},
  author={Dalzell, Alexander M and McArdle, Sam and Berta, Mario and Bienias, Przemyslaw and Chen, Chi-Fang and Gily{\'e}n, Andr{\'a}s and Hann, Connor T and Kastoryano, Michael J and Khabiboulline, Emil T and Kubica, Aleksander and others},
  journal={arXiv preprint arXiv:2310.03011},
  year={2023}
}

@article{dalzell2024shortcut,
  title={A shortcut to an optimal quantum linear system solver},
  author={Dalzell, Alexander M},
  journal={arXiv preprint arXiv:2406.12086},
  year={2024}
}

@article{dalzell2026faster,
  title={Faster quantum linear system solver beyond the condition number},
  author={Dalzell, Alexander M and Li, Jianqiang and Su, Yuan},
  journal={arXiv preprint arXiv:2607.07691},
  year={2026}
}

@article{darwish2009coupled,
  title={A coupled finite volume solver for the solution of incompressible flows on unstructured grids},
  author={Darwish, Marwan and Sraj, Ihab and Moukalled, Fadl},
  journal={Journal of computational physics},
  volume={228},
  number={1},
  pages={180--201},
  year={2009},
  publisher={Elsevier}
}

@book{george1981computer,
  title={Computer solution of large sparse positive definite systems},
  author={George, Alan and Liu, Joseph W},
  year={1981},
  publisher={Prentice Hall Professional Technical Reference}
}

@article{gidney2018halving,
  title={Halving the cost of quantum addition},
  author={Gidney, Craig},
  journal={Quantum},
  volume={2},
  pages={74},
  year={2018},
  publisher={Verein zur F{\"o}rderung des Open Access Publizierens in den Quantenwissenschaften}
}

@article{gilyen2018quantum,
  title={Quantum singular value transformation and beyond: exponential improvements for quantum matrix arithmetics [{Full version]}},
  author={Gily{\'e}n, Andr{\'a}s and Su, Yuan and Low, Guang Hao and Wiebe, Nathan},
  journal={arXiv preprint arXiv:1806.01838},
  year={2018}
}

@inproceedings{gilyen2019quantum,
  title={Quantum singular value transformation and beyond: exponential improvements for quantum matrix arithmetics},
  author={Gily{\'e}n, Andr{\'a}s and Su, Yuan and Low, Guang Hao and Wiebe, Nathan},
  booktitle={Proceedings of the 51st Annual ACM SIGACT Symposium on Theory of Computing},
  pages={193--204},
  year={2019},
  howpublished = {\url{https://dl.acm.org/doi/abs/10.1145/3313276.3316366}}
}

@article{golden2022quantum,
  title={Quantum computing and preconditioners for hydrological linear systems},
  author={Golden, John and O’Malley, Daniel and Viswanathan, Hari},
  journal={Scientific Reports},
  volume={12},
  number={1},
  pages={22285},
  year={2022},
  publisher={Nature Publishing Group UK London}
}

@article{goldfriend2026approximate,
  title={Approximate Quantum Linear Solvers for Hybrid CFD: End-to-End Analysis with a Chebyshev-LCU Approach},
  author={Goldfriend, Tomer and Lapworth, Leigh and Yoran, Nadav and Naveh, Amir},
  journal={arXiv preprint arXiv:2606.01067},
  year={2026}
}

@inproceedings{goldstein2026compas,
  title={COMPAS: A Distributed Multi-Party SWAP Test for Parallel Quantum Algorithms},
  author={Goldstein-Gelb, Brayden and Liu, Kun and Martyn, John M and Zhou, Hengyun and Ding, Yongshan and Liu, Yuan},
  booktitle={Proceedings of the 31st ACM International Conference on Architectural Support for Programming Languages and Operating Systems, Volume 2},
  pages={426--441},
  year={2026}
}

@article{harrow2009quantum,
  title={Quantum algorithm for linear systems of equations},
  author={Harrow, Aram W and Hassidim, Avinatan and Lloyd, Seth},
  journal={Physical Review Letters},
  volume={103},
  number={15},
  pages={150502},
  year={2009},
  publisher={APS}
}

@article{hosaka2023preconditioning,
  title={Preconditioning for a variational quantum linear solver},
  author={Hosaka, Aruto and Yanagisawa, Koichi and Koshikawa, Shota and Kudo, Isamu and Alifu, Xiafukaiti and Yoshida, Tsuyoshi},
  journal={arXiv preprint arXiv:2312.15657},
  year={2023}
}

@article{jennings2025randomized,
  title={Randomized adiabatic quantum linear solver algorithm with optimal complexity scaling and detailed running costs},
  author={Jennings, David and Lostaglio, Matteo and Pallister, Sam and Sornborger, Andrew T and Suba{\c{s}}{\i}, Yi{\u{g}}it},
  journal={PRX Quantum},
  volume={6},
  number={4},
  pages={040373},
  year={2025},
  publisher={APS}
}

@article{khattar2025rise,
  title={Rise of conditionally clean ancillae for efficient quantum circuit constructions},
  author={Khattar, Tanuj and Gidney, Craig},
  journal={Quantum},
  volume={9},
  pages={1752},
  year={2025},
  publisher={Verein zur F{\"o}rderung des Open Access Publizierens in den Quantenwissenschaften}
}

@phdthesis{kothari2014efficient,
  title={Efficient algorithms in quantum query complexity},
  author={Kothari, Robin},
  year={2014},
  school={University of Waterloo},
  url = {https://uwspace.uwaterloo.ca/handle/10012/8625}
}

@article{laneve2025generalized,
  title={Generalized quantum signal processing and non-linear Fourier transform are equivalent},
  author={Laneve, Lorenzo},
  journal={arXiv preprint arXiv:2503.03026},
  year={2025}
}

@article{lapworth2022hybrid,
  title={A Hybrid Quantum-Classical {CFD} Methodology with Benchmark {HHL} Solutions},
  author={Lapworth, Leigh},
  journal={arXiv preprint arXiv:2206.00419},
  year={2022}
}

@article{lapworth2022implicit,
  title={Implicit Hybrid Quantum-Classical {CFD} Calculations using the {HHL} Algorithm},
  author={Lapworth, Leigh},
  journal={arXiv preprint arXiv:2209.07964},
  year={2022}
}

@article{lapworth2024qles,
  title={{L-QLES}: Sparse laplacian generator for evaluating quantum linear equation solvers},
  author={Lapworth, Leigh},
  journal={arXiv preprint arXiv:2402.12266},
  year={2024}
}

@article{lapworth2025preconditioned,
  title={Preconditioned block encodings for quantum linear systems},
  author={Lapworth, Leigh and S{\"u}nderhauf, Christoph},
  journal={Quantum Science and Technology},
  volume={10},
  number={4},
  pages={045064},
  year={2025},
  publisher={IOP Publishing}
}

@article{lapworth2026probability,
  title={The probability of success in solving linear systems of equations using quantum singular value transformation},
  author={Lapworth, Leigh},
  journal={Acta Mechanica Sinica},
  volume={42},
  number={6},
  year={2026},
  publisher={Springer}
}

@article{lemieux2024quantum,
  title={Quantum sampling algorithms for quantum state preparation and matrix block-encoding},
  author={Lemieux, Jessica and Lostaglio, Matteo and Pallister, Sam and Pol, William and Seetharam, Karthik and Sim, Sukin and {\c{S}}ahino{\u{g}}lu, Burak},
  journal={arXiv preprint arXiv:2405.11436},
  year={2024}
}

@article{li2025new,
  title={A New Quantum Linear System Algorithm Beyond the Condition Number and Its Application to Solving Multivariate Polynomial Systems},
  author={Li, Jianqiang},
  journal={arXiv preprint arXiv:2510.05588},
  year={2025}
}

@article{lin2022lecture,
  title={Lecture notes on quantum algorithms for scientific computation},
  author={Lin, Lin},
  journal={arXiv preprint arXiv:2201.08309},
  year={2022}
}

@article{low2024trading,
  title={Trading T gates for dirty qubits in state preparation and unitary synthesis},
  author={Low, Guang Hao and Kliuchnikov, Vadym and Schaeffer, Luke},
  journal={Quantum},
  volume={8},
  pages={1375},
  year={2024},
  publisher={Verein zur F{\"o}rderung des Open Access Publizierens in den Quantenwissenschaften}
}

@article{martyn2021grand,
  title={Grand unification of quantum algorithms},
  author={Martyn, John M and Rossi, Zane M and Tan, Andrew K and Chuang, Isaac L},
  journal={PRX Quantum},
  volume={2},
  number={4},
  pages={040203},
  year={2021},
  month = {Dec},
  publisher = {American Physical Society},
  doi = {10.1103/PRXQuantum.2.040203},
  url = {https://link.aps.org/doi/10.1103/PRXQuantum.2.040203}
}

@article{mottonen2004transformation,
  title={Transformation of quantum states using uniformly controlled rotations},
  author={M{\"o}tt{\"o}nen, Mikko and Vartiainen, Juha J and Bergholm, Ville and Salomaa, Martti M},
  journal={arXiv preprint quant-ph/0407010},
  year={2004}
}

@article{nguyen2022block,
  title={Block-encoding dense and full-rank kernels using hierarchical matrices: applications in quantum numerical linear algebra},
  author={Nguyen, Quynh T and Kiani, Bobak T and Lloyd, Seth},
  journal={Quantum},
  volume={6},
  pages={876},
  year={2022},
  publisher={Verein zur F{\"o}rderung des Open Access Publizierens in den Quantenwissenschaften}
}

@article{o2025quantum,
  title={Quantum state preparation via piecewise QSVT},
  author={O'Brien, Oliver and S{\"u}nderhauf, Christoph},
  journal={Quantum},
  volume={9},
  pages={1786},
  year={2025},
  publisher={Verein zur F{\"o}rderung des Open Access Publizierens in den Quantenwissenschaften}
}

@article{patankar1972calculation,
  title={A calculation procedure for heat, mass and momentum transfer in three-dimensional parabolic flows},
  author={Patankar, SV and Spalding, DB},
  journal={Int. J. Heat Mass Transfer},
  volume={15},
  pages={1787--1806},
  year={1972},
}

@article{patel2026near,
  title={A near-optimal joint lower bound for sparse quantum linear system solvers},
  author={Patel, Dhrumil},
  journal={arXiv preprint arXiv:2609.22445},
  year={2026}
}

@article{pechan2026block,
  title={Block encoding of the three-dimensional heterogeneous Poisson equation with application to fracture flow},
  author={Pechan, Austin and Golden, John and O’Malley, Daniel},
  journal={Physical Review Applied},
  volume={25},
  number={4},
  pages={044038},
  year={2026},
  publisher={APS}
}

@article{rupprecht2026sparse,
  title={Sparse quantum state preparation with improved Toffoli cost},
  author={Rupprecht, Felix and W{\"o}lk, Sabine},
  journal={arXiv preprint arXiv:2601.09388},
  year={2026}
}

@book{saad2003iterative,
  title={Iterative methods for sparse linear systems},
  author={Saad, Yousef},
  year={2003},
  publisher={SIAM}
}

@article{sanders2019black,
  title={Black-box quantum state preparation without arithmetic},
  author={Sanders, Yuval R and Low, Guang Hao and Scherer, Artur and Berry, Dominic W},
  journal={Physical Review Letters},
  volume={122},
  number={2},
  pages={020502},
  year={2019},
  publisher={APS}
}

@article{setty2026block,
  title={Block encoding of sparse matrices via coherent permutation},
  author={Setty, Abhishek},
  journal={Physical Review Research},
  volume={8},
  number={3},
  pages={033282},
  year={2026},
  publisher={APS}
}

@article{shao2018quantum,
  title={Quantum circulant preconditioner for a linear system of equations},
  author={Shao, Changpeng and Xiang, Hua},
  journal={Physical Review A},
  volume={98},
  number={6},
  pages={062321},
  year={2018},
  publisher={APS}
}

@article{sunderhauf2024block,
  title={Block-encoding structured matrices for data input in quantum computing},
  author={S{\"u}nderhauf, Christoph and Campbell, Earl and Camps, Joan},
  journal={Quantum},
  volume={8},
  pages={1226},
  year={2024},
  publisher={Verein zur F{\"o}rderung des Open Access Publizierens in den Quantenwissenschaften}
}

@article{tong2021fast,
  title={Fast inversion, preconditioned quantum linear system solvers, fast Green's-function computation, and fast evaluation of matrix functions},
  author={Tong, Yu and An, Dong and Wiebe, Nathan and Lin, Lin},
  journal={Physical Review A},
  volume={104},
  number={3},
  pages={032422},
  year={2021},
  publisher={APS}
}

@article{wan2018asymptotic,
  title={Asymptotic quantum algorithm for the Toeplitz systems},
  author={Wan, Lin-Chun and Yu, Chao-Hua and Pan, Shi-Jie and Gao, Fei and Wen, Qiao-Yan and Qin, Su-Juan},
  journal={Physical Review A},
  volume={97},
  number={6},
  pages={062322},
  year={2018},
  publisher={APS}
}

@article{yang2025dictionary,
  title={Dictionary-based Block Encoding of Sparse Matrices with Low Subnormalization and Circuit Depth},
  author={Yang, Chunlin and Li, Zexian and Yao, Hongmei and Fan, Zhaobing and Zhang, Guofeng and Liu, Jianshe},
  journal={Quantum},
  volume={9},
  pages={1805},
  year={2025},
  publisher={Verein zur F{\"o}rderung des Open Access Publizierens in den Quantenwissenschaften}
}

@article{zhang2024circuit,
  title={Circuit complexity of quantum access models for encoding classical data},
  author={Zhang, Xiao-Ming and Yuan, Xiao},
  journal={npj Quantum Information},
  volume={10},
  number={1},
  pages={42},
  year={2024},
  publisher={Nature Publishing Group UK London}
}

@article{zhou2017efficient,
  title={Efficient quantum circuits for dense circulant and circulant like operators},
  author={Zhou, SS and Wang, JB},
  journal={Royal Society open science},
  volume={4},
  number={5},
  pages={160906},
  year={2017},
  publisher={The Royal Society Publishing}
}

@article{malvetti_quantum_2021,
    title = {Quantum {Circuits} for {Sparse} {Isometries}},
    volume = {5},
    issn = {2521-327X},
    url = {https://quantum-journal.org/papers/q-2021-03-15-412/},
    doi = {10.22331/q-2021-03-15-412},
    language = {en},
    urldate = {2026-07-20},
    journal = {Quantum},
    author = {Malvetti, Emanuel and Iten, Raban and Colbeck, Roger},
    month = mar,
    year = {2021},
    pages = {412},
}

@article{berry_qubitization_2019,
    title = {Qubitization of {Arbitrary} {Basis} {Quantum} {Chemistry} {Leveraging} {Sparsity} and {Low} {Rank} {Factorization}},
    volume = {3},
    issn = {2521-327X},
    url = {http://arxiv.org/abs/1902.02134},
    doi = {10.22331/q-2019-12-02-208},
    language = {en},
    urldate = {2025-08-07},
    journal = {Quantum},
    author = {Berry, Dominic W. and Gidney, Craig and Motta, Mario and McClean, Jarrod R. and Babbush, Ryan},
    month = dec,
    year = {2019},
    note = {arXiv:1902.02134 [quant-ph]},
    pages = {208},
}

@article{shende_synthesis_2006,
    title = {Synthesis of {Quantum} {Logic} {Circuits}},
    volume = {25},
    issn = {0278-0070, 1937-4151},
    url = {http://arxiv.org/abs/quant-ph/0406176},
    doi = {10.1109/TCAD.2005.855930},
    number = {6},
    urldate = {2025-09-09},
    journal = {IEEE Transactions on Computer-Aided Design of Integrated Circuits and Systems},
    author = {Shende, Vivek V. and Bullock, Stephen S. and Markov, Igor L.},
    month = jun,
    year = {2006},
    note = {arXiv:quant-ph/0406176},
    pages = {1000--1010},
}

@misc{sunderhauf_matrix_2025,
    title = {Matrix inversion polynomials for the quantum singular value transformation},
    url = {http://arxiv.org/abs/2507.15537},
    doi = {10.48550/arXiv.2507.15537},
    urldate = {2026-10-06},
    publisher = {arXiv},
    author = {Sünderhauf, Christoph and Németh, Zalán and Walayat, Adnaan and Patterson, Andrew and Berntson, Bjorn K.},
    month = jul,
    year = {2025},
    note = {arXiv:2507.15537 [quant-ph]},
}
\bibliographystyle{ieeetr}
%\bibliographystyle{apalike}

% use instead of \appendix to change references from section to appendix
% ----------------------------------------------------------------------
\begin{appendices}
  \crefalias{section}{appendix}
  
\section{Multiplexing Sparse State Preparation and Permutation}
\label{app-multiplex-ssp}

Whereas the main text of the paper has focused on rotation gates, in this appendix we perform an analysis of the block encodings in terms of non-Clifford resources (Toffoli and T gates). We focus on the new 1-sparse decomposition encoding and its closest contender, the q-norm encoding.
These block encoding methods rely strongly on multiplexed permutations (1-sparse decomposition) as well as multiplexed sparse state preparation ($q$-norm encoding). Multiplexed in the context of quantum circuits means that, depending on the state $\ket{i}$ of an address register, a different unitary $U_i$ (permutation or sparse state preparation) is applied to a target register. A permutation is meant to be an isometry effecting a permutation of the computational basis.

In this appendix, we study multiplexed permutations and multiplexed sparse state preparation. This allows us to compare the implication on gate cost of our new 1-sparse and the $q$-norm block encoding.
For simplicity, this exposition focuses on dimensions that are powers of 2, which can usually be relaxed. Rather than giving exact gate count costs, we will describe the main circuit elements required. In turn, these can have different costs depending on chosen implementations, for example lookup tables (LUTs, also called QROM, QROAM, Select/SWAP networks) allow data lookup with a tunable space time tradeoff \cite{low2024trading}. More recently, it has been shown that also unary iteration (QROM) can be implemented with a variety of tradeoffs in ancilla qubit and Toffoli count \cite{khattar2025rise}.

\subsection{Multiplexed permutations}
A multiplexed permutation is defined by a set of permutations $π_i$. In a quantum circuit, it permutes the computational basis:
\begin{equation}
    \text{MultiPerm}\ket{l}\ket{i} = \ket{l}\ket{π_l(i)}.
\end{equation}
In this appendix, we call $l<L$ the address register, with $\log L$ qubits. The system register is $i<N$, and each permutation $π_l$ is a bijection on the computational basis $[0,N-1]$ (denoting the set $\{0,\ldots,N-1\}$).

\subsubsection{Writing the permutation into an ancilla register}

\paragraph{General permutation}
A promising way of implementing a general permutation is using $\log N$ ancilla qubits.  A lookup table indicated as LUT in the below quantum circuit with both $l$ and $i$ as address register loads $π_l(i)$ into the ancilla register. After swapping with Cliffords, the ancilla register must be uncomputed. This can be performed by measurement based uncomputation, where a $\pm1$ phase must be fixed depending on the X-basis measurement outcomes \cite{berry_qubitization_2019}:
\begin{equation}
\begin{tikzpicture}
\begin{yquant}
qubit {$\ket{l}$} l;
qubit {$\ket{i}$} i;
qubit {anc $\ket{0}$} anc;

["north:$\log_2 L$" {font=\protect\footnotesize, inner sep=0pt}]    
slash l;

["north:$\log_2 N$" {font=\protect\footnotesize, inner sep=0pt}]    
slash i;

["north:$\log_2N$" {font=\protect\footnotesize, inner sep=0pt}]    
slash anc;

%[operator style={sharp corners}]
controlbox {$l$} l;
controlbox {$i$} i;
%[operator style={rounded corners=5pt}]
box {LUT $π_l(i)$} anc | l,i;

swap (i,anc);

[direct control]
measure {$X$} anc;
box {fixup} (l,i) | anc;
discard anc;

output {$\to \ket{l}$} l;
output {$\to \ket{π_l(i)}$} i;

\end{yquant}
\end{tikzpicture}
\label{eq:general perm circuit}
\end{equation}
The lookup-table (LUT) can be implemented with unary iteration, or, using more ancillas but saving on Toffolis, with a QROAM/Select-Swap variant. Unchanged input qubit registers of gates have been indicated with rounded corners. The fixup essentially loads one bit indicating $\pm1$ (depending on the measurement outcomes) for each of the computational basis states $LN$ and imparts that phase. As shown in \cite[Appendix~C]{berry_qubitization_2019}, the cost of the fixup is almost identical to that of a LUT with inDimension=$LN$ and outBits=1, incl.~possible space time tradeoffs. Note that any uncomputation that may be required for any of the initial LUT's internal ancillas can be included for free in the fixup via measurement based uncomputation.

The non-Clifford cost is that of the LUT and fixup-LUT in the circuit:
\begin{equation}
    \text{LUT}(\text{inDim=$LN$, outBits=$\log_2 N$}) + \text{LUT}(\text{inDim=$LN$, outBits=1}).
\end{equation}
The ancilla cost is $\log_2 N$ plus that needed for the LUTs, which could be minimal (eg.~using a unary iteration with $\log^* LN$ ancilla qubits \cite{khattar2025rise}) or make use of a more substantial number of clean or dirty ancillas to reduce Toffoli count \cite{low2024trading}.

A great advantage of this permutation implementation is that the matrix element values can be loaded as part of the same LUT, leading to efficiencies in data loading.

\paragraph{Permutations close to the identity}
In some cases, all permutations could be close to the identity, i.e.~
\begin{equation}
    D_0\le π_l(i) -i < D_1\ \forall l<L,i<N
\end{equation}
with $-N \le D_0 < D_1 <N$ and maximum bandwidth $D:=D_1-D_0$. This situation happens in the 1-sparse decomposition encoding if all matrix values are located near the diagonal. In that case, the circuit cost can be reduced by using a smaller ancilla register of $\log_2D$ qubits only. The LUT looks up the difference $π_l(i)-i$ (if an offset from 0 is required to fit into the $\log_2D$ register, it can be included as quantum addition of a constant in the circuit). The difference is then added\footnote{Negative numbers can be loaded in 2s complement by the LUT and will give the anticipated result in the $i$ register. The required run of 1s in the prefix for negative numbers beyond the $\log_2 D$ qubits present can be simulated by reusing the most significant bit for all higher $\log_2 N - \log_2 D$ bits when performing the addition.} into the $i$ register with a quantum addition.
\begin{equation}
\begin{tikzpicture}
\begin{yquant}
qubit {$\ket{l}$} l;
qubit {$\ket{i}$} i;
qubit {anc $\ket{0}$} anc;

["north:$\log_2 L$" {font=\protect\footnotesize, inner sep=0pt}]    
slash l;

["north:$\log_2 N$" {font=\protect\footnotesize, inner sep=0pt}]    
slash i;

["north:$\log_2D$" {font=\protect\footnotesize, inner sep=0pt}]    
slash anc;

%[operator style={sharp corners}]
controlbox {$l$} l;
controlbox {$i$} i;
%[operator style={rounded corners=5pt}]
box {LUT $π_l(i) -i$} anc | l,i;

controlbox {$x$} anc;
box {$+x$} i | anc;

[direct control]
measure {$X$} anc;
box {fixup} (l,i) | anc;
discard anc;

output {$\to \ket{l}$} l;
output {$\to \ket{π_l(i)}$} i;

\end{yquant}
\end{tikzpicture}
\end{equation}
Compared to the general circuit \eqref{eq:general perm circuit}, it has an extra addition costing $\log N$ Toffolis \cite{gidney2018halving}. Yet, the ancilla register and LUT outBits can be potentially significantly smaller if $D<N$. The non-Clifford cost is
\begin{equation}
    \text{LUT(inDim=$LN$, outBits=$\log_2D$)} + \text{LUT(inDim=$LN$, outBits=1)} + \text{Add(bits=$\log_2N$)}
\end{equation}
Ancilla qubit requirement, beyond internal ancillas in above circuit elements, is the $\log_2D$ register.

\paragraph{Permutations with a limited number of offsets}
Perhaps, the permutation only has a small number ($D$) of distinct offsets $\{Δ_0,...Δ_{D-1}\}$:
\begin{equation}
    |\{π_l(i)-i,\ \text{for}\ l<L,i<N\}| = |\{Δ_0,…,Δ_{D-1}\}| = D.
\end{equation}
This situation corresponds to matrices that have a band structure, with only certain diagonals populated. 

Then the corresponding offset indices can be loaded, and a unary iteration sums the corresponding offset into the system register.
\begin{equation}
\begin{tikzpicture}
\begin{yquant}
qubit {$\ket{l}$} l;
qubit {$\ket{i}$} i;
qubit {anc $\ket{0}$} anc;

["north:$\log_2 L$" {font=\protect\footnotesize, inner sep=0pt}]    
slash l;

["north:$\log_2 N$" {font=\protect\footnotesize, inner sep=0pt}]    
slash i;

["north:$\log_2D$" {font=\protect\footnotesize, inner sep=0pt}]    
slash anc;

%[operator style={sharp corners}]
controlbox {$l$} l;
controlbox {$i$} i;
%[operator style={rounded corners=5pt}]
box {LUT $x$ with $Δ_x =π_l(i) -i$} anc | l,i;

controlbox {$x$} anc;
box {$+Δ_x$} i | anc;

[direct control]
measure {$X$} anc;
box {fixup} (l,i) | anc;
discard anc;

output {$\to \ket{l}$} l;
output {$\to \ket{π_l(i)}$} i;

\end{yquant}
\end{tikzpicture}
\end{equation}
The cost is as above except for an additional unary iteration over the ancilla register, and addition of the classical integer $Δ_x$ into the system register at each step of the unary iteration:
\begin{align}
    &\text{LUT(inDim=$LN$, outBits=$\log_2D$)} + \text{LUT(inDim=LN, outBits=1)}\\
    &{}+ \text{UnaryIt(inDim=$D$)} + D\cdot\text{Add(bits=$\log_2N$)}
\end{align}

\subsubsection{Using controlled transpositions}
An alternative approach to synthesising permutations is decomposing them into
transpositions (or other small cycles) and implementing all required transpositions in turn. The transpositions must then be multiplexed. This is generally more expensive than the methods presented above: A generic
permutation of $N$ points requires $N-\#\text{cycles}\approx N$ transpositions, and
each transposition of two arbitrary computational basis states requires a multi-controlled Toffoli gate with
$\log_2N-1$ controls. Together, for a single unmultiplexed permutation, $\Theta(N\log N)$ non-Clifford gates are required, see for example \cite{malvetti_quantum_2021,shende_synthesis_2006}.
%App.~D in \cite{malvetti_quantum_2021}, Theorem 33 in \cite{shende_synthesis_2006}.

In order to implement multiplexed permutations by transpositions, $\approx LN$  total transposition gates are required. In addition to the Toffoli with $\log_2N-1$ controls, they must also each be controlled on the multiplex address register (or a unary iterator thereof). Both introduce extra logarithmic factors that are absent in the above method. It also loses the opportunity of LUT (QROAM) space--time tradeoff as each transposition requires its own decomposition.

Transposition-based synthesis can become preferable when each
permutation moves only a small number of basis states, i.e.~there are many cycles of length 1 that do not have to be implemented.

\subsection{Sparse state preparation}
\label{subapp-ssp}
Before turning to multiplexed sparse state preparation, let us review a single sparse state preparation. A sparse state is a state $\ket{ψ}$ on $\log N$ qubits that only has $S<N$ coefficients in the computational basis that are non-zero:
\begin{equation}
\ket{\psi} = \sum_{x=0}^{S-1}a_{x}\ket{π(x)},\ π(x)\in[0,N-1]
\end{equation}
Here, $π(x)$ is a permutation whose action is only relevant on $x\in[0,S-1]\subsetneq[0,N-1]$.
The state of the art for (unmultiplexed) sparse state preparation \cite{rupprecht2026sparse} combines a dense state preparation with $S$ amplitudes on a subset of the $\log N$ qubits with a permutation:
\begin{equation}
\begin{tikzpicture}
\begin{yquant}
qubit {$\ket{0}$} a;
qubit {$\ket{0}$} b;

["north:$\log_2 (N/S)$" {font=\protect\footnotesize, inner sep=0pt}]    
slash a;

["north:$\log_2 S$" {font=\protect\footnotesize, inner sep=0pt}]    
slash b;

box {Dense state prep} b;

box {Permutation $π$} (a,b);

output {$\to$ sparse state $\ket{\psi}$} (a,b);
\end{yquant}        
\end{tikzpicture}
\label{eq:sparse state preparation circuit}
\end{equation}
The circuit first prepares a dense state on the first $S$ amplitudes and then permutes the computational basis:
\begin{equation}
    \ket{0}^{\otimes\log_2N} \xrightarrow{\text{Dense state prep}} \sum_{x=0}^{S-1}a_x\ket{x} \xrightarrow{\text{Permutation } π} \sum_{x=0}^{S-1}a_x\ket{π(x)} = \ket{\psi}.
\end{equation}
The circuit ``Permutation $π$'' must only be constructed up to a unitary on the bottom $\log_2S$ qubits, which can be absorbed into the sparse state preparation. In \cite{malvetti_quantum_2021} and \cite{rupprecht2026sparse}, the Permutation isometry circuit is constructed backwards, i.e.~$π^{-1} =π^\dag$, and the circuit then reversed.
Both methods use CNOT and multicontrolled NOT gates to clear the extra $\log_2(N/S)$ qubits to $\ket{0}$, basis state by basis state. The latter method further uses a batching procedure \cite{rupprecht2026sparse}. Multiple multicontrolled not gates are replaced by a partial unary iteration (PUI) circuit, saving on Toffolis. The achieved headline Toffoli cost (neglecting lower order terms) for the permutation is independent of $N$:
\begin{equation}
    \text{SSPPerm($N,S$)} \approx 2S.
    \label{eq:SSPPerm cost}
\end{equation}
This SSPPerm (Sparse-State-Preparation-Permutation) is not amenable to a space-time tradeoff as in QROAM, as there is no LUT involved.

Implementing the permutation by writing $π(x)$ to an ancilla register with a LUT as in \eqref{eq:general perm circuit} would be more expensive. The fixup would acquire a LUT(inDim=$N$,outBits=1) dependence on $N>S$.

\subsection{Multiplexed sparse state preparation}
Finally, we show a new method to prepare multiplexed sparse states, i.e.~a family $\ket{ψ_l},\ l<L$ of sparse states. The dense state preparation in \eqref{eq:sparse state preparation circuit} can easily be multiplexed, by multiplexing all the LUTs inside the dense state preparation.

The permutation of \cite{rupprecht2026sparse} in \eqref{eq:sparse state preparation circuit} can be straightforwardly multiplexed to $π_l$ by running a global unary iteration on the $l$ address. In each step of the unary iteration, the usual circuit for the permutation $π_l$ can be performed, controlled on the unary iterator. This further control promotes CNOT gates to Toffoli gates, leading to expensive non-Clifford cost.
Instead, we adopt an approach where we have a global unary iteration on the $l$ register, but only control the partial unary iterations (PUI) on the unary iterator qubit. This control only adds an extra Toffoli gate per PUI; CNOT gates stay CNOT gates. When determining the required circuit for $π_{l}$, the effect of all previous gates from prior $π_{l'}, l'<l$ must be taken into account, as they can have changed the extra $\log_2(N/S)$ qubits' state. However, conversely, the batch-building circuit for $π_l$ will never change the action of $l'<l$, because the extra qubits are all zero for $l'<l$, and no CNOT/Toffoli gates are fired. The PUIs don't affect $l'<l$ as they are controlled on the unary iterator.

Therefore, the total Toffoli cost of a multiplexed sparse state preparation is
\begin{equation}
    \text{MultiStatePrep(inDim=$L$,stateDim=$S$)} +\text{UnaryIt(inDim=$L$)} + L\lceil S/m\rceil +  L\cdot \text{SSPPerm($N,S$)},
\end{equation}
where $L\lceil S/m\rceil$ is the total number of PUIs, and $m$ is the number of batches in \cite{rupprecht2026sparse}.

The multiplexed permutations SSPPerm implemented with this method are not amenable to a space-time tradeoff like QROAM, as all $L$ permutations $π_l$ must be implemented in sequence. 

\subsubsection{Multiplexed sparse state preparation for banded matrices}

If the desired block-encoded matrix is banded, i.e.~only $S=D$ bands ($D_0 \le i-j < D_1, D=D_1-D_0$ along the diagonal are populated, the sparse states required for the $q$-norm encoding are structured. In this case, the $S$ populated computational basis states form a continuous run of integers. The multiplexed permutation $π_l$ can be implemented more efficiently than as above. In fact, they can be implemented by addition of $j+D_0$ or $i-D_0$, respectively. Then the cost of multiplexed sparse state preparation is only
\begin{equation}
    \text{MultiStatePrep(inDim=$L$,stateDim=$D$)} + 2\text{Add(bits=$\log_2 N$})
\end{equation}

\subsection{Comparing the two block encodings}
\label{sec:comparison}

Both block encodings are built from the primitives above. In this section, we analyse and compare their total cost, focusing on gate count. Subnormalisation has been studied numerically (see Fig.~\ref{fig:qnorm ratio}), and it appears 1-sparse decomposition has roughly a factor of 1.5 higher subnormalisation.

Table~\ref{tab:be-comparison} provides a summary of the comparison. It shows that the 1-sparse decomposition has roughly $4\times$ lower Toffoli cost than the $q$-norm encoding. Its advantage increases as Toffolis are traded for (clean or dirty) ancilla qubits, as the $q$-norm encoding does not allow such a trade-off.

For banded real matrices that are dense within their bandwidth, both block encoding methods have a similar leading order $T$ cost and allow a space-time tradeoff. The 1-sparse decomposition is therefore especially suited to unstructured matrices.

First, we will give a reminder of the circuits of the two block encoding methods. Then we will analyse the dominant costs, explaining each line in Table~\ref{tab:be-comparison} in detail.

\begin{table}[t]
\centering
\begin{tabular}{l|ll}
 & 1-sparse decomposition & $q$-norm encoding \\\hline
Dominant element
  & multiplexed permutation
  & $2\times$ multiplexed sparse state prep. \\
Toffolis, minimal ancillas
  & $\approx2SN$
  & $\approx 8SN$ (general matrix) \\
  &&$\approx4SN$ (real matrix and optimistic) \\
  \qquad for banded matrices & $\approx 2SN$ & $\approx 3SN$ \\
  & & $\approx 2SN$ (real matrix) \\
Extra ancillas
  & $\log_2N$ (or $\log_2D$)
  & - \\
Space--time tradeoff possible
  & all dominant elements
  & dense state preparation layer only \\
\end{tabular}
\caption{Leading-order comparison of the two block encodings. $S$ denotes the
sparsity and $N$ the dimension. $D$ denotes a possible restricted bandwidth of
the matrix. Optimistic refers to using SSPPerm$=S$ instead of \eqref{eq:SSPPerm cost}, as is usually observed in \cite{rupprecht2026sparse}.}
\label{tab:be-comparison}
\end{table}

\paragraph{1-sparse decomposition encoding}
In order to compare the main circuit costs, we first give a reminder of the circuit implementations.

For the 1-sparse decomposition encoding into $L=S$ 1-sparse matrices, the arbitrary permutation circuit is supplemented by loading the values of matrix elements within the same LUT, and using the phase gradient technique to implement required rotations.

\begin{equation}
\begin{tikzpicture}
\begin{yquant}
qubit {flag (LCU)} l;
qubit {system $\ket{i}$} i;
qubit {anc $\ket{0}$} anc;
qubit {anc $\ket{0}$} angle;
qubit {flag} rot;

["north:$\log_2 L$" {font=\protect\footnotesize, inner sep=0pt}]    
slash l;

["north:$\log_2 N$" {font=\protect\footnotesize, inner sep=0pt}]    
slash i;

["north:$\log_2N$" {font=\protect\footnotesize, inner sep=0pt}]    
slash anc;

["north:$\log_21/\epsilon$" {font=\protect\footnotesize, inner sep=0pt}]    
slash angle;

box {Prepare $\ket{\psi}$} l;

%[operator style={sharp corners}]
controlbox {$l$} l;
controlbox {$i$} i;
%[operator style={rounded corners=5pt}]
box {\Ifcase\idx LUT $\pi_l(i)$\Else LUT  $a_{l,i}/a_l$\Fi} anc, angle | l, i;

controlbox {angle} angle;
box {$R(\text{angle})$} rot | angle;

swap (i,anc);

[direct control]
measure {$X$} anc,angle;
box {fixup} (l,i) | anc,angle;
discard anc;
discard angle;

box {Unprepare $\ket{\psi}$} l;

output {$\to \ket{π_l(i)}$} i;

\end{yquant}
\end{tikzpicture}
\end{equation}
If the LUT is implemented with a unary iteration, ancilla qubits can be saved by directly performing $R($angle$)$ at each step of the unary iteration.

The dominant cost of the 1-sparse decomposition circuit are the LUTs for implementing the multiplexed permutation and loading the angles, and inside the fixup.
\begin{equation}
    \underbrace{\text{LUT(inDim=$SN$, outBits=$\log_2N$+$\log_21/ε$)}}_{\text{permuted index and angles}}
    +\ \underbrace{\text{LUT(inDim=$SN$, outBits=1)}}_{\text{fixup}}
\end{equation}
For minimum ancilla count, unary iteration (QROM) can be used for the LUTs. The cost is then independent of outBits; the total 1-sparse circuit Toffoli cost is then $2SN$.
A space time tradeoff is possible for the LUTs, reducing Toffoli cost at the expense of additional ancillas.

\paragraph{$q$-norm encoding}

For the $q$-norm encoding, the circuit consists of two multiplexed sparse state preparations.

\begin{equation}
\begin{tikzpicture}
\begin{yquant}
qubit {extra flag} extra2;
qubit {flags} flag;
qubit {extra flag} extra;
qubit {sys $\ket{j}$} sys;

["north:$\log_2 N$" {font=\protect\footnotesize, inner sep=0pt}]    
slash flag;
["north:$\log_2 N$" {font=\protect\footnotesize, inner sep=0pt}]    
slash sys;

controlbox {$j$} sys;
box {Prepare $\ket{\psi_j}$} (flag, extra2) | sys;

swap (sys, flag);

controlbox {$i$} sys;
box {Unprepare $\ket{\phi_i}$} (flag, extra) | sys;

output {$\ket{i}$} sys;
\end{yquant}
\end{tikzpicture}
\end{equation}

As discussed above, the multiplexed sparse state preparations are split into multiplexed dense state preparations followed by a multiplexed permutation.

The multiplexed dense state preparation is a Grover--Rudolph style circuit with
one angle loading LUT per qubit of the $\log_2S$ subregister, each multiplexed on the
$N$ addresses, so that the $j$th lookup has inDim $N2^j$. In each step the angle is imparted using a phase-gradient addition. A final LUT is required for the phases. Measurement based uncomputation of the angle register can be included in the PUIs of the permutation \cite{rupprecht2026sparse}. At minimum ancilla count, this leads to a Toffoli cost of 
\begin{equation}
    \text{MultiStatePrep(inDim=$N$,stateDim=$S$)} = \sum_{j=0}^{\log_2S-1}\left(
    \underbrace{\text{LUT(inDim=$N2^j$)}}_{\text{load angle}}
    + \underbrace{\text{cADD(bits=$b$)}}_{\text{rotation}}\right)
    \approx 2SN
\end{equation}
per multiplexed state preparation.
One multiplexed sparse state preparation
therefore costs
\begin{equation}
    \underbrace{\text{MultiStatePrep(inDim=$N$, stateDim=$S$)}}_{\approx\,2SN}
    + \underbrace{N\cdot\text{SSPPerm($N,S$)}}_{\approx\,2SN}
    + \underbrace{\text{UnaryIt(inDim=$N$)} + N\lceil S/m\rceil}_{\text{subleading}}
    \;\approx\;4SN ,
\end{equation}
and the $q$-norm encoding needs two of them. At minimal ancilla count the
1-sparse decomposition is thus cheaper by a factor of roughly 4, with the
dense and permutation layers contributing almost equally to the gap. It pays for this
with $\log_2N$ additional ancilla qubits for the permutation register, or
$\log_2D$ when the matrix has bandwidth $D$.

The estimate $\text{SSPPerm}(N,S)\approx2S$ is a worst case, and on random states the
observed cost is closer to $S$ \cite{rupprecht2026sparse}. For real states, the MultiStatePrep cost is also reduced to $SN$, as the final layer for phases can be omitted. These two effects can reduce the 1-sparse encoding's advantage to a factor of $2$.

The comparison is no longer a constant factor once clean or dirty ancillas are
available, because only one of the two encodings can use them throughout. In
the 1-sparse decomposition every dominant element is a LUT, so both the
permutation LUT and its fixup accept the QROAM/Select-Swap tradeoff
\cite{low2024trading}. In the $q$-norm encoding the multiplexed dense state
preparations accept it as well, but the permutation layer does not. As established in the previous section, neither the (unmultiplexed) SSPPerm nor the multiplexed version allows for a space-time tradeoff.

One might try to escape this by replacing the permutations optimised for sparse
state preparation with the LUT-based construction
\eqref{eq:general perm circuit}, which does admit the tradeoff. However, then the $q$-norm encoding would clearly be more expensive: A single multiplexed sparse state preparation built that way
already costs more than a single multiplexed permutation as it also needs the dense state preparation. And the $q$-norm
encoding needs two of the multiplexed sparse state preparations. 

\paragraph{Banded matrices}

In the case of banded matrices, the circuits can be improved.

In the 1-sparse decomposition, the output size of the LUT is reduced. This does not affect the dominant cost in the minimal ancilla case, but improves the space-time tradeoff.

In the $q$-norm encoding, the multiplexed permutations can be replaced by quantum additions. This makes them subdominant, and at highest order, only the dense state preparation costs remain. However, the uncomputation (fixup) must now be performed explicitly, as they cannot be wrapped into PUI circuits in the now removed permutations. 
A single multiplexed sparse state preparation cost at minimum ancilla count is therefore $\approx 3SN = 2SN$ (multiplexed dense state preparation) $+SN$ (uncomputation fixup).
In the case of real matrices, this reduces to $\approx 2SN = SN + SN$. The dominant T estimate is then on par with the 1-sparse decomposition encoding.
Further, all dominant elements of the $q$-norm encoding then also support a space-time tradeoff.

From this analysis, it appears that the 1-sparse decomposition encoding can be advantageous to the $q$-norm encoding  for unstructured matrices that do not possess a band structure.

In practice, it could be prudent to try multiple approaches, as subnormalisation factors can also vary from instance to instance.

  \section{Generalised circuit derivations}
\label{sec-app-deriv}

In this appendix we give a detailed derivation of the proposed 1-sparse decomposition block encoding circuit variants.
All circuits and derivations use big-endian ordering.

\subsection{Row-column encoding}
\label{subsec-app-rowcol}
Here we derive the general form encoding based on row and column indices in
\Cref{subsec-index-ij}.

\begin{figure}[htb]
  \centering
  \includegraphics[width=.6\textwidth]{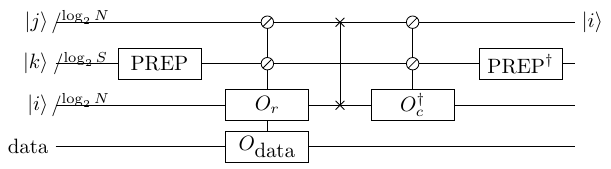}
  \captionsetup{justification=centering}
  \caption{Circuit description for encoding a sequence of 1-sparse matrices using row-column addressing.}
  \label{fig:app-deriv02}
\end{figure}

As shown in \Cref{fig:app-deriv02}, four registers are needed:
a column register $j$ (system qubits), a sparse register $k$ (flag qubits), a row register $i$ (ancilla qubits) and a data (flag) qubit.
The decomposition into 1-sparse matrices is as in \Cref{eqn_1sp_sum} and
the \textsc{Prep} operator as in \Cref{eqn_1sp_prep}.
The row and column registers contain $n = \lceil log_2N \rceil$ qubits each
and the sparse register has $s= \lceil log_2S \rceil$ qubits. 

To simplify the notation, we use:
\begin{equation}
    \hat{A}^{(k)}_{i,j} = A^{(k)}_{i,j}/||A^{(k)}||_{\max}
\end{equation}
for the $(i,j)^{th}$ entry of the $k^{th}$ 1-sparse matrix. 
We further condense our notation by setting $\beta_k=||A^{(k)}||_\text{max}$ to be the max norm of the
$k^{th}$ 1-sparse matrix.

Due to the 1-sparsity, the mappings $(k,j)\mapsto i$ and $(k,i)\mapsto j$ describing the non-zero elements of the 1-sparse matrix $A^{(k)}$ are injective. Empty rows and columns are ignored and the 1-1 nature (for each $k$) of the mappings ensure the ancilla qubits are uncomputed to the $\ket{0}$ state.
Thus,
the three oracles in \Cref{fig:app-deriv02} are:
\begin{equation}
\begin{split}
    O_{r}\ket{0}\ket{k}\ket{j} &= \ket{i}\ket{k}\ket{j} \\
    O_{c}\ket{0}\ket{k}\ket{i} &= \ket{j}\ket{k}\ket{i}\\
    O_{\text{data}}\ket{0}\ket{i}\ket{k}\ket{j} &= \left(\hat{A}^{(k)}_{i,j}\ket{0} + \sqrt{1 - (\hat{A}^{(k)}_{i,j})^2}\ket{1}\right) \ket{i}\ket{k}\ket{j}
    \label{eqn_rc_oracles}
\end{split}
\end{equation}

The full multiplexed compute oracle can be written as:
\begin{equation}
    O_{\text{dr}} = \sum_{k=0}^{S-1} \sum_{j \in A^{(k)}} O_{\text{data}}\otimes O_{r} \otimes\ket{k}\bra{k} \otimes \ket{j}\bra{j}
    \label{eqn_multiplex_Or}
\end{equation}

Applying this oracle to the state $\ket{0} \ket{0}_i \ket{k} \ket{j}$:
\begin{equation}
\begin{split}
    O_{\text{dr}}\ket{0} \ket{0}_i \ket{k} \ket{j} &= 
    \sum_{k'=0}^{S-1} \sum_{j' \in A^{(k)}} O_{\text{data}}\otimes O_{r} \otimes\ket{k'}\bra{k'} \otimes \ket{j'}\bra{j'} . \ket{0}\ket{0}\ket{k}\ket{j} \\ 
    &= \sum_{k'=0}^{S-1} \sum_{j' \in A^{(k)}} O_{\text{data}} \ket{0}\ket{i}\ket{k'}\ket{j'} \delta_{k'k} \delta_{j'j} \\ 
    &= \left(\hat{A}^{(k)}_{i,j}\ket{0} + \sqrt{1 - (\hat{A}^{(k)}_{i,j})^2}\ket{1}\right)\ket{i}\ket{k}\ket{j}
\end{split}
\label{eqn-apply-Or}
\end{equation}

The multiplexed uncompute oracle can be similarly derived. We note that optionally its cost can be reduced by using measurement based uncomputation techniques \cite{berry_qubitization_2019}.
To recover the matrix from the block encoding, the data qubit and the qubits in the $k,i$ registers are initialised to $\ket{0}$. The data qubit and the flag register $k$ are post-selected as $\ket{0}$, 
whereas the uncompute step returns the ancilla $i$ register to the $\ket{0}$ state.
Similarly for the flag and ancilla registers in the following subsections.
The matrix entry, $A^{(k)}_{i,j}$ is 
obtained by initialising the column register to $\ket{j}$ and post-selecting on $\ket{i}$.
Applying the circuit in \Cref{fig:app-deriv02} gives:
\begin{flalign*}
    &\bra{0}\bra{0}_i\bra{0}_k\bra{i}
     \text{PREP}^{\dagger} O^{\dagger}_c \text{SWAP}_{ij} O_{\text{dr}} \text{PREP}
     \ket{0} \ket{0}_i \ket{0}_k \ket{j} \\ 
     = 
    &\frac{1}{\alpha} \sum_{k'=0}^{S-1} \sum_{k=0}^{S-1}
     \sqrt{\beta_{k'}\beta_{k}}
     \bra{0}\bra{j}\bra{k'}\bra{i}
      O^{\dagger}_c \text{SWAP}_{ij} O_{\text{dr}}
     \ket{0} \ket{0}_i \ket{k} \ket{j} \\ 
     =
     &\frac{1}{\alpha} \sum_{k'=0}^{S-1} \sum_{k=0}^{S-1}
     \sqrt{\beta_{k'}\beta_{k}}
     \bra{0}\bra{j}\bra{k'}\bra{i}
      \text{SWAP}_{ij}
     \left(\hat{A}^{(k)}_{i,j}\ket{0} + \sqrt{1 - (\hat{A}^{(k)}_{i,j})^2}\ket{1}\right) 
     \ket{i} \ket{k} \ket{j} \\
     =
     & \frac{1}{\alpha} \sum_{k'=0}^{S-1} \sum_{k=0}^{S-1}
     \sqrt{\beta_{k'}\beta_{k}}
     \bra{i}\bra{k'}\bra{j}
     \hat{A}^{(k)}_{i,j}
     \ket{j} \ket{k} \ket{i} \\
     =
     & \frac{1}{\alpha} \sum_{k'=0}^{S-1} \sum_{k=0}^{S-1}
     \sqrt{\beta_{k'}\beta_{k}}
     \hat{A}^{(k)}_{i,j} \delta_{k'k} \\
     =
     & \sum_k \frac{1}{\alpha} ||A^{(k)}||_{\max} \hat{A}^{(k)}_{i,j}
     = \sum_k \frac{1}{\alpha}A^{(k)}_{i,j} = \frac{1}{α} A_{i,j}
\end{flalign*}

To reduce the number of tensor products with the identity matrix, the
above derivation has implicitly assumed the qubits associated with each
operation.

%
% diag-col encoding
% -----------------
\subsection{Offset-column encoding}
\label{subsec-app-diagcol}
Here we derive the general form of the encoding based on offsets from the
diagonal in \Cref{subsec-diag-offset}.

\begin{figure}[htb]
  \centering
  \includegraphics[width=.6\textwidth]{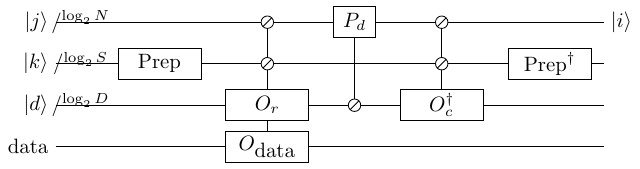}
  \captionsetup{justification=centering}
  \caption{Circuit description for encoding a sequence of 1-sparse matrices using diagonal-column addressing.}
  \label{fig:app-deriv03}
\end{figure}

In this encoding the row and column oracles in \Cref{eqn_rc_oracles} are replaced
by oracles that give diagonal offsets for entries in the $i^{th}$ row and $j^{th}$
column. The SWAP gates are replaced by a permutation operator, $P_d$ which applies the arithmetic offsets listed in the $d$ register.  This gives:
\begin{equation}
\begin{split}
    O_{r}\ket{0}\ket{k}\ket{j} &= \ket{d}\ket{k}\ket{j} \\
    O_{c}\ket{0}\ket{k}\ket{i} &= \ket{d}\ket{k}\ket{i}\\
    \sum_{d'=0}^{D-1}  \ket{d'}\bra{d'} \otimes I^{\otimes^k} \otimes P_{d'} \ket{d}\ket{k}\ket{j} &= \ket{d}\ket{k}\ket{j+d}
    \label{eqn_dc_oracles}
\end{split}
\end{equation}
where it is taken that the $i$ and $j$ in the above equation refer to the same
entry in the $k^{th}$ 1-sparse matrix and, hence, both implement the same
value of $d$.

The full multiplexed compute oracle, $O_{\text{dr}}$, is defined analogously to \Cref{eqn-apply-Or} giving:
\begin{equation}
\begin{split}
    O_{\text{dr}}\ket{0} \ket{0}_d \ket{k} \ket{j} 
    &= \left(\hat{A}^{(k)}_{i,j}\ket{0} + \sqrt{1 - (\hat{A}^{(k)}_{i,j})^2}\ket{1}\right)\ket{d}\ket{k}\ket{j}
\end{split}
\label{eqn-apply-Ord}
\end{equation}

As above, the matrix entry, $A^{(k)}_{i,j}$ is 
obtained by initialising the column register to $\ket{j}$ and post-selecting on $\ket{i}$, noting that $i=j+d$.
Applying the circuit in \Cref{fig:app-deriv03} gives:
\begin{flalign*}
    &\bra{0}\bra{0}_d\bra{0}_k\bra{i}
     \text{PREP}^{\dagger} O^{\dagger}_c 
     \left(\sum_{d'=0}^{D-1} \ket{d'} \bra{d'} P_{d'}\right) 
     O_{\text{dr}} \text{PREP}
     \ket{0} \ket{0}_d \ket{0}_k \ket{j} \\ 
     = 
    &\frac{1}{\alpha} \sum_{k'=0}^{S-1} \sum_{k=0}^{S-1}
     \sqrt{\beta_{k'}\beta_{k}}
     \bra{0}\bra{0}_d\bra{k'}\bra{i} 
      O^{\dagger}_c 
      \left(\sum_{d'=0}^{D-1} \ket{d'} \bra{d'} P_{d'}\right)
      O_{\text{dr}} \ket{0} \ket{0}_d \ket{k} \ket{j} \\ 
     =
     &\frac{1}{\alpha} \sum_{k'=0}^{S-1} \sum_{k=0}^{S-1}
     \sqrt{\beta_{k'}\beta_{k}}
     \bra{0}\bra{d}\bra{k'}\bra{i}
     \left(\sum_{d'=0}^{D-1} \ket{d'} \bra{d'} P_{d'}\right)
     \left(\hat{A}^{(k)}_{i,j}\ket{0} + \sqrt{1 - (\hat{A}^{(k)}_{i,j})^2}\ket{1}\right) 
     \ket{d} \ket{k} \ket{j} \\
     =
     & \frac{1}{\alpha} \sum_{k'=0}^{S-1} \sum_{k=0}^{S-1}
     \sqrt{\beta_{k'}\beta_{k}}
     \bra{d}\bra{k'}\bra{i}
     \hat{A}^{(k)}_{i,j}
     \ket{d} \ket{k} \ket{j+d} \\
     =
     & \frac{1}{\alpha} \sum_{k'=0}^{S-1} \sum_{k=0}^{S-1}
    \sqrt{\beta_{k'}\beta_{k}}
     \hat{A}^{(k)}_{i,j} \delta_{k'k} \delta_{i,j+d}\\
     =
     & \frac{1}{\alpha} \sum_{k=0}^{S-1}||A^{(k)}||_{\max} \hat{A}^{(k)}_{j+d,j}
     = \frac{1}{\alpha}A_{j+d,j}
\end{flalign*}

%
% list-col encoding
% -----------------
\subsection{List-column encoding}
\label{subsec-app-listcol}

The encoding based on a list of offsets (\Cref{subsec-diag-counter}) is a simple extension of the offset-column
encoding. All occurring offsets $\{d_0,…,d_{L-1}\}$ are indexed by $l\in [0,L-1]$. Rather than loading the offset $d$ of each entry as in the previous circuit, the index $l$ corresponding to the appropriate offset is loaded. This reduces the qubit requirement for $L<D$, when not all offsets occur in the matrix, i.e.~some intermediate diagonals are zero.   The oracles implement:
\begin{equation}
\begin{split}
    O_{r}\ket{0}\ket{k}\ket{j} &= \ket{l}\ket{k}\ket{j} \\
    O_{c}\ket{0}\ket{k}\ket{i} &= \ket{l}\ket{k}\ket{i}\\
    P_{l}\ket{k}\ket{j} &= \ket{k}\ket{j+d(l)}
    \label{eqn_lc_oracles}
\end{split}
\end{equation}
and $d(l)$ is the diagonal offset of the $l^{th}$ item in the list.
The encoded matrix entry is:
\begin{equation}
    \frac{1}{\alpha}A_{j+d(l),j}
\end{equation}

%
% list-col encoding with angle lists
% ----------------------------------
\subsection{List-column encoding with list of angles}
\label{subsec-app-listcol-ang}

Here, we derive the general form of the encoding based on lists for the diagonal offsets and 
for the unique rotation angles from \Cref{subsec_list_rot}.
Similarly to the previous subsection which used a dictionary for the offset values, here we additionally use a dictionary for the values of the matrix elements, using an additional $r$ register to index them by $r=0…(R-1)$.
The lengths of each list are $L$ and $R$ respectively.

\begin{figure}[htb]
  \centering
  \includegraphics[width=.6\textwidth]{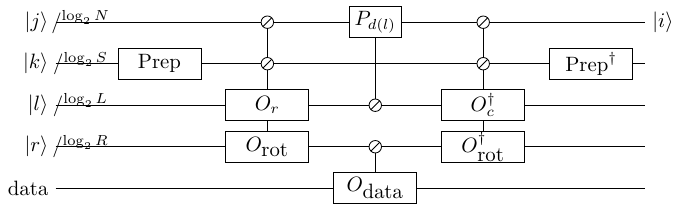}
  \captionsetup{justification=centering}
  \caption{Circuit description for encoding a sequence of 1-sparse matrices using lists for the diagonal offsets and digitised rotation angles.}
  \label{fig:app-deriv04}
\end{figure}

The oracles shown in \Cref{fig:app-deriv04} are:
\begin{equation}
\begin{split}
    O_{\text{rot}}\otimes O_r\ket{0}\ket{0}\ket{k}\ket{j} &= \ket{r}\ket{l}\ket{k}\ket{j} \\
    O_{\text{rot}}\otimes O_c\ket{0}\ket{0}\ket{k}\ket{i} &= \ket{r}\ket{l}\ket{k}\ket{i} \\
    O_{\text{data}}\ket{0}\ket{r} &= \left(\hat{A}_{r}\ket{0} + \sqrt{1 - \hat{A}_{r}^2}\ket{1}\right) \ket{r} \\
    P_{d(l)}\ket{l}\ket{k}\ket{j} &= \ket{l}\ket{k}\ket{j+d(l)}
    \label{eqn_lr_oracles}
\end{split}
\end{equation}
and
\begin{equation}
    \hat{A}_r = A^{(k)}_r/||A^{(k)}||_{\max} = A^{(k)}_r/\beta_k
\end{equation}
where $r$ is the $r^{th}$ repeated entry normalised by the max norm of the 1-sparse matrix $k$ in which it occurs.
The multiplexers on the $O_{\text{rot}}$ oracle ensure that $\hat{A}_r$ is associated with the correct 1-sparse matrix.
Note that some of the unique entries in $A$ are duplicated if they occur in more than one sparse matrix.
Since the max-search places repeated entries in $A$ into contiguous blocks, it will try to collect as many repeated entries into each 1-sparse matrix as possible.
Note that $P_{d(l)}$ in \Cref{fig:app-deriv04} performs the same arithmetic operations as $+d(l)$ in \Cref{fig:sparse-count-rot-gen}.

Using the oracles from \Cref{eqn_lr_oracles} and writing $O_{\text{rr}}$ and $O_{\text{rc}}$ as the multiplexed compute and uncompute oracles, we have:
\begin{flalign*}
    &\bra{0}\bra{0}_r\bra{0}_l\bra{0}_k \bra{i}
     \text{PREP}^{\dagger} 
     O^{\dagger}_{\text{rr}}  
     \left(\sum_{r''=0}^{R-1} \ket{r''} \bra{r''} O_{\text{data}}\right) \otimes
     \left(\sum_{l''=0}^{L-1} \ket{l''} \bra{l''} P_{d(l'')}\right) \\
     & O_{\text{rc}} 
     \text{PREP}
     \ket{0}\ket{0}_r \ket{0}_l \ket{0}_k \ket{j} \\ 
     = 
    &\frac{1}{\alpha} \sum_{k'=0}^{S-1} \sum_{k=0}^{S-1} \sqrt{\beta_{k'}\beta_{k}}
     \bra{0}\bra{r'}\bra{l'}\bra{k'} \bra{i} 
     \left(\sum_{r''=0}^{R-1} \ket{r''} \bra{r''} O_{\text{data}}\right) \otimes
     \left(\sum_{l''=0}^{L-1} \ket{l''} \bra{l''} P_{d(l'')}\right)
      \ket{0} \ket{r} \ket{l} \ket{k} \ket{j} \\ 
     =
    &\frac{1}{\alpha} \sum_{k'=0}^{S-1} \sum_{k=0}^{S-1} \sqrt{\beta_{k'}\beta_{k}}
     \bra{r'}\bra{l'}\bra{k'}\bra{i} 
     \left(\sum_{r''=0}^{R-1} \ket{r''} \bra{r''} \hat{A}_{r''}\right) \otimes
     \left(\sum_{l''=0}^{L-1} \ket{l''} \bra{l''} P_{d(l'')}\right)
      \ket{r} \ket{l} \ket{k} \ket{j}\\    
    =
    &\frac{1}{\alpha} \sum_{k'=0}^{S-1} \sum_{k=0}^{S-1} \sqrt{\beta_{k'}\beta_{k}}
     \bra{l'}\bra{k'}\bra{i} 
     \hat{A}_{r} \otimes
     \left(\sum_{l''=0}^{L-1} \ket{l''} \bra{l''} P_{d(l'')}\right)
    \ket{l} \ket{k} \ket{j}\\    
    =
    &\frac{1}{\alpha} \sum_{k'=0}^{S-1} \sum_{k=0}^{S-1} \sqrt{\beta_{k'}\beta_{k}}
     \bra{k'}\bra{i} 
     \frac{A^{(k)}_{r}}{\beta_k} 
     \ket{k} \ket{j+d(l)}\\    
    =
    & \frac{1}{\alpha} \sum_{k=0}^{S-1} A^{(k)}_{r} \delta_{i,j+d(l)} \\
    =
    & \frac{1}{\alpha} A_{j+d(l),j}
\end{flalign*}

Effectively, the circuit selects the $l^{th}$ value from the list of
values and the $\delta$ function places it in the correct position in
the matrix, determined by the offset $d(l)$. 
Since each entry in $A$ occurs in one and only one $A^{(k)}$, 
$\delta_{i,j+d(l)}$ uniquely identifies the correct 1-sparse matrix.

  \section{Frontal behaviour of SPAI}
\label{sec-spai-front}

In the main text, we have made two observations regarding increasing levels of SPAI preconditioning. Firstly, the subnormalisation stays approximately constant, see Figs.~\ref{fig_subn_af2D}--\ref{fig_subn_duct3d}. Secondly, the matrix element values fall into fewer numbers of bins, see Table~\ref{tab-smooth-digit}, where $n_b$ decreases as the infill level increases. This appendix provides theoretical justifications for both of these observations. 

For a SPAI preconditioner, $P_k$ with $k$ levels of infill and a sparsity pattern $S_k$ we have \cite{saad2003iterative}:

\begin{equation}
    P_k = \underset{P\in S_k} {\arg\min} ||I-PA||_F
    \label{eqn-app-spai-min}
\end{equation}

By construction, adding an additional level of infill gives $S_{k} \subseteq S_{k+1}$.
Since minimisation over $S_{k+1}$ includes all entries in $S_k$, we have:

\begin{equation}
    ||I-P_{k+1}A||_F \le ||I-P_{k}A||_F
    \label{eqn-app-spai-ineq}
\end{equation}

Since the diagonal entries of $P_{k+1}A$ and $P_{k}A$ are unity by construction, this reduces to:

\begin{equation}
    \sqrt{\sum_{i \ne j} |(P_{k+1}A)_{i,j}|^2} \le \sqrt{\sum_{i \ne j} |(P_{k}A)_{i,j}|^2}
    \label{eqn-app-spai-off}
\end{equation}

\begin{figure}[ht]
  \centering
  \begin{subfigure}[b]{0.3\textwidth}
      \centering
      \includegraphics[width=0.9\textwidth]{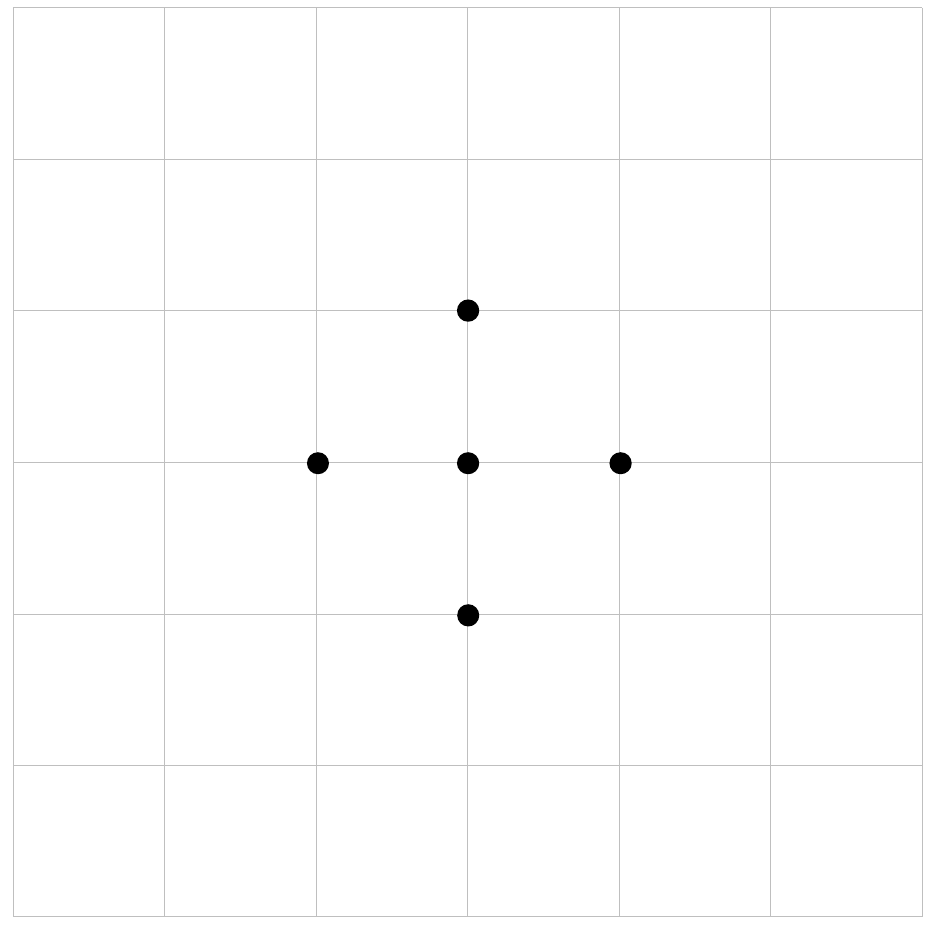}
      \caption{\small CFD Matrix, $A$.}
      \label{fig-infill01}
  \end{subfigure}
  \begin{subfigure}[b]{0.3\textwidth}
      \centering
      \includegraphics[width=0.9\textwidth]{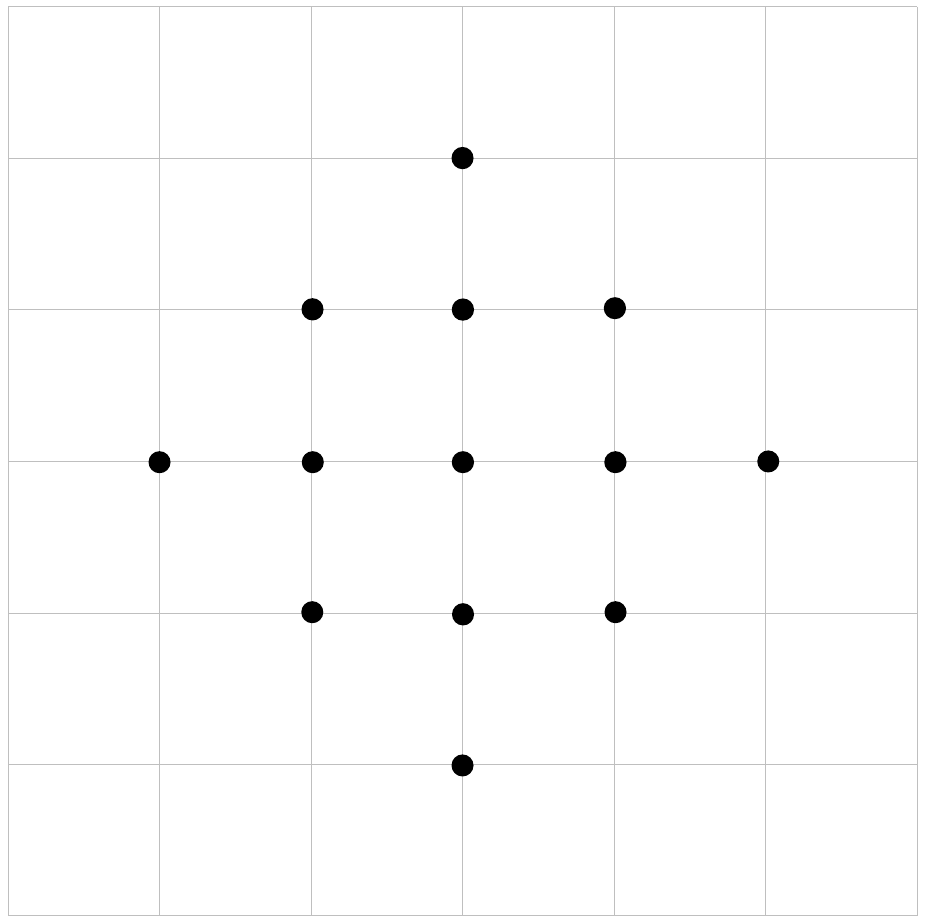}
      \caption{\small Preconditioner, $P$.}
      \label{fig-infill02}
  \end{subfigure}
    \begin{subfigure}[b]{0.3\textwidth}
      \centering
      \includegraphics[width=0.9\textwidth]{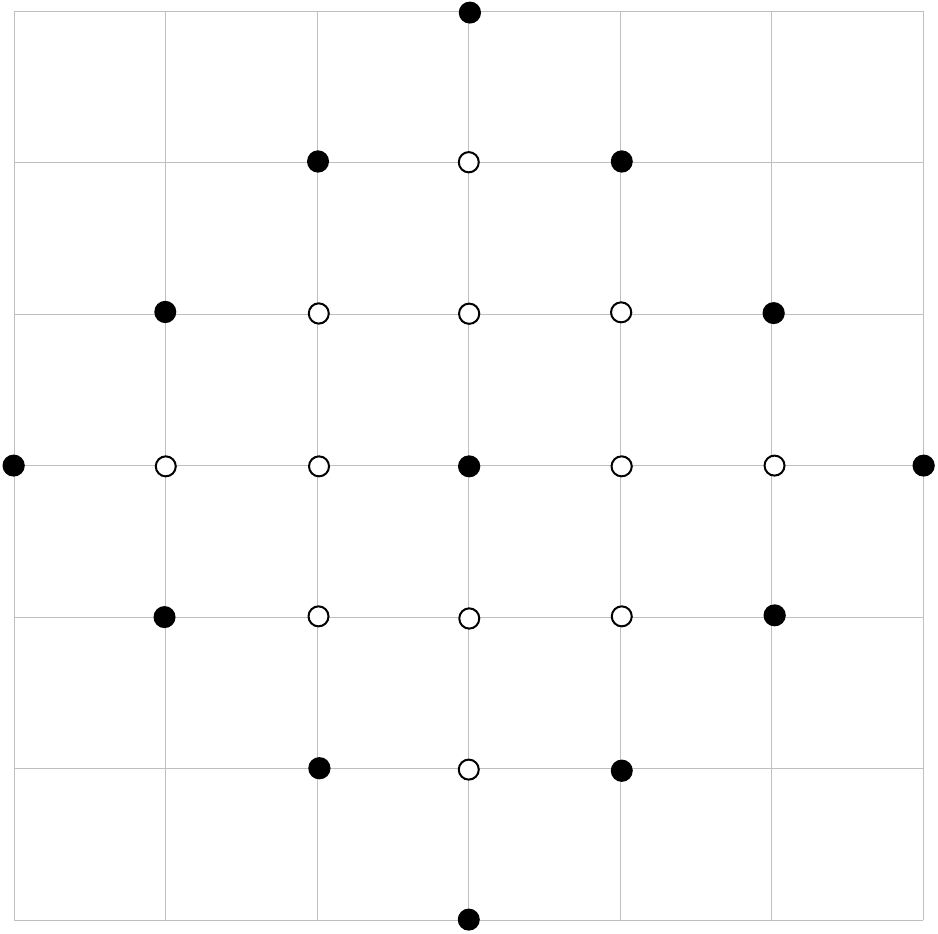}
      \caption{\small Product $PA$.}
      \label{fig-infill03}
  \end{subfigure}
  \caption{\centering Difference stencil for a single node in a 2D structured CFD mesh, the SPAI preconditioner, $P$, with one level of infill and the product $PA$. Solid circles denote non-zero entries in the stencil, open circles denote entries in the stencil for $PA$ that are zero by construction.
  Reproduced from \cite{lapworth2025preconditioned}.}
  \label{fig-infill}
\end{figure}

\Cref{fig-infill}, reproduced from \cite{lapworth2025preconditioned}, shows the
frontal nature of $PA$ when direct matrix inversion is used for the minimisation.
From \cite{lapworth2025preconditioned}, the number of matrix entries in the front
is $4i+8$ where $i$ is the number of infill levels.
For 3D structured meshes the number of matrix entries in the front is $4i^2 + 18i + 19$. 
For unstructured meshes there are not such simple scaling formulae but the
scaling with $\mathcal{O}(i)$ in 2D and $\mathcal{O}(i^2)$ in 3D is the same.

Defining $f_k \subseteq S_k$  as the set of non-diagonal entries in $P_{k}A$ that are non-zero,
\Cref{eqn-app-spai-off} can be written as:

\begin{equation}
    \sqrt{\sum_{(i,j)\in f_{k+1}} |(P_{k+1}A)_{i,j}|^2} \le \sqrt{\sum_{(i,j)\in f_k} |(P_{k}A)_{i,j}|^2}
    \label{eqn-app-spai-front}
\end{equation}

From the above scaling arguments, we expect the cardinality of $f_k$ to be such that:
\begin{equation}
   |f_{k+1}| \ge |f_{k}|
   \label{eqn-app-spai-modf}
\end{equation}

There is one further consideration which is how the front behaves when it
reaches an extremity of the mesh. In this case, the point on the boundary cannot
progress any further but, because it is not enclosed, it remains part of the front.
Strict equality $|f_{k+1}| = |f_{k}|$ is only achieved when the fronts for every mesh point have reached the boundaries. Otherwise the inequality in \Cref{eqn-app-spai-modf} holds. Hence,

\begin{equation}
    \sqrt{\frac{1}{|f_{k+1}|}\sum_{(i,j)\in f_{k+1}} |(P_{k+1}A)_{i,j}|^2} \le
    \sqrt{\frac{1}{|f_k|}\sum_{(i,j)\in f_{k+1}} |(P_{k+1}A)_{i,j}|^2} \le 
    \sqrt{\frac{1}{|f_k|}\sum_{(i,j)\in f_k} |(P_{k}A)_{i,j}|^2}
    \label{eqn-app-spai-front2}
\end{equation}

From \Cref{eqn-app-spai-front} we have that whilst $P_{k+1}A$ has more entries than
$P_{k}A$, and hence more matrices in the 1-sparse decomposition, the sum of the squares of the
entries is lower which mitigates the greater number of 1-sparse matrices and keeps
the subnormalisation value, $\alpha$, roughly constant.
Strictly, the 2-norms in \Cref{eqn-app-spai-front} do not guarantee a reduction in the sum of the 1-norms used to set $\alpha$. However, larger 1-norms in some positions must be offset by lower values elsewhere. If an increased 1-norm value happens to be the largest entry in one of the 1-sparse matrices, this will increase $\alpha$.
Similarly if a lower 1-norm value is the largest entry it will lower $\alpha$. 
In each case, if they are not the maximal entry, they do not affect $\alpha$.

From \Cref{eqn-app-spai-front2} we have that the root mean square (RMS) average of the entries in $P_{k+1}A$ 
is lower than in $P_{k}A$.
The use of RMS values is a common convergence monitor in CFD calculations. Larger values are emphasised by being squared. Hence, a reducing RMS indicates a reduced range of values. For binning, with the same smoothing factor, the reduced range leads to fewer bins
with a larger number of entries in each bin. 
For digitisation with the same number of bits, there are two effects.
Firstly, the reduction in range means that higher value bit patterns are no longer populated and, secondly, the clustering effect leads to more entries with the same bit pattern.

\end{appendices}

\end{document}